\documentclass{cmspaper}
\pdfoutput=1
\usepackage{graphicx}
\usepackage{rotating}

\usepackage{color}

\begin{document}

\newcommand{\degree}{\ensuremath{^\circ}}



  \cmsnote{2008/032}
 \date{18 March 2008}

\title{Performance studies of the CMS Strip Tracker before installation}
 






\begin{Authlist}

W.~Adam, T.~Bergauer, M.~Dragicevic, M.~Friedl, R.~Fr\"{u}hwirth, S.~H\"{a}nsel, J.~Hrubec, M.~Krammer, M.Oberegger, M.~Pernicka, S.~Schmid, R.~Stark, H.~Steininger, D.~Uhl, W.~Waltenberger, E.~Widl
\Instfoot{vienna}{Institut f\"{u}r Hochenergiephysik der \"{O}sterreichischen Akademie der Wissenschaften (HEPHY), Vienna, Austria}
% ...
   \Anotfoot{a}{Also at CERN, European Organization for Nuclear Research, Geneva, Switzerland}
   \Anotfoot{b}{Now at University of Hamburg, Institute for Experimental Physics, Hamburg, Germany}
   \Anotfoot{c}{On leave from IFIN-HH, Bucharest, Romania}
   \Anotfoot{d}{Now at LLR-Ecole Polytechnique, France}
   \Anotfoot{e}{Now at Northeastern University, Boston,  USA}
   \Anotfoot{f}{Now at IFIC, Centro mixto U. Valencia/CSIC, Valencia, Spain}
   \Anotfoot{g}{Now at Universiteit Antwerpen, Antwerpen, Belgium}
  \Anotfoot{h}{Now at ETH Zurich, Zurich, Switzerland}
   \Anotfoot{i}{Also Peking University, China}
   \Anotfoot{j}{Now at Imperial College, London, UK}
   \Anotfoot{k}{Now at Universit\'{e} catholique de Louvain, UCL, Louvain-la-Neuve, Belgium}
   \Anotfoot{l}{Now at Eidgen\"{o}ssische Technische Hochschule, Z\"{u}rich, Switzerland}
   \Anotfoot{m}{Now at University of California, Davis, California, USA}
   \Anotfoot{n}{Now at Kansas University, USA}
   \Anotfoot{o}{Also at Research Institute for Particle and Nuclear Physics, Budapest, Hungary}
   \Anotfoot{p}{Now at University of Liverpool, UK}
   \Anotfoot{q}{Now at Massachusetts Institute of Technology, Cambridge, Massachusetts, USA}
   \Anotfoot{r}{Now at Stanford University, Stanford, California, USA}
   \Anotfoot{s}{Now at Harvard University, Cambridge, Massachusetts, USA}  
   \Anotfoot{t}{Now at University of California, Los Angeles, California, USA}


 

\end{Authlist}


 \begin{abstract}

In March 2007 the assembly of the Silicon Strip Tracker was completed
at the Tracker Integration Facility  at CERN.  Nearly 15\% of the detector
 was instrumented using cables, fiber optics, power supplies, and
electronics intended for the operation at the LHC. A local chiller was used to
circulate the coolant for low temperature operation. In order
to understand the efficiency and alignment of the strip tracker
modules, a cosmic ray trigger was implemented. From March
through July 4.5 million triggers were recorded. This period,
referred to as the Sector Test, provided practical experience with the
operation of the Tracker, especially safety, data acquisition, power, and cooling
systems. This paper describes the performance of the strip system
during the Sector Test, which consisted of five distinct periods
defined by the coolant temperature. Significant emphasis is placed on
comparisons between the data and results from Monte Carlo studies.
   \end{abstract}





\setcounter{page}{2}

\pagebreak
\tableofcontents
\pagebreak

\section{Introduction}

  The CMS Tracker~\cite{TK} is the inner tracking detector built for the CMS experiment at the CERN Large Hadron Collider. 
It is a unique instrument in both size and complexity: it contains two systems based on silicon 
sensor technology, one employing pixels and another using silicon microstrips.  
 The Pixel Detector, which surrounds the beam pipe, consists of three barrel layers and 
four disk detectors, two on each side of the barrel. It contains 64 million detector channels. 
The Silicon Strip  Tracker, the subject of this paper, surrounds the pixel system and consists 
of four major subsystems: the Inner Barrel (TIB), Inner Disks (TID), an Outer Barrel (TOB) 
and two End Caps (TEC). It is the largest silicon detector ever built, 
with almost 10 million sensor channels covering a surface area of over 200\,m$^2$. 
The Silicon Strip Tracker (referred to simply as the Tracker in 
this note) was designed to measure charged particles with 
high efficiency and spatial resolution over wide range of momenta, and to operate with 
minimal intervention for the nominal LHC lifetime of 10 years. Consequently, the quality of 
the construction, and the performance of the components contained within it, are of vital 
importance and must be thoroughly evaluated under realistic conditions before operation at 
LHC.

First experience of the operation and detector performance of the Tracker
was gained during summer 2006, with a small fraction ( 1\%) of the detector
inserted in the CMS experiment. Cosmic rays were detected in the presence of  a 4\,T solenoidal
 field and all
the CMS sub-detectors were read out. The results are summarized in Ref.~\cite{MTCC}. 

Silicon Strip Tracker construction proceeded during 2006 and 2007: a brief account of the integration and assembly processes can 
also be found in~\cite{CERNCourrier}; it was an endeavour shared by the entire 
Tracker collaboration and tasks were distributed throughout much of the world. 
   The final assembly of the Tracker  was carried out in a large, purpose-built, clean area at CERN: the Tracker 
Integration Facility (TIF). This facility was instrumented with a significant fraction of the final 
infrastructure and services needed to operate, control and read out a sector of the Tracker that corresponds to 
15\% of the entire detector: cables, patch panels, optical 
fibers, cooling manifolds and rack mounted power supplies and off-detector electronics were 
installed to allow connection to data acquisition; cooling, control and safety systems were implemented; 
a trigger system was devised to detect cosmic rays traversing the Tracker, and dedicated 
computing resources were provided.

 Following assembly of the sub-system and prior to the installation into CMS, the detector was 
commissioned and operated for several months at the TIF, a period referred to as the Sector Test.   The goals of this test
 included commissioning the active sector of the Tracker under 
realistic cabling and grounding conditions, establishing the data acquisition, and confirming  
stable and safe operation under the supervision of the dedicated monitoring systems. The 
objectives were also to develop and validate monitoring tools, to identify issues in service 
routing or connections, to establish stable and safe running at low temperature, and 
 to demonstrate operating procedures. Finally, the acquisition of cosmic ray data allowed 
 the measurement of the detector performance, the understanding of tracking algorithms, and to perform an initial 
alignment of the active modules. 

 The test progressed in an incremental way, beginning with testing of parts of the 
sub-systems, then proceeding to a test of the barrel systems, and finally incorporating one 
endcap. There were also short system tests to check for any interference between the 
Tracker and a small fraction of the forward pixels: these tests, which are not
described in this note, did not show any interference effects.

 The Sector Test  provided an important  opportunity to evaluate Tracker performance in some detail 
and address any weak points. Several million cosmic ray events were taken under a range of conditions, 
operating the Tracker over a range of temperatures, from $15\,^{\circ}\mathrm{C}$ down to 
$-15\,^{\circ}\mathrm{C}$.

The results obtained from the Sector Test
are described in three dedicated papers. This note documents  results on 
the detector performance while results on track reconstruction and  
alignment are described in Ref.~\cite{TIFTrackingNote} and Ref.~\cite{TIFAlignmentNote}
respectively.

 The Sector Test setup  is described in Section 2. The data sets, reconstruction and processing
are presented in Section 3, while Sections 4 and 5 are concerned with the detector 
performance. Section 6 covers the generation of the cosmic muons
in the Tracker and the simulation of the detector response.
The tuning of key parameters in Monte Carlo simulation studies to match the data is 
the subject of Section 7.


\section{ Sector Test description}
 The Silicon Strip Tracker has a barrel region with four layers in the TIB and six layers in the TOB, and a forward region with three disks in the TID and nine in the TEC on each side of the Tracker. Each TID and each TEC disk is organized in three and seven rings respectively, where each ring corresponds to a different radial range and  different module geometries. For the Sector Test a fraction of the  Tracker was powered and read out using the APV25 front-end chip~\cite{APV,APVres}, selected so as to maximize the crossing of all the different layers or rings  by  cosmic rays. The first quadrant in the Z+ side of the Tracker was chosen. Details of Sector Test sub-detectors are shown in Table\,\ref{tklayout}. Sensors were biased with a voltage of
300\,V for TIB, TID and TOB, and  250\,V for TEC. 

The analog signal from the silicon sensors is amplified, shaped and stored in a 192 element deep analogue pipeline every 25\,ns. A subsequent stage can either pass directly the pipeline signal (Peak mode) or form a weighted sum of three consecutive samples effectively reducing the shaping time to 25\,ns (Deconvolution mode):  only results obtained with APV25 in Peak mode have been thoroughly studied at the Sector Test and therefore no Deconvolution mode results will be
described.  The signal is then multiplexed by the front-end chip APV25. In order to approach the behavior of an ideal CR-RC circuit with a 50\,ns shaper, the APV25 settings  can be optimized
 depending on the  value of input capacitance and operating temperature. Laboratory studies~\cite{APVres} were made to evaluate the optimal settings for specific  geometries over a range of temperatures which were used during the Sector Test.

The signals from each APV25 channel are amplified, converted to light by an Analog Opto Hybrid (AOH)~\cite{AOH} and sent via optical fibers to the Front End Driver board (FED)~\cite{FED, fedres} where they are digitized and further processed, prior to transmission to the central DAQ system. Data can be taken in two different modes: Virgin Raw (VR) or Zero Suppressed (ZS).  In VR mode, all channels are read out with the full 10 bits ADC resolution.  In ZS mode, the FED applies pedestal subtraction, common mode rejection and a fast clustering algorithm, using signal height with a reduced 8 bits resolution: only channels forming a cluster are output.  Almost all  cosmic ray data were taken in Virgin Raw mode since the low cosmic trigger rate did not require  data reduction, and VR running allowed offline optimization of thresholds.

  Clock, trigger signals, and slow control communication with the front-end electronics are managed by the Front End Controller (FEC) boards~\cite{fec} and sent via optical fibers to the Digital Opto Hybrid (DOH)~\cite{doh,dohres}  for each control ring of the Tracker: signals are  distributed by the DOH to every  Communication Control Unit (CCU)~\cite{ccu} in a  control ring. Finally each CCU sends signals to Tracker modules, in particular clock and triggers via a Phase Locked Loop (PLL)  circuit on each module~\cite{pll}.  It receives monitoring  data provided by each module from the Detector Control Unit (DCU) chip~\cite{dcu}. 

FEDs and FECs were read out during the Sector Test via  VME bus, rather than the high speed S-Link interface to be utilised in CMS. The DAQ software developed to configure and manage readout of Tracker data is based on XDAQ applications~\cite{xdaq} which also provided algorithms to commission the detector~\cite{commissioning}.

The services needed  consisted of 277 LV/HV cables, 52 control cables, 73 optical fiber cables  and 32 cooling loops. The electronics racks required  145 power supply modules, 17 power supply control modules, all of which were units from the final system~\cite{paoletti}, 65 FEDs and 8 FECs for a total of 41 control rings.  

\begin{table}
\caption{ \label{tklayout} Composition of the Sector Test with the different subdetector layers/disks.  }
\begin{center}
 \begin{tabular}{|c|c|c|c|} 
\hline
Tracker      & Number of  & Number of &  Fraction \\
Sub-detector &  Modules   &  Channels       & of Tracker  \\
\hline

 TIB    & 438   & 282\,624  & 16\% \\
 TID    & 204   & 141\,312  & 25\% \\
 TOB    & 720   & 476\,460  & 15\% \\
 TEC    & 800   & 483\,328  & 13\% \\

\hline
Total   & 2162   &  1\,383\,424 & 15\%  \\
\hline
\end{tabular}
\end{center}
\end{table}

The cooling plant used for the Sector Test was simpler than the final system and its cooling power limited. 
A minimum operating temperature of $-10\,^{\circ}\mathrm{C}$ was obtained, compared to  $-25\,^{\circ}\mathrm{C}$ in the final system. The temperatures measured at the cooling tubes proved to be very stable with variations of less than $0.1\,^{\circ}\mathrm{C}$.  A test was possible at a temperature of $-15\,^{\circ}\mathrm{C}$ but only by limiting power to half of the Sector Test modules. 

Dry air was supplied inside to maintain a sufficiently low dew point to avoid condensation; it flowed in a semi-hermetic tent  that contained the Tracker.

\subsection{The Tracker Safety and Control Systems}

The Tracker Safety system (TSS)~\cite{Frank} is designed to guarantee protection of the Tracker. It is  self-contained and operates  on  information provided by a thousand hardwired sensors. A system based on Programmable Logic Controllers (PLC) handles the process of monitoring those sensors and taking actions depending on the
monitored values, which consist of digital values of temperatures, humidities and
TIF, or CMS cavern, information. The TSS will interlock the
Tracker power according to  a user defined scheme when either a single temperature sensor or
a group of  sensors varies outside user-defined safety
limits, or when the TIF or cavern systems fail to respond. During the
Sector Test two TSS sections, out of the final six, were fully functional and the Tracker was 
reliably interlocked on several occasions on over-temperature, cooling plant failure,
 power cuts, or other alarm conditions.


The Tracker Control System (TCS)~\cite{Frank,Frey} handles all interdependencies of
control, low and high voltages, as well as fast ramp-downs in case of
higher than allowed temperatures and currents in the detector, or in
case of general unsafe conditions detected in the experiment cavern.  
The size and complexity of the CMS Tracker imposes several demands on the control system to ensure the safety 
and operation of the modules; the  TCS evaluates about  10\,000 power supply parameters, nearly 1000 parameters from TSS and 100\,000 parameters from the DCUs situated on all front hybrids and CCU modules. The TCS should intervene before the TSS system enters into action, 
allowing a gentle switch off of parts under risk, and with a finer granularity than the TSS.
 During the Sector Test about 20-25\% of the final system
was controlled by the TCS/TSS. 

The basic software building block is the commercial SCADA (Supervisory
Control And Data Acquisition) software PVSS (Prozess- visualisierungs-
und Steuerungssystem by ETM~\cite{ETM}). This has been greatly extended by CERN
in a common LHC framework, adapting  it to  the needs of 
individual experiments by incorporating local modifications.  The LHC features also extend
functionalities such as archiving of values and the treatment of
alarms, warnings and error messages.

The Sector Test enabled final implementation of the full TCS to PLC
communication including the downloading of parameters, such as sensor
limits and sensor-group to interlock-grouping, to the PLC.  Complete calibrations inside PVSS 
were established, including all ADC to
physical value conversions. Several additional checks were
introduced from PVSS to PLC programming: for example, a minimal
number of sensors should  participate in interlock voting or limits
required to stay in a sensible region.  A full checkout interlock
routine was developed.

The complexity of the TCS system required a network of PCs for the
distribution of requests. The Sector Test was
the perfect test bed to establish, check, and tune the distribution
needed for the final system as performance needs for the Sector Test were
very close to those of the final system. The main bottleneck was
found to be memory but the OPC performance of the CAEN system was also 
identified as a critical item.

\subsection{Cosmic Muon Trigger}
The cosmic trigger configuration was designed to allow
studies of tracking performance and detector alignment. 
The trigger design was constrained by  space above and below the Tracker;
 in particular clearance below the Tracker  allowed 
only 5 cm lead bricks for filtering low momentum tracks.
 Six scintillators (T1, T2, T3, T4, T5, T6) were placed above the
 Tracker, in a fixed position; below the
Tracker there was initially only one scintillator (B0) mounted on a
movable support structure; later a further set of four
scintillators was added (B1, B2, B3, B4) to increase the trigger acceptance.
The dimensions and the single trigger rates of all scintillators are shown in
Table~\ref{tab:scint-dim}.

The upper scintillator signals were synchronized   by  simple NIM logic using cosmic
rays and put in a logical-OR to obtain a Top-scintillator signal.
This was put in coincidence with the lower scintillator signal.
 The coincidence rates are shown  in Table ~\ref{tab:scint-dim}. 
From the entries it can be seen that efficiency of the upper scintillators was not uniform.

In the same way from  synchronized logical-OR signals of scintillators B1, B2, B3, B4 a Bottom-scintillator
signal was obtained and was put in coincidence with the Top-scintillator signal.
The rates are shown  in Table ~\ref{tab:scint-dim}. They show good
performance: the discriminator threshold allowed to obtain a
very uniform coincidence rate. 

\begin{figure}[p]
	\begin{center}
		(a)
		\includegraphics[width=5.2cm]{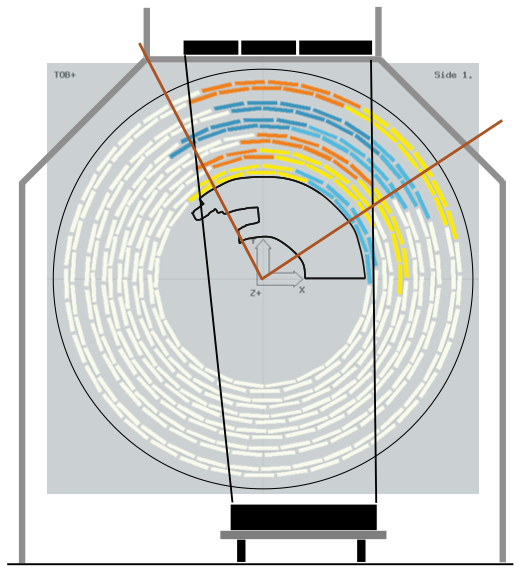}
		\includegraphics[width=10cm]{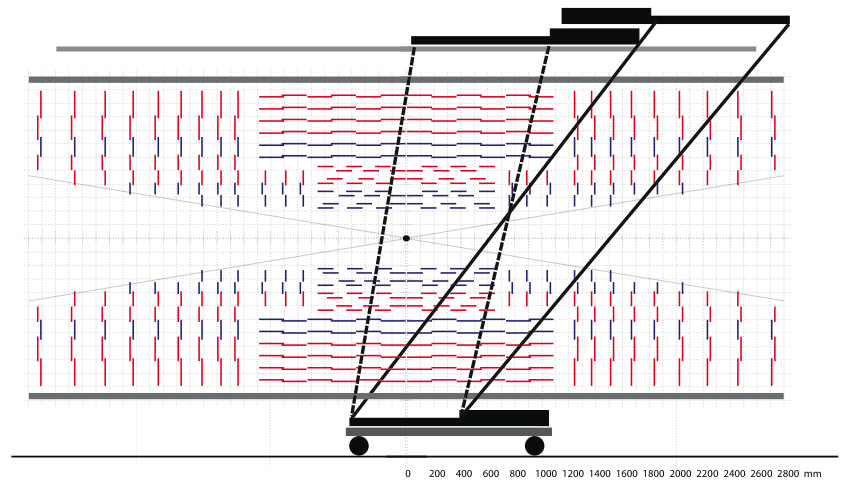} \\
		\vspace{1cm}
		(b)
		\includegraphics[width=5.2cm]{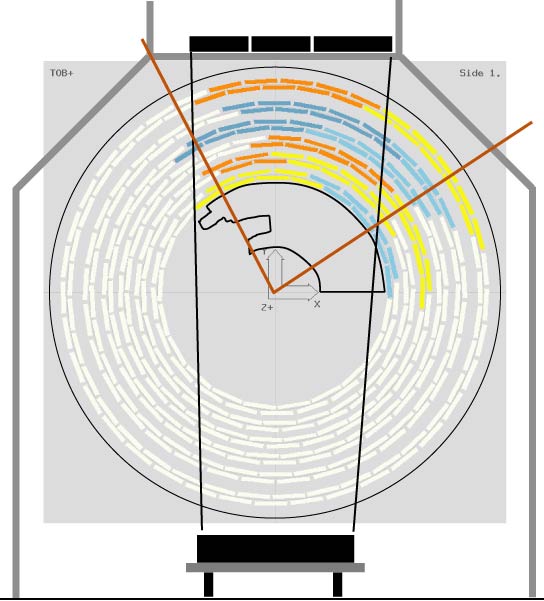}
		\includegraphics[width=10cm]{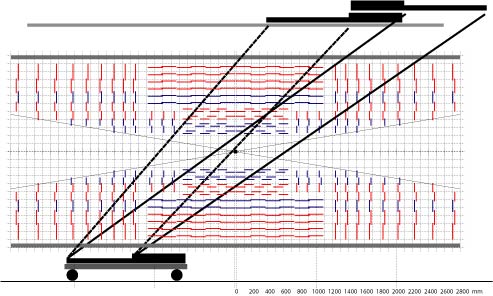} \\
		\vspace{1cm}
		(c)
		\includegraphics[width=5.2cm]{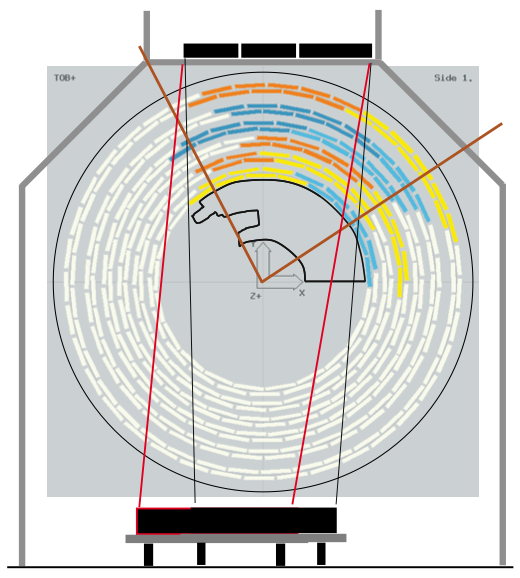}
		\includegraphics[width=10cm]{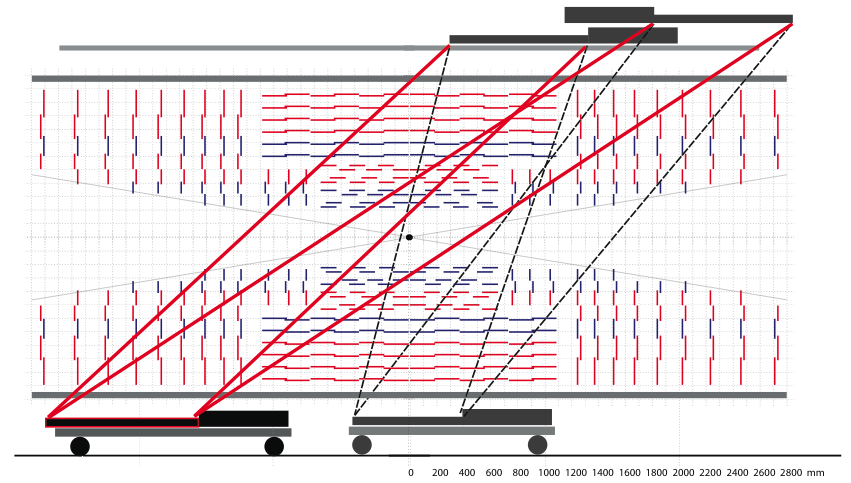}
		\caption{{ Layout of the various trigger scintillator
		configurations used during the cosmic data taking at the
		TIF (in chronological order): (a) configuration
		TA; (b) configuration TB; (c)
		configuration TC. The $xy$ view is shown on the
		left side, the $rz$ view is shown on the right. The
		straight lines connecting the active areas of the top
		and bottom scintillation counters indicate the
		acceptance region.} }
		\label{fig:trigger_layouts}
	\end{center}
\end{figure}

Three main trigger configurations have been used at the Sector Test: (TA)$=$  (Top-Scintillator $\&$ B0) vertical
position; (TB)$=$ (Top-Scintillator $\&$ B0) slanted position; (TC) $=$ (TA) plus 
(Top-Scintillator $\&$ Bottom-scintillator) slanted position. The schematic representations of these
 three configurations are shown in
Fig.~\ref{fig:trigger_layouts}. The rate of spurious coincidences
relative to the true level was of order $10^{-9}$. The trigger rates
achieved were: 3.5\,Hz (TA), 1.5\,Hz (TB), and 6.5\,Hz (TC). Since the
DAQ rate was limited to about 3 Hz by the FED readout via VME,
 a trigger veto was implemented to keep the rate under this level.

\begin{table}
\begin{center}
\caption{
\label{tab:scint-dim}
 { Dimensions of scintillators used at Sector Test; single 
 and coincidence rates.} }
\begin{tabular}{|c|c|c|c|c|c|}
\hline
Scintillator  & DX  & DY   & Thickness & Single Rate& Rate on coincidence  \\
              & (cm)& (cm) & (cm)  & (Hz)   & (T$_i$ $\&$ B0) (Hz) \\
\hline
T1 & 30 & 100  & 0.8 & 80 &  0.42 \\ 
T2 & 30 & 100 & 0.8 & 60 &   0.93 \\
T3 & 40 & 80  & 0.8 & 72 &   1.67 \\
T4 & 30 & 100 & 0.8 & 85 &   0.42 \\
T5 & 30 & 100 & 0.8 & 97  &   0.18 \\
T6 & 40 &  80 & 0.8 & 110 &  0.62 \\
\hline
B0 & 75 & 85 & 2 & 235  & - \\
\hline
Scintillator  & DX   & DY  & thickness & Single Rate & Rate on coincidence        \\
              & (cm) & (cm)& (cm) & (Hz)   & (B$_i$ $\&$ Top-Scint.)(Hz)\\
\hline 
B1 & 40 & 110 & 2 & 260 & 0.45 \\
B2 & 40 & 110 & 2 & 275 & 0.45 \\
B3 & 40 & 110 & 2 & 290 & 0.45 \\
B4 & 40 & 110 & 2 & 240 & 0.45 \\
\hline
\end{tabular}
\end{center}
\end{table}

\section{Data sets and reconstruction}

The division of the data in different sets, is detailed in
Table~\ref{tab:dsets}, where run number intervals are specified for
the participating detectors, the trigger configurations, and the
operating temperatures.  The APV25 parameters listed in the table are for specific hybrid temperatures ($\rm{T}_{APV25}$)
and were studied for TIB and TOB detector modules only. For the TID and the TEC (where a large number of variants exist)
 the APV25 parameters have not been optimized for each module type. Instead  the TID used the
TIB parameters while the TEC used parameters very close to those of
the TOB.

Each run was checked using online and offline data quality monitoring tools. If a run did not meet the quality requirements or if a configuration or hardware problem was discovered, it was flagged as bad and excluded from the offline analysis. 

\begin{table} [htb]
   \begin{center}
  \caption{{  Sector Test data sets.} }
  \label{tab:dsets}
      \begin{tabular} {|c|c|c|c|c|c|c|}
         \hline
         Run          &  Detector     &  Trigger  &  Operating\,T ($^{\circ}\mathrm{C}$) & Total Events & Good Events  &  $\rm{T}_{APV25}$($^{\circ}\mathrm{C}$)                         \\ \hline
         6203-6930    &  TIB TID TOB      &     A     &  15  & 703\,996 &  665\,409        &  $30$   \\ 
         7277-7296    &  TIB TID TOB TEC  &     A     &  14  & 191\,154 &  189\,925        &  $30$      \\ 
         7635-8055    &  TIB TID TOB TEC  &     B     &  14  & 193\,337 &  177\,768        &  $30$      \\ 
         9255-9341    &  TIB TID TOB TEC  &     C     &  14  & 132\,311 &  129\,378        &  $30$      \\ 
         10145-10684  &  TIB TID TOB TEC  &     C     &  10  & 992\,997 &  534\,759        &  $30$   \\ 
         10848-11274  &  TIB TID TOB TEC  &     C     &  -1  & 893\,474 &  886\,801        &  $10$      \\ 
         11316-11915  &  TIB TID TOB TEC  &     C     & -10  & 923\,571 &  902\,881        &  $10$      \\ 
         12045-12585  &  TIB TID TOB TEC  &     C     & -15  & 656\,923 &  655\,301        &  $0$       \\ 
         12599-12656  &  TIB TID TOB TEC  &     C     &  14  & 112\,139 &  112\,134        &  $30$      \\  \hline
      \end{tabular}
    \end{center}
\end{table}


\subsection{Reconstruction}
Event reconstruction, event selection, data quality monitoring, simulation and analysis of the Tracker Sector Test at the TIF were performed within the CMS software framework known as CMSSW\cite{PTDR1}. 
Hit and track reconstruction was performed  offline taking the raw data 
or the simulation data as  input. A reconstruction job is composed of a series of applications executed for each event in the order specified by a configuration file. 
A fundamental part of the processing is the availability of non-event data such as cable map, pedestal, alignment constants and calibration information. 





The first step consists in mapping ADC counts for individual strips as they are coming from the FED output, into objects that are uniquely 
assigned to a specific detector module exploiting the cabling map information stored in the configuration database. 
In the case of data collected without performing online zero-suppression (VR), the necessary pedestal information 
 must be acquired from the database. 
At this point, the input data files, whether real data or Monte Carlo simulated events, contain the same information 
and can be further processed with identical code. 



A three-threshold algorithm, described in detail in the CMS Physics TDR\cite{PTDR1}, is used to form clusters. 
The cluster seed is defined as a strip whose charge is at least three times greater than the strip noise, while  
neighboring strips are added if their charge exceeds  twice their strip noise. 
The cluster is kept if the total cluster charge is more than  five times the cluster noise level, 
defined as the quadratic sum of all the strip noise values.
Finally, the position of the cluster is calculated as the centroid of the individual strip charges.

Three tracking algorithms have been applied: the Combinatorial Track Finder, the Road Search and the  Cosmic Track Finder. The Combinatorial Track Finder and the Road Search are reconstruction algorithms developed for $p-p$ collisions, while the Cosmic Track Finder is a specialized code for  reconstruction of single cosmic track events\,\cite{TIFTrackingNote}.
The input to the tracking algorithms are  reconstructed hits as described above and information from the alignment constants (either from survey or from alignment calibration studies). 

All the tracking algorithms employ the same three steps: seeding, pattern recognition and track fitting. 
The seeding and the pattern recognition are specific to each  algorithm, while the final fit is the same for all of them. 
 Since no magnetic field was present  the tracks are extrapolated as straight lines. Material effects, energy loss and multiple Coulomb scattering are estimated each time a track crosses an active layer. The amount of material at normal incidence is estimated via the reconstruction geometry developed  for $p-p$ collisions.   
Since the momentum of the track is not measured, a constant value of 1\,GeV/$c$ is assigned, close to the expected average from simulation. 

More detailed information about the algorithms and their performance can be found in Ref.\,\cite{TIFTrackingNote}. 
All the results presented in sections 5, 6 and 7 of this paper refer to tracks obtained only with the Combinatorial Track Finder algorithm to avoid duplication of information.


\subsection{Data Quality Monitoring }
The Data Quality Monitoring\,(DQM) system for the Tracker is designed
to ensure that good quality data are recorded and detector problems are
spotted very early on.
The system is based on the
``Physics and Data Quality Monitoring'' framework~\cite{cms_dqm} of
CMS. The task is fulfilled in three steps\,: (a) histograms, called 
Monitoring Elements\,(MEs),  are defined and filled with relevant event
information by the ``Producer''\,(DQM Source) application; (b) a
``Consumer''\,(DQM Client) application accesses the MEs, performs
further analysis and generates alarms; and (c) the ``Graphical User
Interface''\,(GUI) provides tools for visualization of the MEs.

The MEs are defined in the DQM source at various levels of data
reconstruction chain. Starting from the level of pedestal, noise,
digitization, cluster reconstruction and finally track-related
properties are defined and filled in various MEs. 

The DQM Client accesses low-level MEs and performs further analysis on
them and creates summary MEs, performs quality tests comparing MEs
with references to generate alarms. The summary plots are important as
it would be too time consuming to check each of the huge set of MEs
in the Tracker, which consists of over fifteen thousand detector
modules. The information from detector level MEs is accessed and
summarized in MEs at higher levels following the geometrical
structure. Similarly the detector level MEs are compared with
reference MEs or parameters and ``Ok'', ``Warning'',  and ``Error'' alarms
are generated, as appropriate, based on the comparisons.


The DQM system was operational with full functionality during
TIF data taking. Events were accessed from the 
on-line data and full reconstruction was performed on an event by event basis and
provided to the DQM sources. The system was stable during operations
and the Tracker DQM GUI, which is a web based application, was accessible from CERN as well as from other institutes
involved in the Tracker activities.

\subsection{Data Processing}
The tracker data processing consisted of the following steps: data archiving,  conversion of data from the raw format to the CMS Event data Model (EDM) format 
based on ROOT\cite{PTDR1},  registration in the CMS official data bookkeeping services,  transfer to remote computing tiers, data reconstruction and  analysis  in a distributed computing environment.

Data taking and analysis needed a large amount of available disk space, since both raw data and on-site local analysis results had to be stored centrally  in a safe way. 
The design chosen for the computing model was a centralized one, where all the data are stored on a single machine which is connected via NFS to the local network.
A room containing all PCs and local storage was equipped to allow the data taking control, monitoring and first analysis of the data and it is referred to as the 
Tracker Analysis Center (TAC). 

All the data collected by the tracker detector were written into a main storage machine at the TAC, which behaved like a temporary data buffer. Data were backed-up to CASTOR 
storage  at CERN every five  minutes, copying data files only when they did not exist on CASTOR side and had not been accessed in the last hour (to prevent the  copy of data still being modified). Once all the files belonging to a run were copied to CASTOR, a  catalog was prepared for that run after a few checks. 
Because of the limited resources at TAC and following  the CMS computing and 
analysis model, most of the processing of the data was done in remote sites as soon as data were officially published and transferred 
to them. The transfer was performed to remote sites using the CMS official data movement tool PhEDEx
. Tracker data were transferred and registered successfully at Bari, Pisa and FNAL.

The datasets have been reprocessed several times to feed back into the reconstruction phase the improvements in the alignment and
calibration corrections obtained from the analysis. The final reprocessing was  performed in FNAL in December 2007. 


\section{Detector Performance based on Calibration Data}

\begin{figure}[ht]
  \begin{center}
    \includegraphics[width=0.4\textwidth,angle=0]{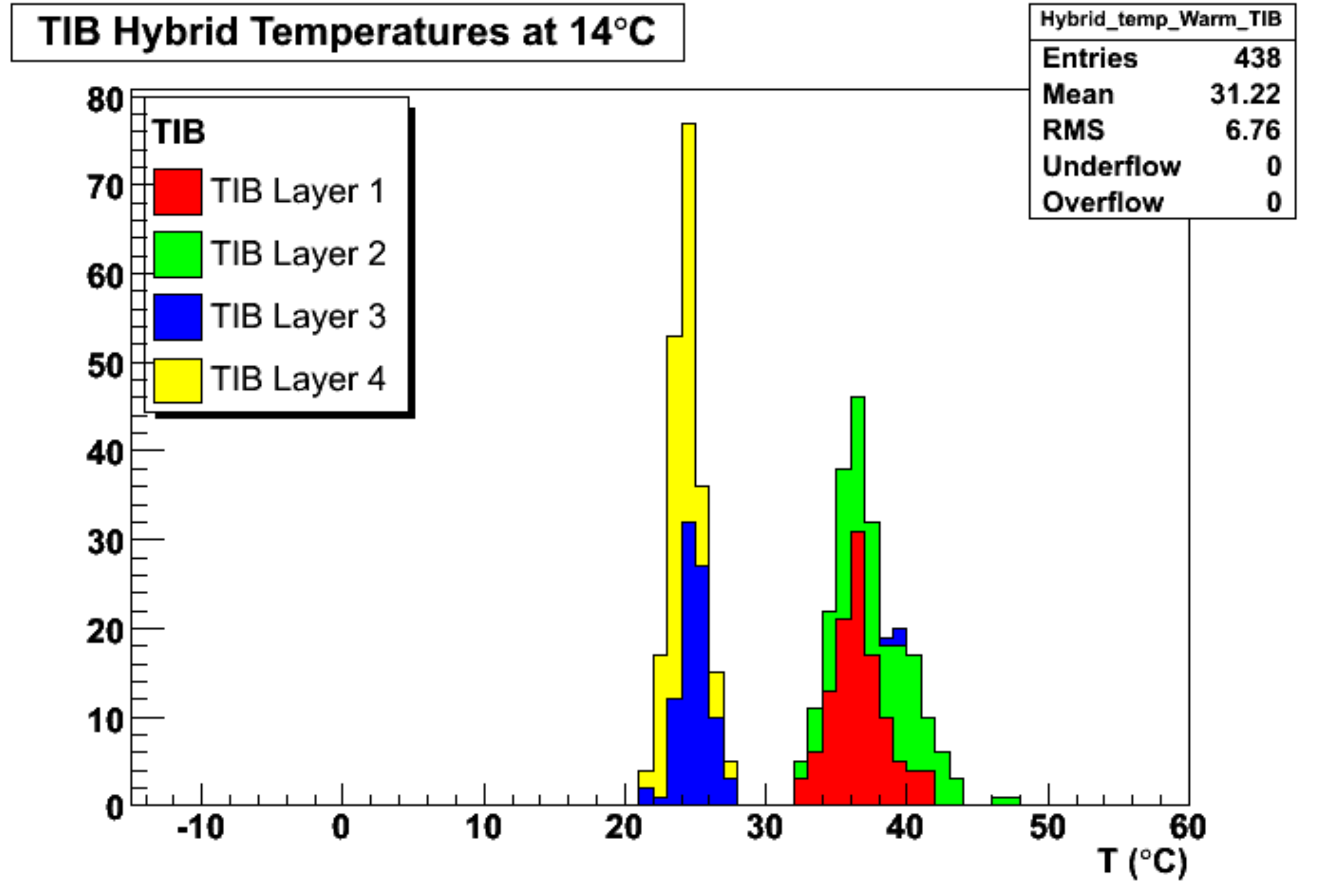}
    \includegraphics[width=0.4\textwidth,angle=0]{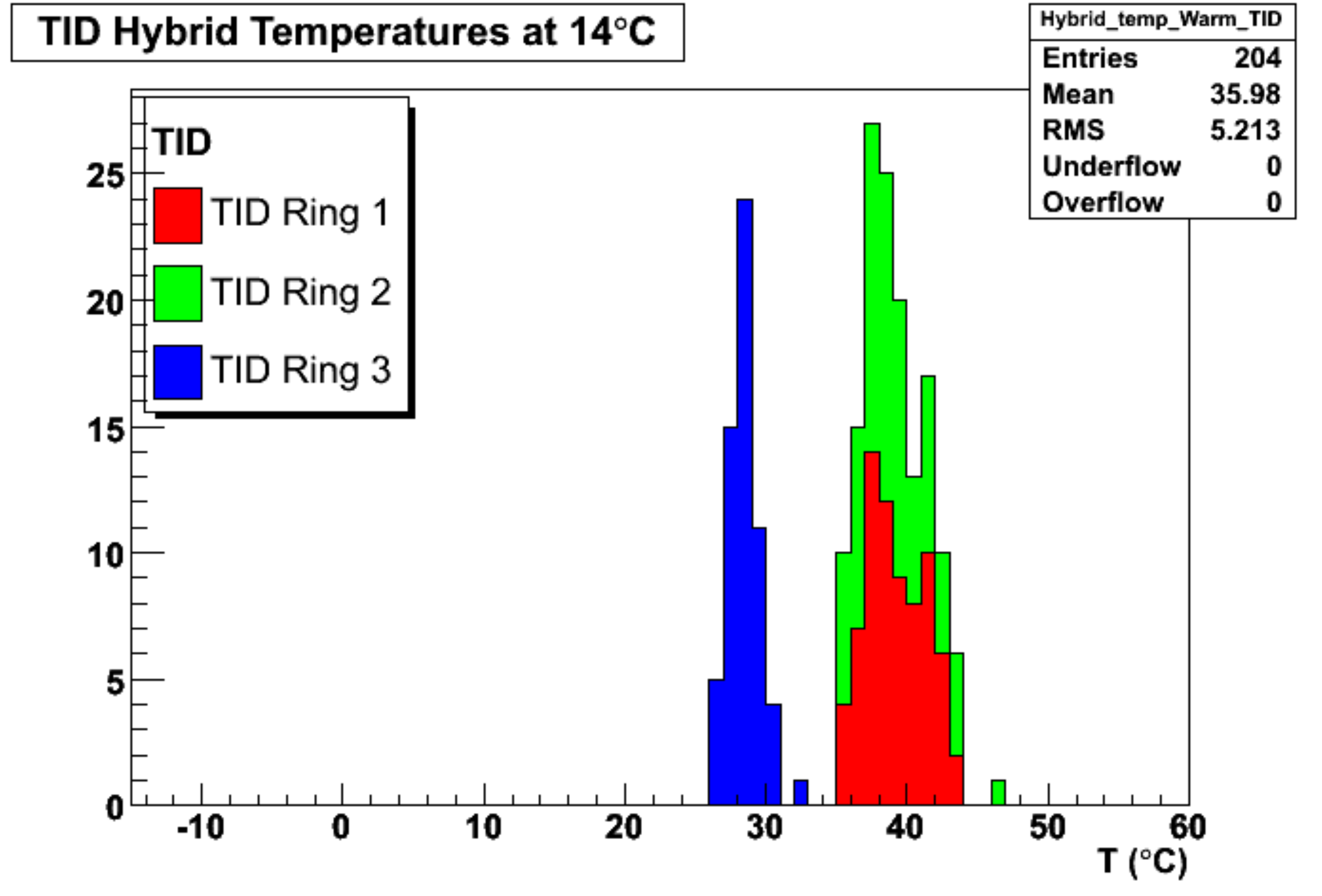}

    \includegraphics[width=0.4\textwidth,angle=0]{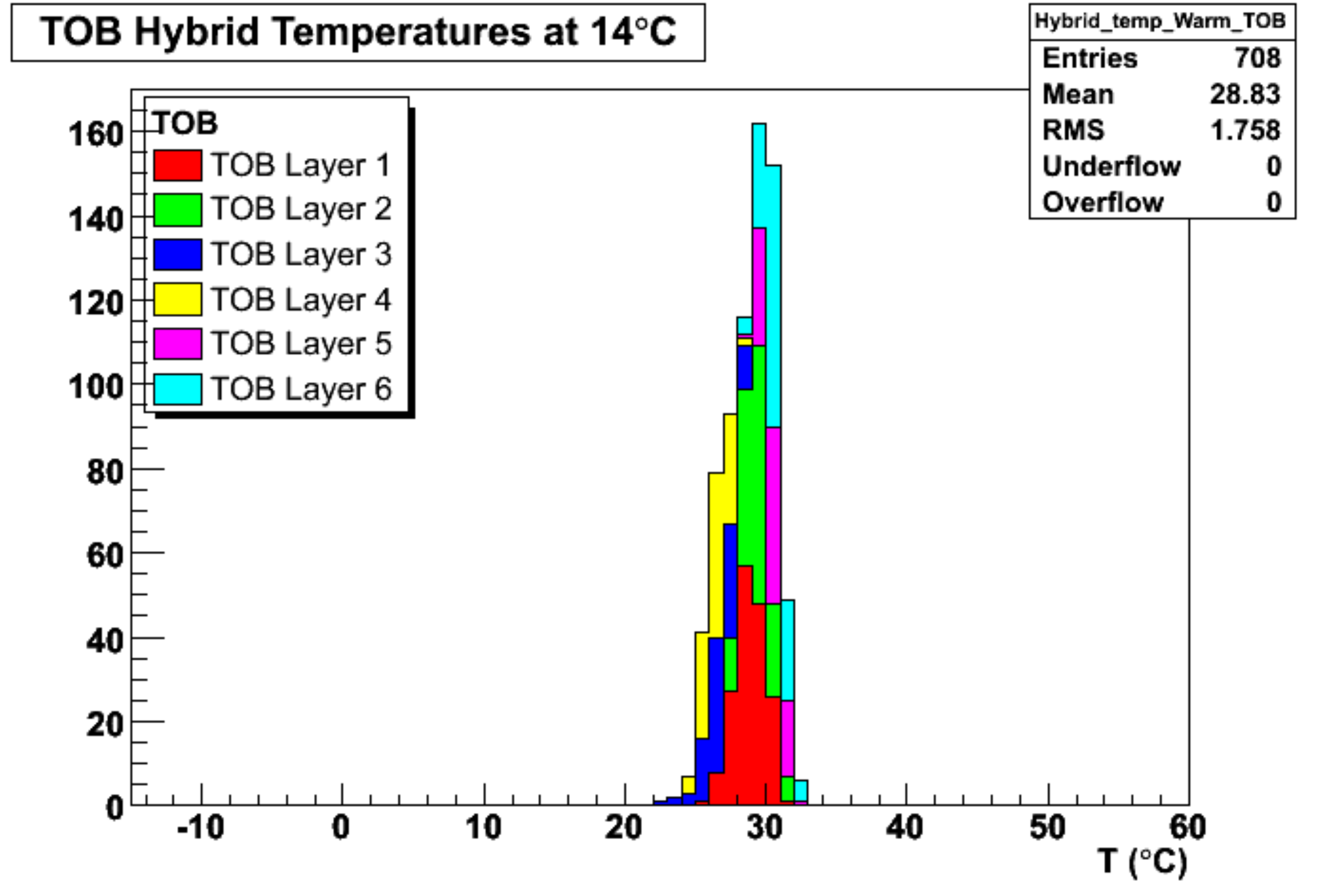}
    \includegraphics[width=0.4\textwidth,angle=0]{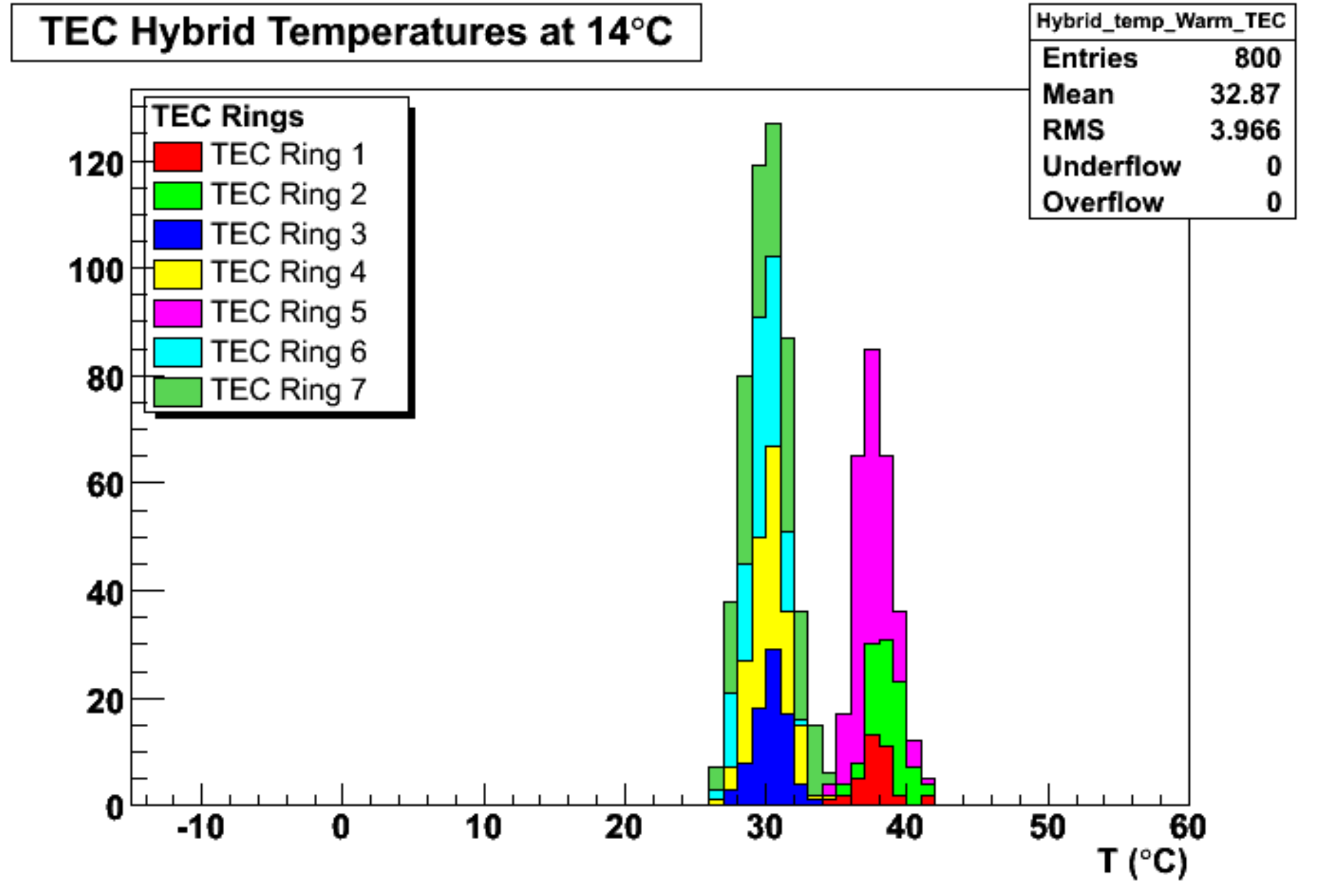}

    \caption{\sl Hybrid temperature distributions for 
    TIB, TID, TOB, and TEC at a coolant temperature of $14\,^{\circ}\mathrm{C}$.}
    \label{fig:Htemp}
  \end{center}
\end{figure}
\begin{figure}[!ht]
  \begin{center}
   \includegraphics[width=0.4\textwidth,angle=0]{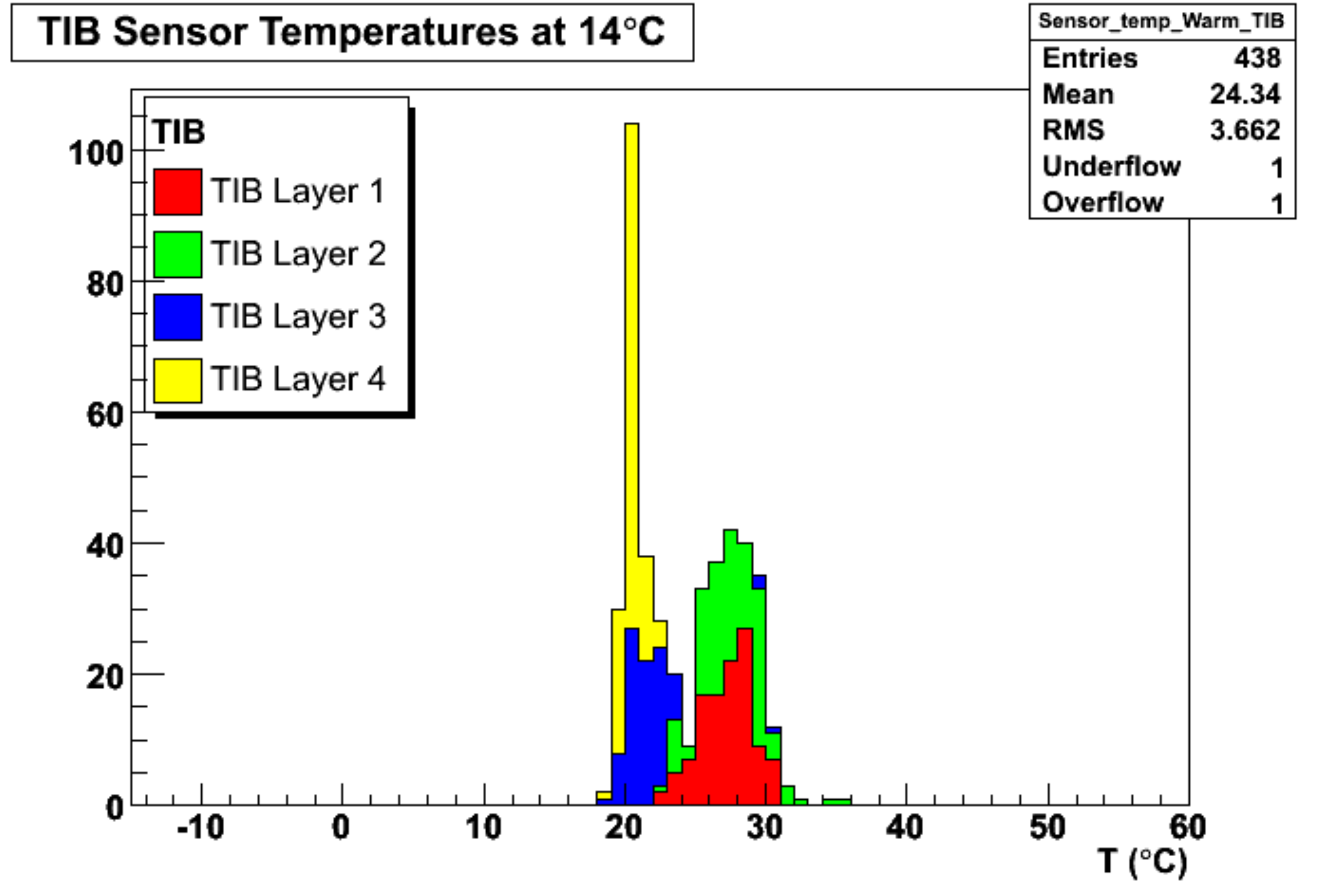}
   \includegraphics[width=0.4\textwidth,angle=0]{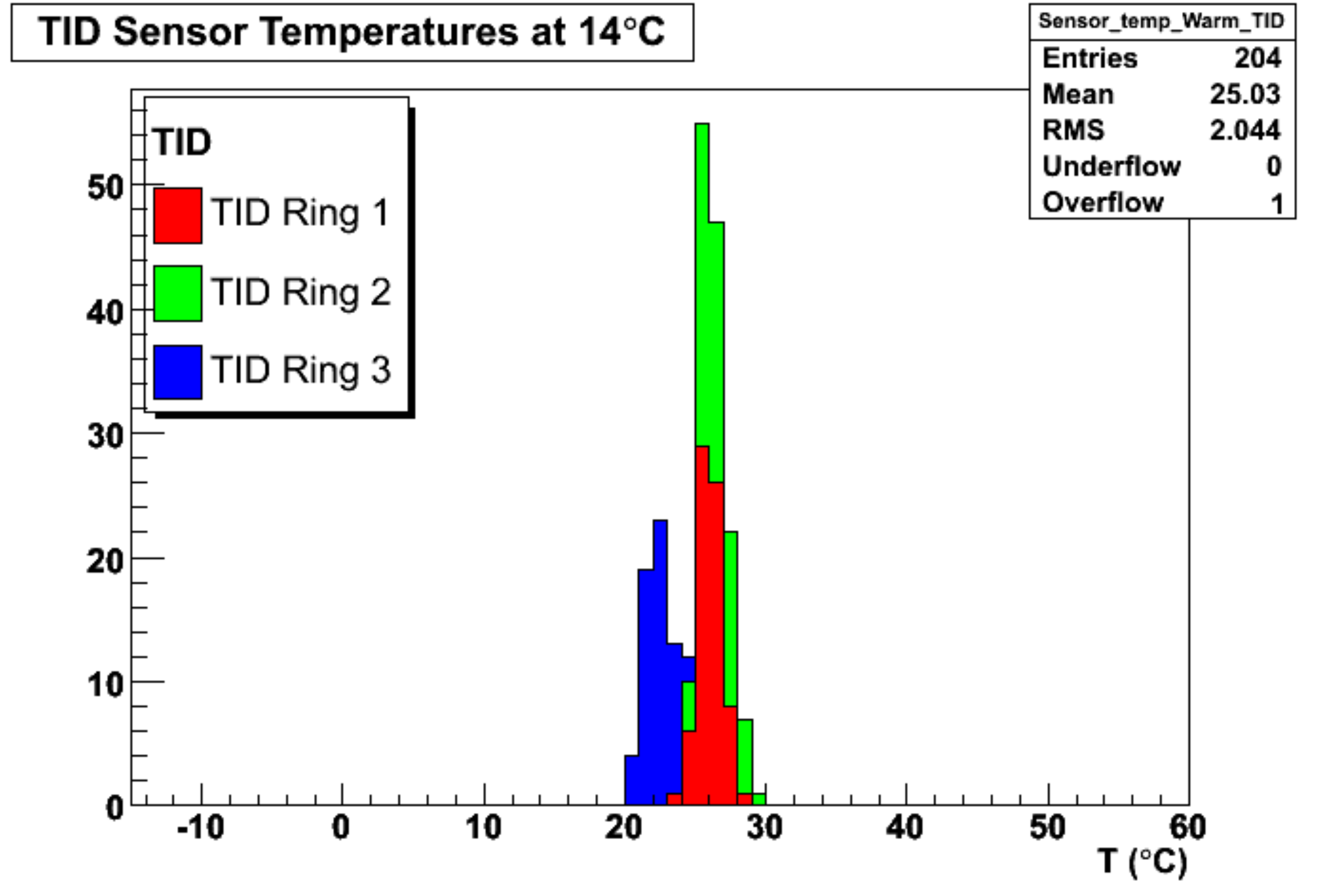}

   \includegraphics[width=0.4\textwidth,angle=0]{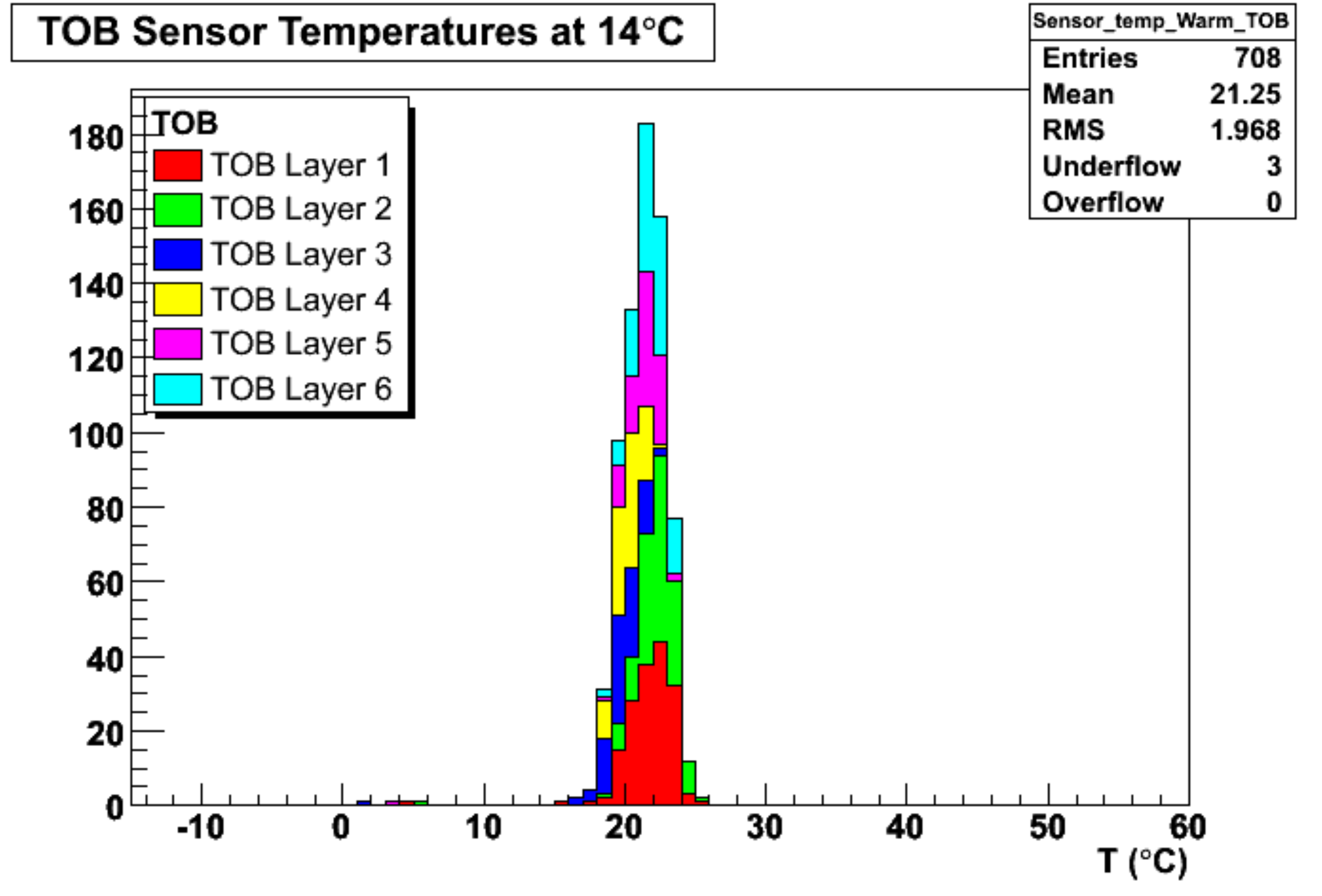}
   \includegraphics[width=0.4\textwidth,angle=0]{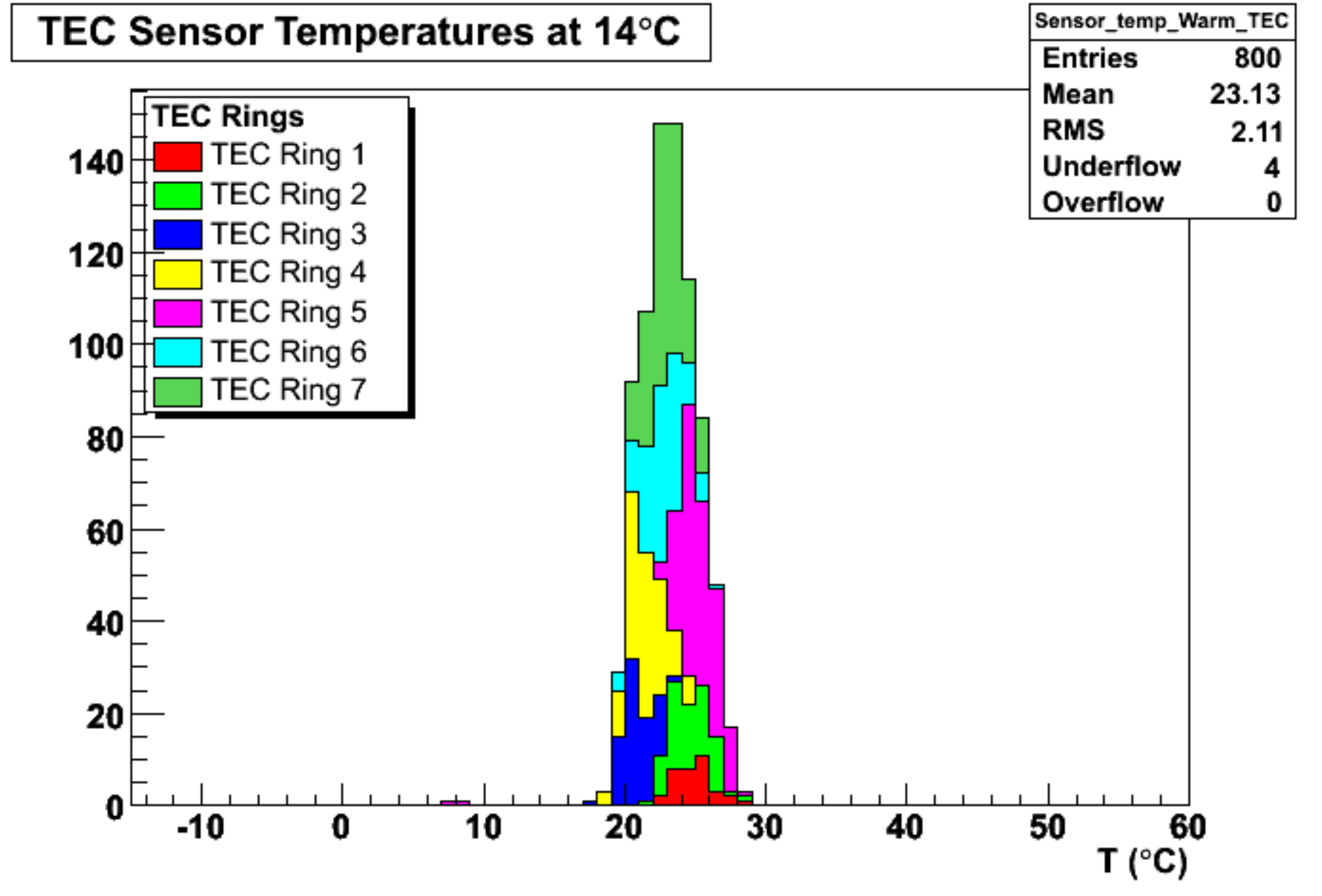}

    \caption{\sl Sensor temperature distributions at a coolant temperature of $15\,^{\circ}\mathrm{C}$ for 
    TIB, TID, TOB, and TEC. }
    \label{fig:Stemp}
  \end{center}  
\end{figure}

The performance of the Tracker depends on front end electronics supply voltage 
and  operating temperature. These quantities are
measured by the DCU chip located on each module hybrid, which can be
read out digitally via the control ring protocol through the
electronic chain (DCU, CCU, DOH, FEC). 

Results obtained for the hybrid temperature are shown in Fig.~\ref{fig:Htemp} and for sensor temperature in
Fig.~\ref{fig:Stemp} for $14\,^{\circ}\mathrm{C}$ coolant temperature. All TOB layers show the same temperature, while for TIB, TID and TEC  differences are visible: TIB layers 1 and 2, TID ring 1 and 2,  and TEC 
ring 1, 2 and 5  are double-sided layers and have a temperature 
substantially higher compared with single-sided layers; the effect is more pronounced for the hybrid than for the sensor.
These differences are expected to be lower with the final cooling system.

It is also apparent that three modules of TIB layer 3 have a higher temperature of about $15\,^{\circ}\mathrm{C}$; they are in a string where the cooling pipe was blocked.  Results from the  other operating temperatures  are consistent with those obtained at $14\,^{\circ}\mathrm{C}$.



\begin{figure}[!ht]
  \begin{center}
    \includegraphics[width=0.4\textwidth]{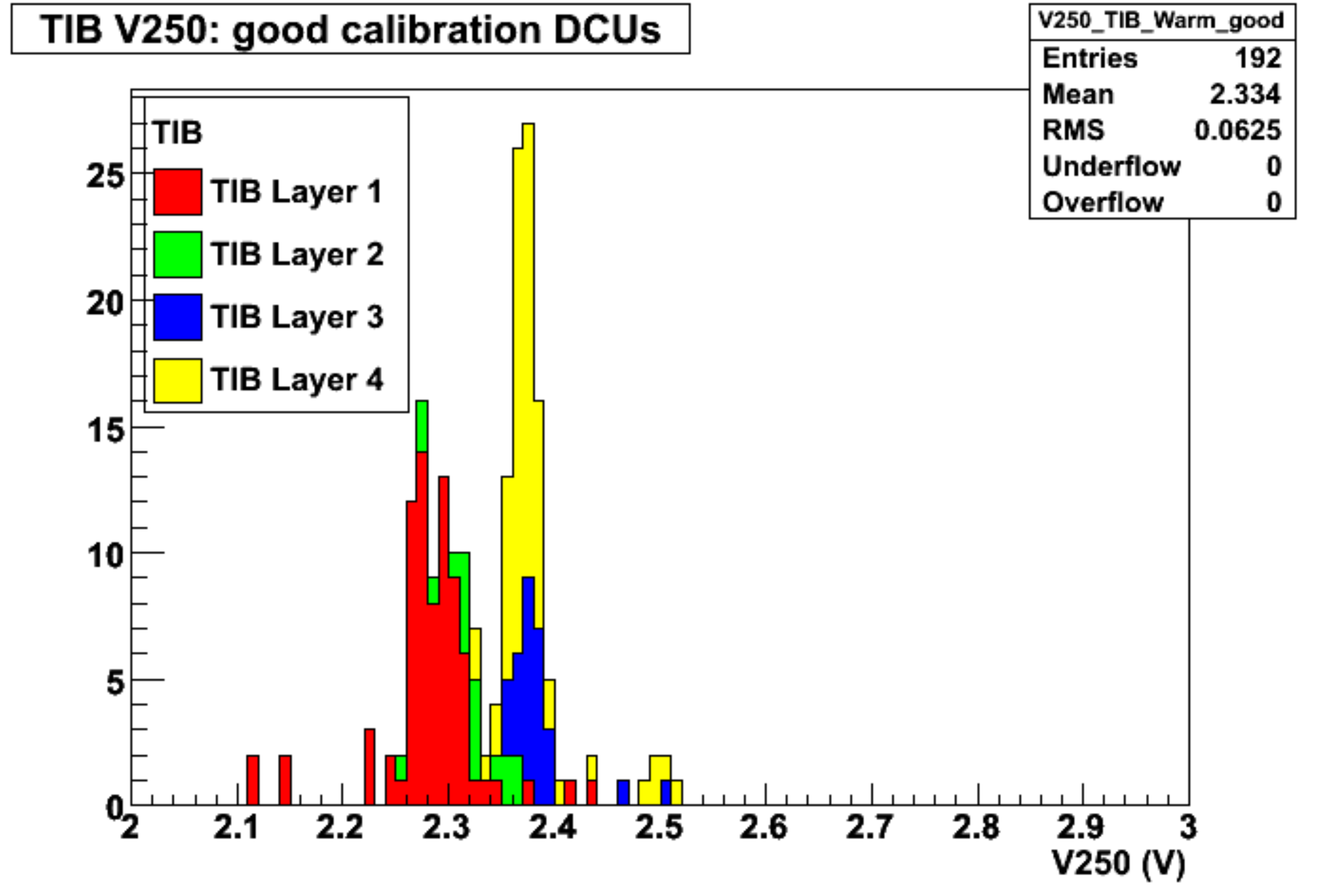}
    \includegraphics[width=0.4\textwidth]{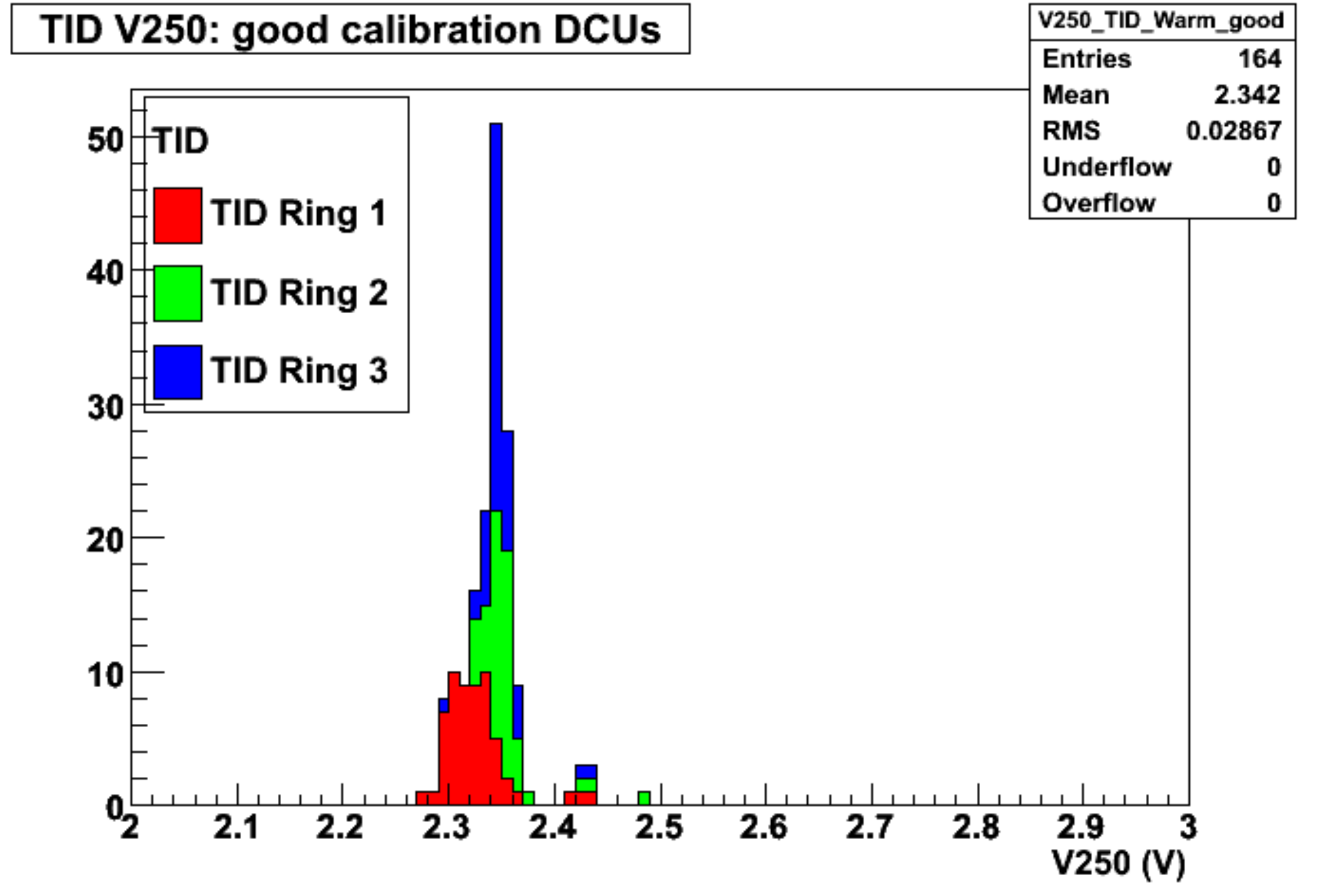}

    \includegraphics[width=0.4\textwidth]{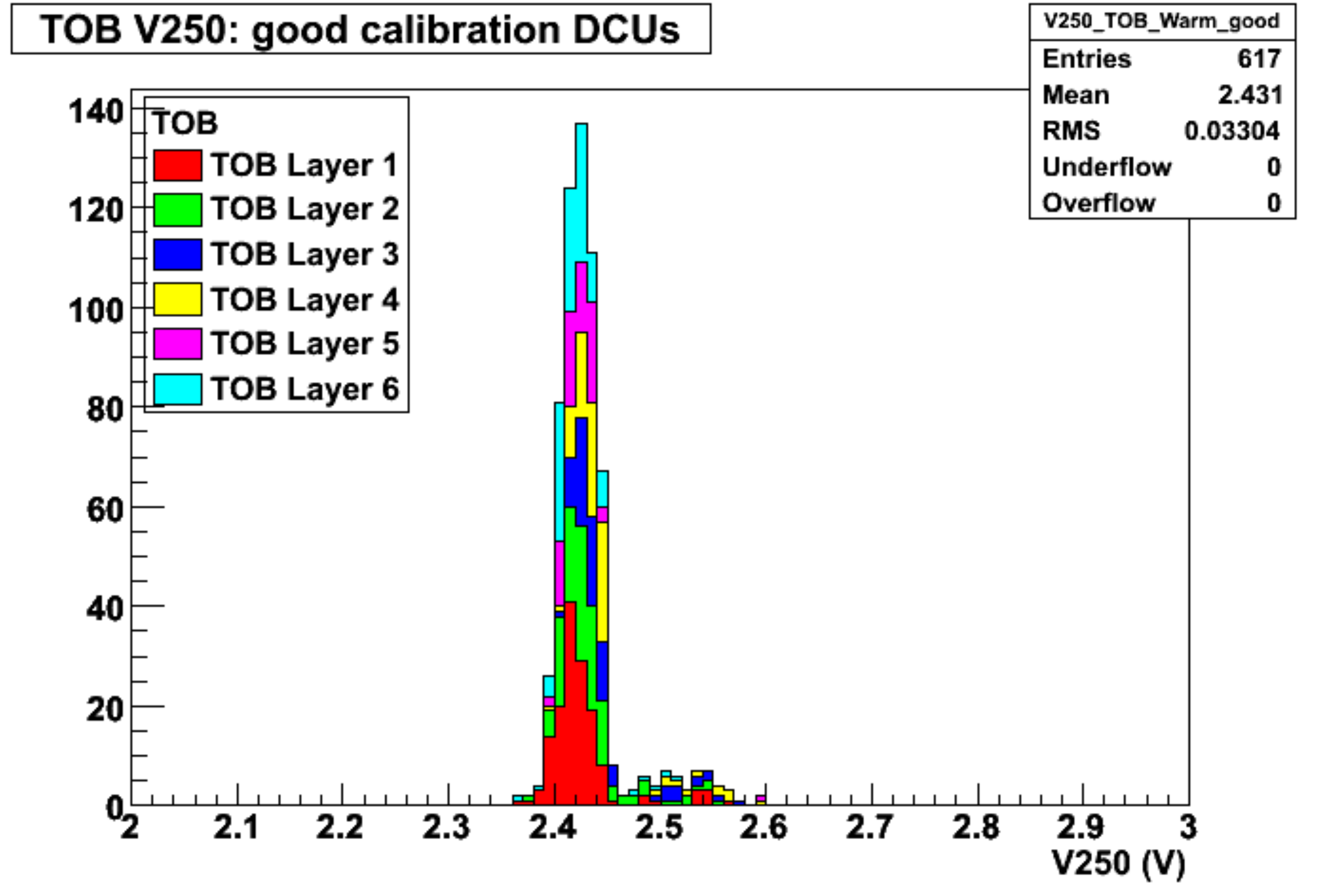}
    \includegraphics[width=0.4\textwidth]{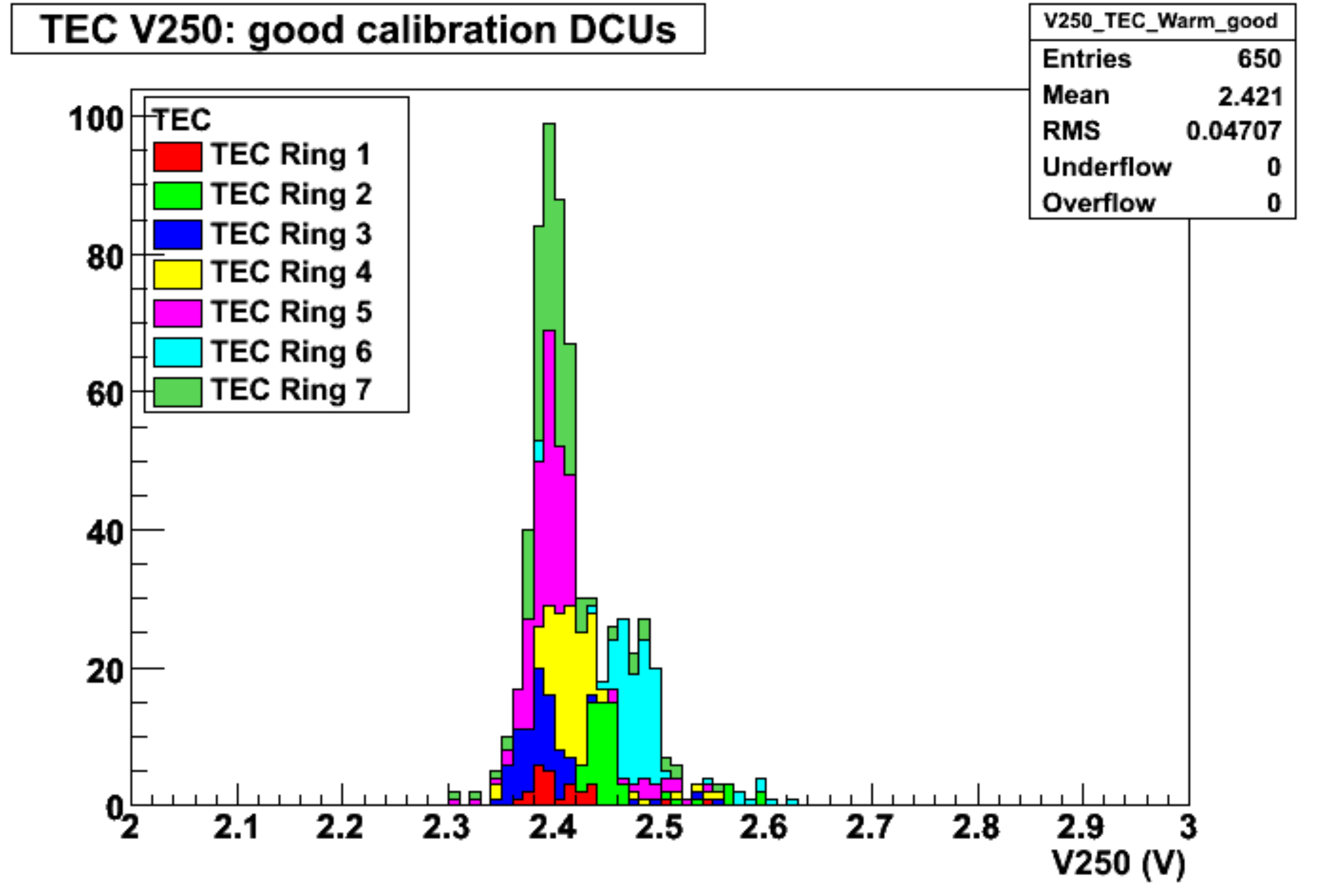}
    \caption{\sl 2.5\,V voltage line value measured at the module for TOB, TIB, TEC, and TID. }
    \label{fig:25V}
  \end{center}
\end{figure}

\begin{figure}[!ht]
  \begin{center}
     \begin{tabular}{cc}
    \includegraphics[width=0.4\textwidth]{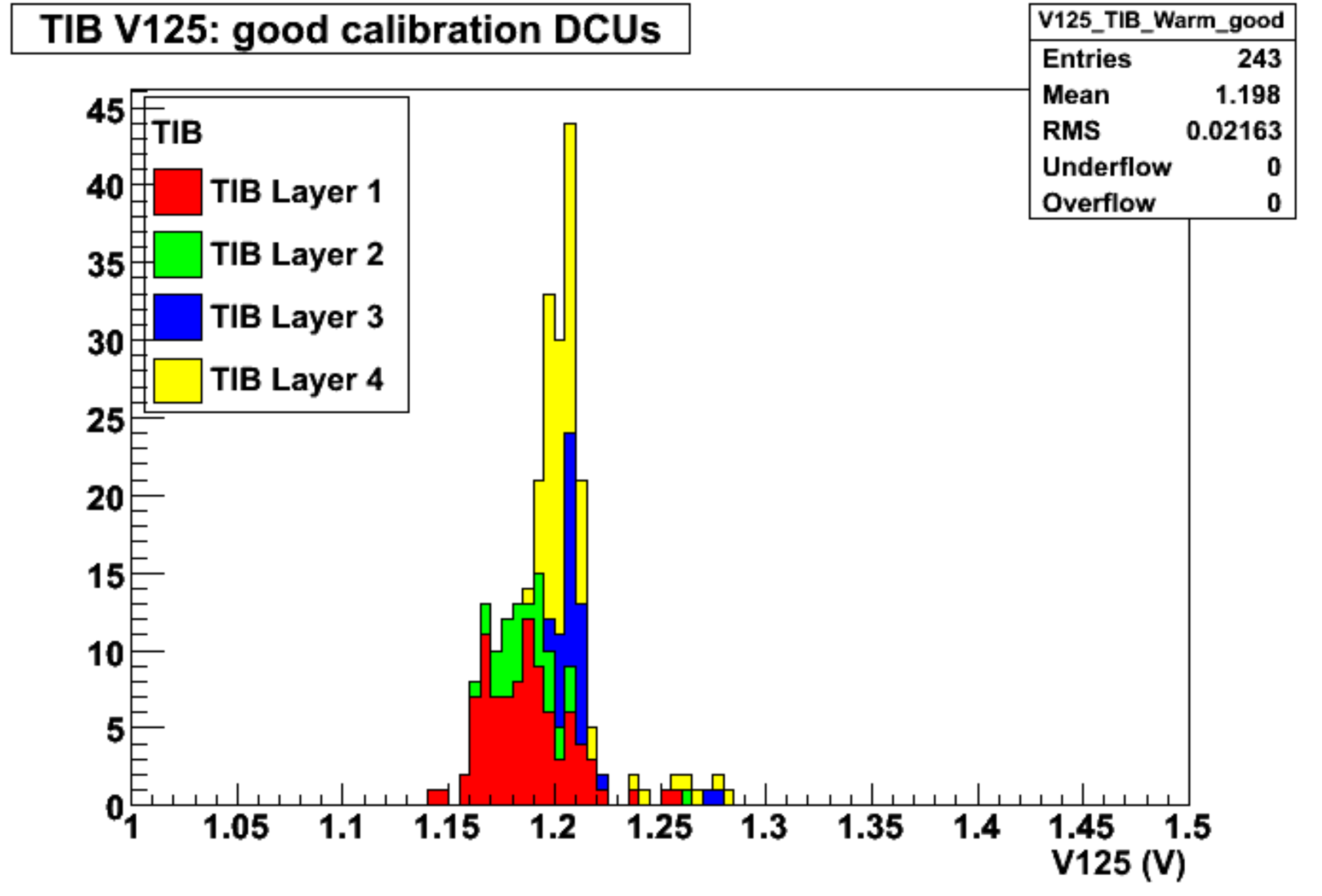} &
    \includegraphics[width=0.4\textwidth]{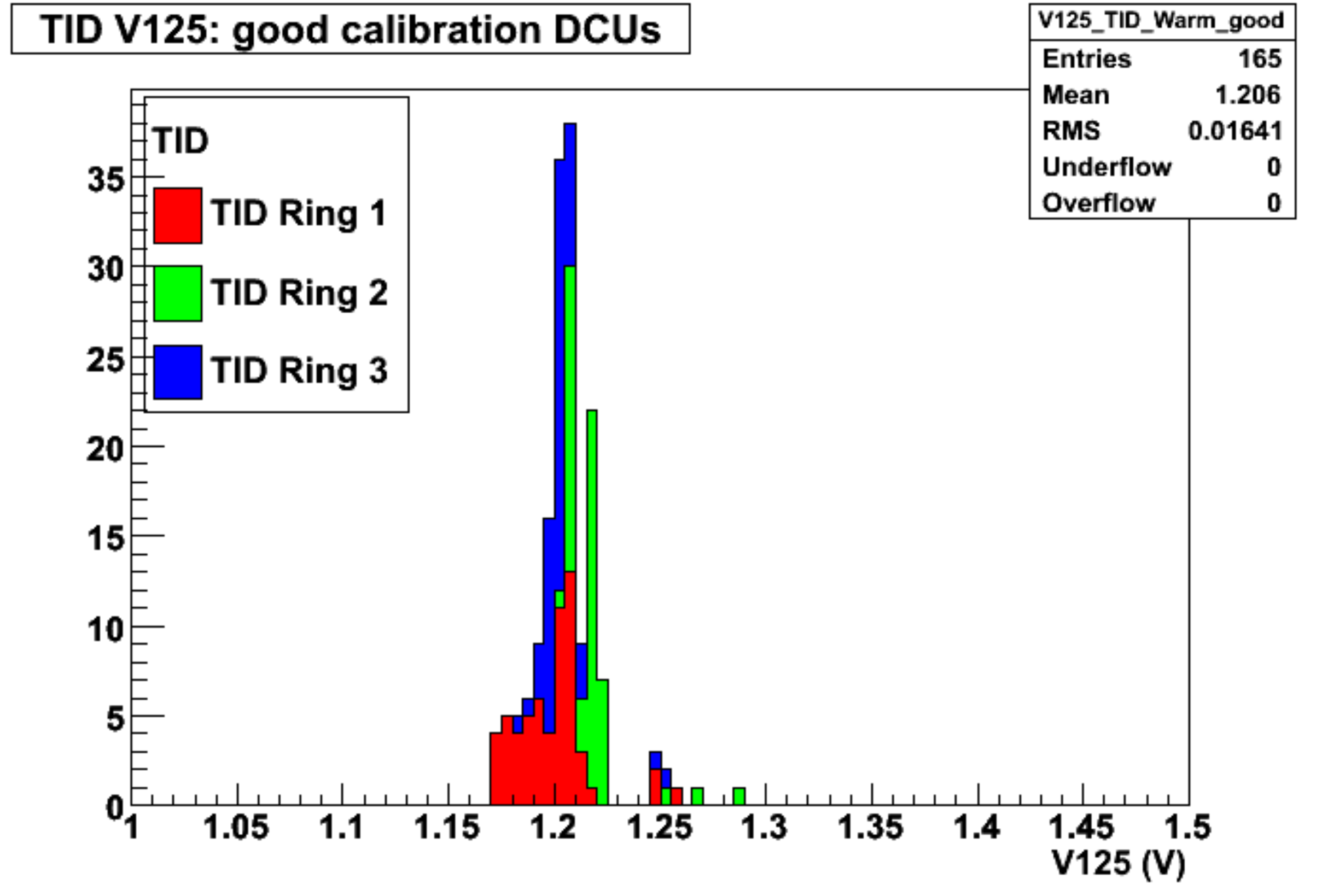} \\
    \includegraphics[width=0.4\textwidth]{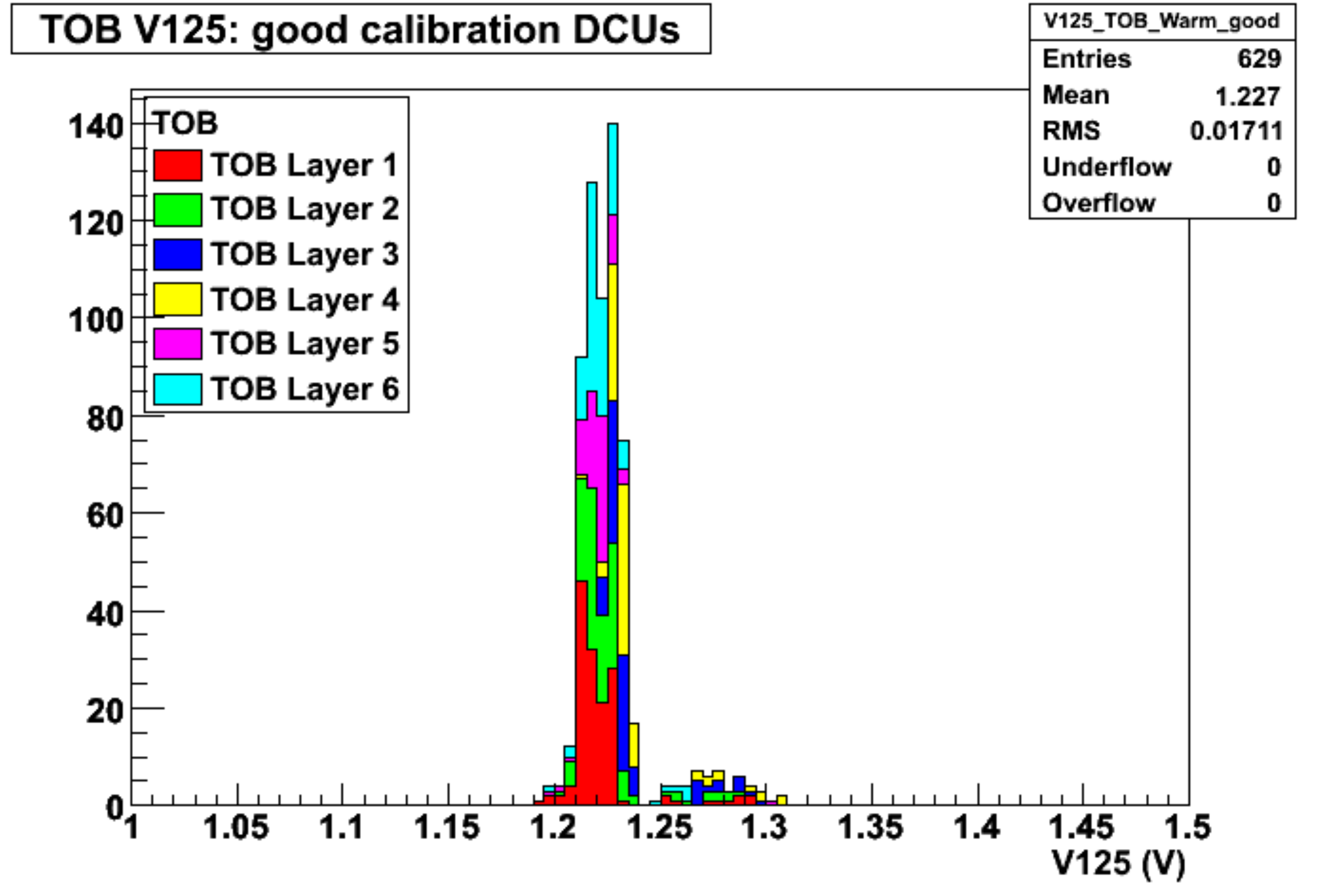} &
    \includegraphics[width=0.4\textwidth]{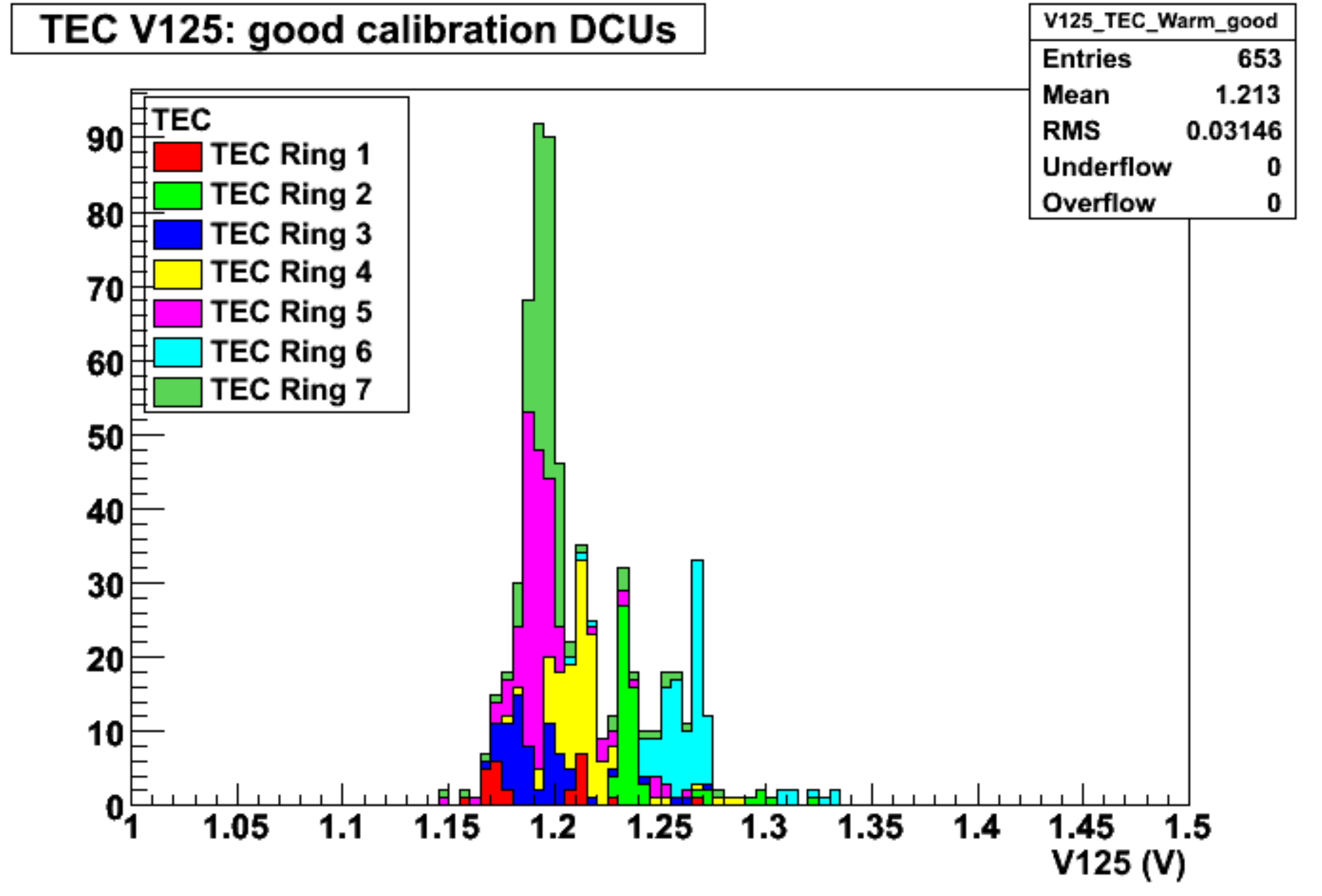} \\
     \end{tabular}
    \caption{\sl 1.25\,voltage line value measured at the module for TOB, TIB, TEC, and TID. }
    \label{fig:125V}

  \end{center}
\end{figure}

The power supply system provides 2.5\,V and 1.25\,V lines to the modules, correcting for voltage drops along the long power supply cables  by
 sensing the voltage levels as close  to the modules as possible.
 Nevertheless, there still remains a small cable resistance  which causes a non-negligible voltage drop. The operating voltage of
the  modules is measured by the DCU  and results are shown
in Fig.~\ref{fig:25V} for the 2.5\,V line and in
Fig.~\ref{fig:125V} for the 1.25\,line. The average is below nominal for all layers of sub-detectors. These measurements show that it is possible to evaluate these voltage drops and therefore to eventually better equalize the 
module voltages to match the requirements.


\subsection{Electronic Gain Measurements}\label{sec:gain}
This section describes how the electrical gain of the Tracker system
is both determined and adjusted, that is, the procedures by which the
gains of individual APV25s are adjusted to a common and known
value. The readout chain for the Tracker involves pairs of adjacent APV25s, the
hybrid MUX, the AOH~\cite{AOH}, an optical fiber, and the FED~\cite{FED}. Instabilities in
the low or high voltages or changes in temperature can affect the
gain. Measuring the electrical stability of the Tracker  as a function of time, voltage, temperature, and other
variables, should lead to an improved understanding of the likely
performance of the full Tracker system during actual LHC operations.

The  commissioning procedure~\cite{commissioning} determines
 optimal settings of the electronics to achieve a uniform electronic gain. 
This is obtained by measuring the value of the height of synchronization
pulses, referred to as tick-marks, generated by the APV25  which provide
a stable value that can be used as reference to equalize the 
response of the full electronic chain. It is  required that
the tick-mark height of each APV25 be within the FED dynamic range and, 
 by varying the AOH offset and gain, should  correspond closely to the target value of 640 ADC channels. 
The full commissioning procedure was consistently applied  whenever there were changes in  coolant
temperature, changes in APV25 parameters, or  in the hardware
configuration. Commissioning  runs, referred to as timing runs, provide precise measurements of the
tick-mark heights, therefore of the electronics gain, and were repeated at
intervals between full commissioning runs to get more statistics on stability of the system. 

\begin{figure}[!hbt]
  \begin{center}
    \includegraphics[width=0.49\textwidth]{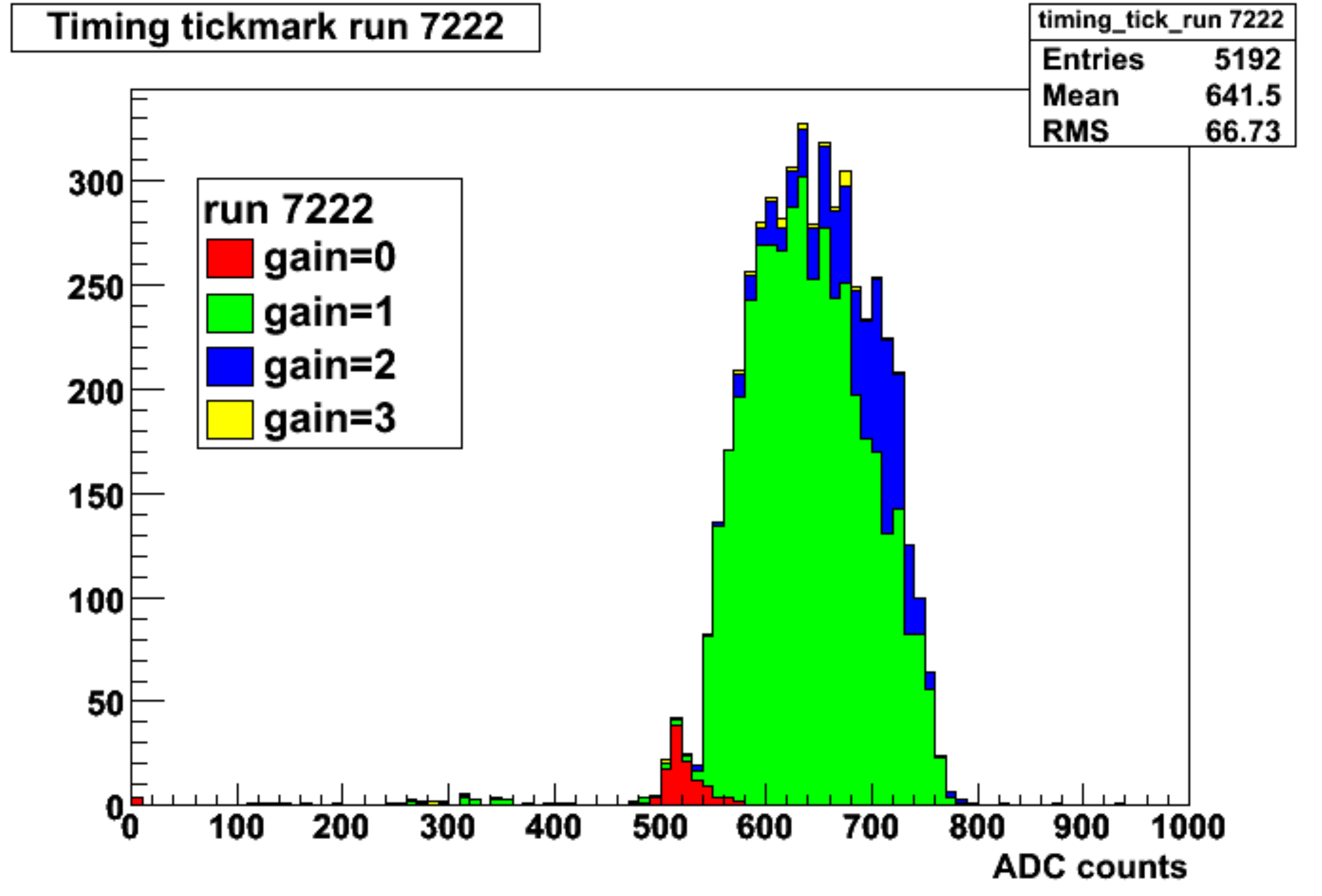}
    \includegraphics[width=0.49\textwidth]{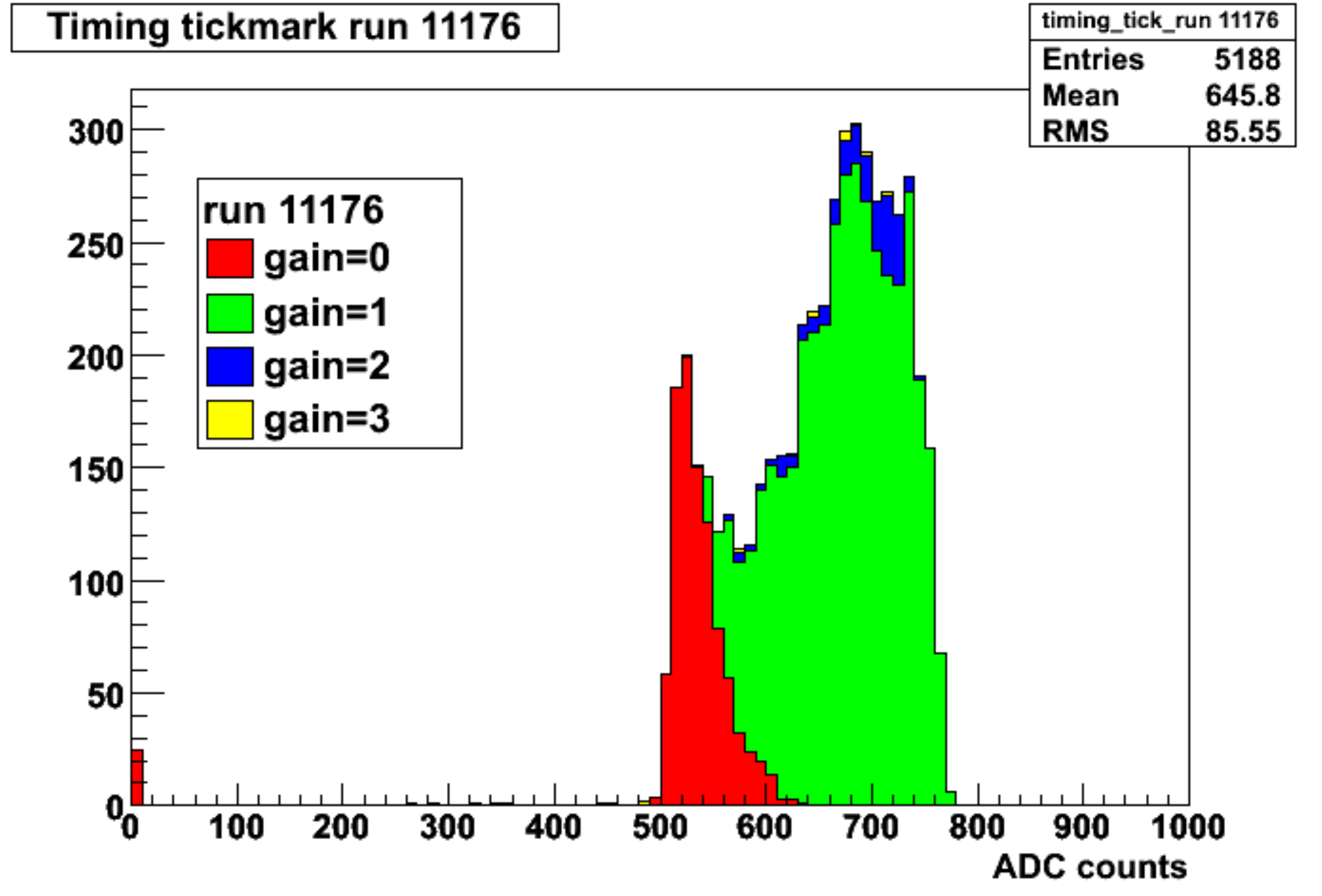}
    \caption{\sl APV25 tickmark height distribution after commissioning
    procedure: run 7096, taken at $T=14\,^{\circ}\mathrm{C}$ (left), and run
    11176, taken at $T=-10\,^{\circ}\mathrm{C}$ (right). }
    \label{fig:elgain-dis}
  \end{center}
\end{figure}

Tick-mark height distributions obtained after the commissioning procedures are
shown in Fig.~\ref{fig:elgain-dis}. The several gray levels represent
the components with different AOH gain settings (called 0, 1, 2, 3 from lower to higher gain): the difference between the left and right distributions is mainly due to the strong dependence  on the temperature of the AOH gain. The effect is manifest by the increased number of AOH with gain equal to 0 at lower temperature.  
This implies that the average values of the tickmark height distribution changes when varying the temperature.

Even after the commissioning procedure the tickmark height 
distributions still indicate about a 10$\%$ spread in electrical
gain. This variation is consistent with the coarse precision of the
AOH laser gain settings. An offline calibration of the electronic gain is necessary to improve the precision of the measurement of noise and signal. This is achieved by  normalizing to an assumed digital
header of 640 ADC counts. In the following sections, the analysis of
signal and noise will take into account this calibration if not otherwise stated, by applying a correction 
factor of
\begin{equation}
 C_{corr}  =   \frac{640}{TickMark} ;
\label{form:corr}
\end{equation}
One limitation in the use of this equation is due to the different module operating voltages for different layers. It was known that the tick-mark amplitude is linearly proportional to the 2.5\,V operating voltage, therefore tickmarks from different modules can be compared only if they operate at the same voltage. The signal at the APV25 amplifier output is not much affected by
changes in the supply voltage. Therefore to make a more precise estimate of the electronic gain it is necessary  either to equalize all the operating voltages or to correct the tick-mark value
 for the difference compared to  2.5\,V. In this paper this correction was not applied
therefore the electronic gain can be considered having a systematic variation of about $5\%$
that affects direct comparison between different layers.

\begin{figure}[ht]
  \begin{center}
    \includegraphics[height=0.75\textwidth,angle=90]{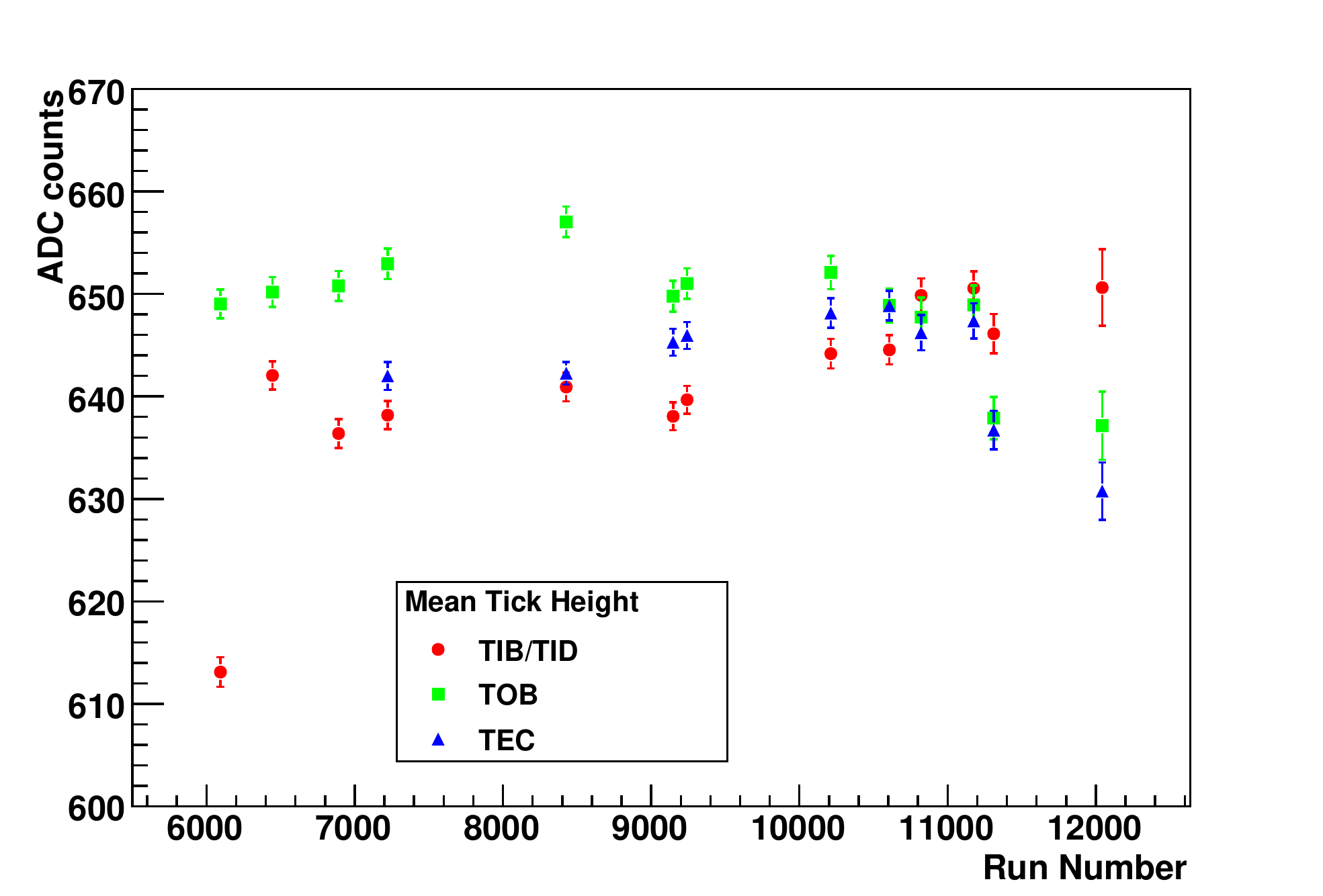}
    \caption{\sl Tickmark height distribution versus run number. Error bars
    represent the statistical error of the average value.}
    \label{fig:elgain-stab}
  \end{center}
\end{figure}

It is important to understand the stability of the electronics gain,
since calibration runs are taken only at selected times. In the
analysis of the timing runs, bad modules have been
discarded and a selection of good commissioning runs have been made.
Figure~\ref{fig:elgain-stab} shows average tick-mark heights for
individual sub-detectors as a function of run number (with the help
of Table~\ref{tab:dsets} the period at different temperatures can be identified).
At least one timing run was taken as part of the commissioning
procedure whenever the coolant temperature changed. During the 
period at $T=14-15\,^{\circ}\mathrm{C}$, in particular, there were several timing runs, which
provide information on the stability of the system. 

 TOB
shows, in 7 runs, variations of $\pm 0.6 \%$. The TIB/TID shows a large
discrepancy in the first run (6094), where results are lower by $5
\%$, but this is due to the fact that the modules were operating at  low 
electrical gain and therefore that was corrected by a very high AOH gain:
after removing this run the TIB shows, in 6 runs, variations of $\pm 0.6
\%$.  The TEC shows, in 4 runs,  $\pm 0.4 \%$ variations.

\begin{figure}[!ht]
  \begin{center}
    \begin{tabular}{cc}
    \includegraphics[width=0.39\textwidth]{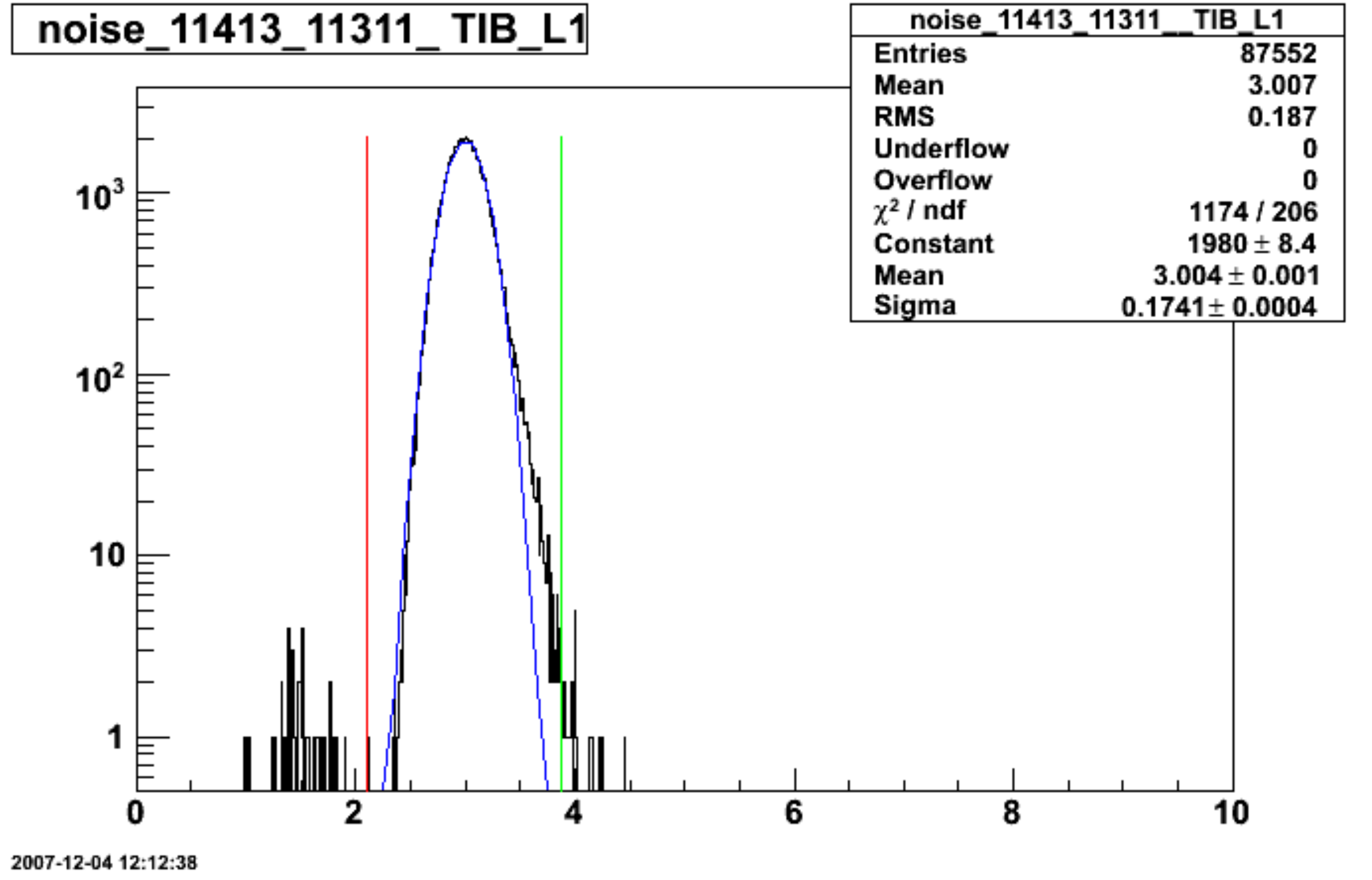} &
    \includegraphics[width=0.39\textwidth]{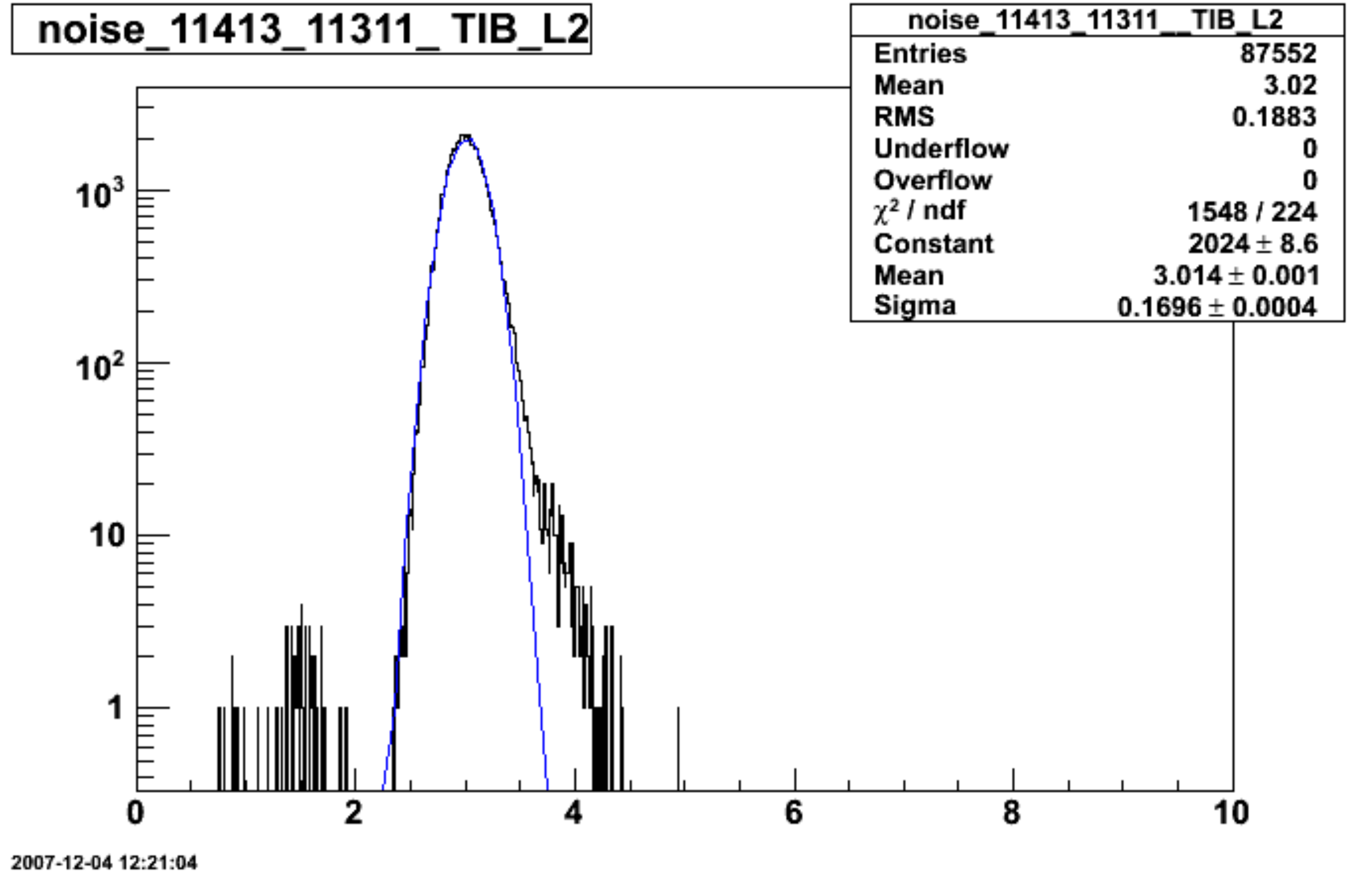} \\
    \includegraphics[width=0.39\textwidth]{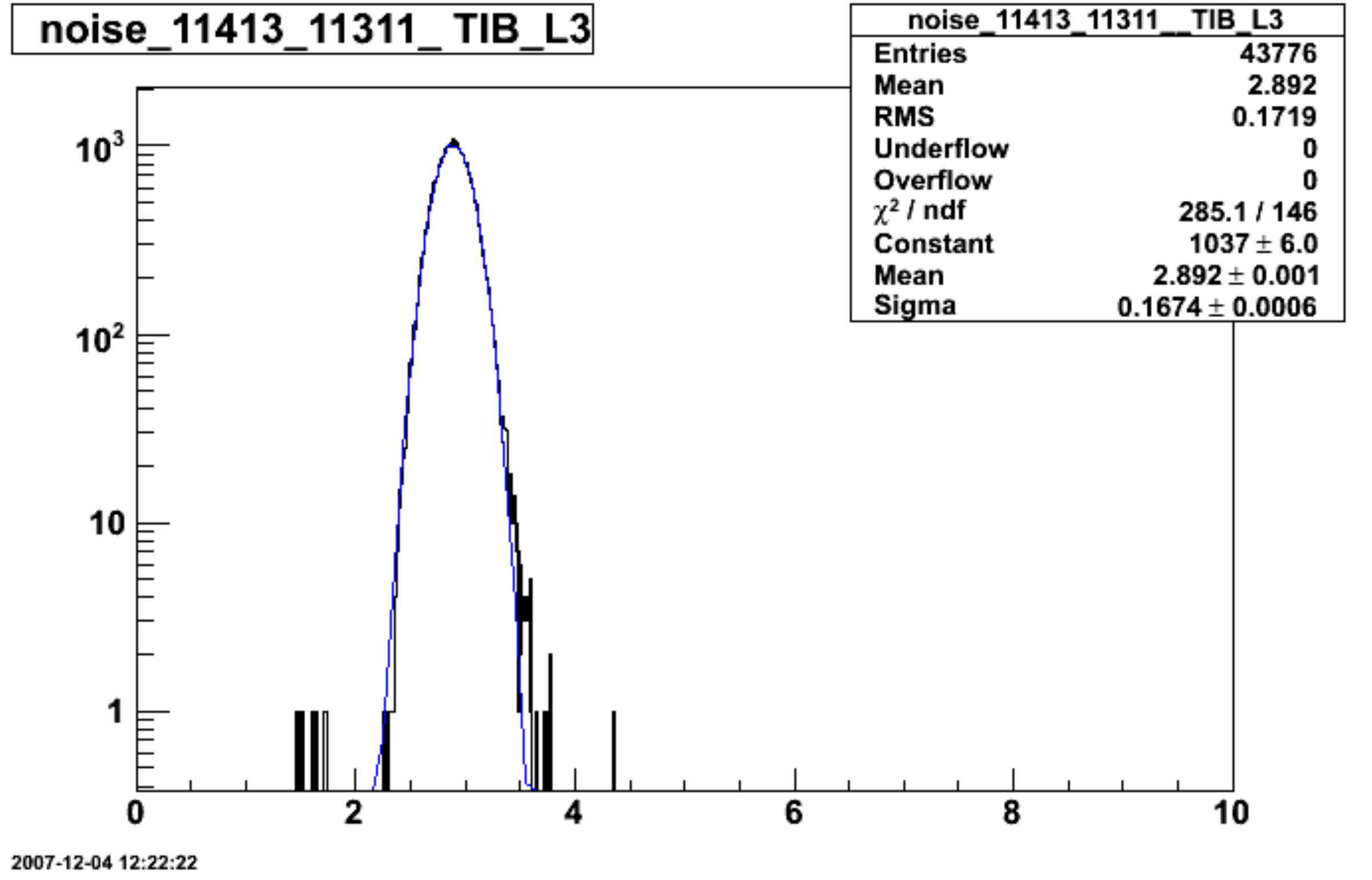} &
     \includegraphics[width=0.39\textwidth]{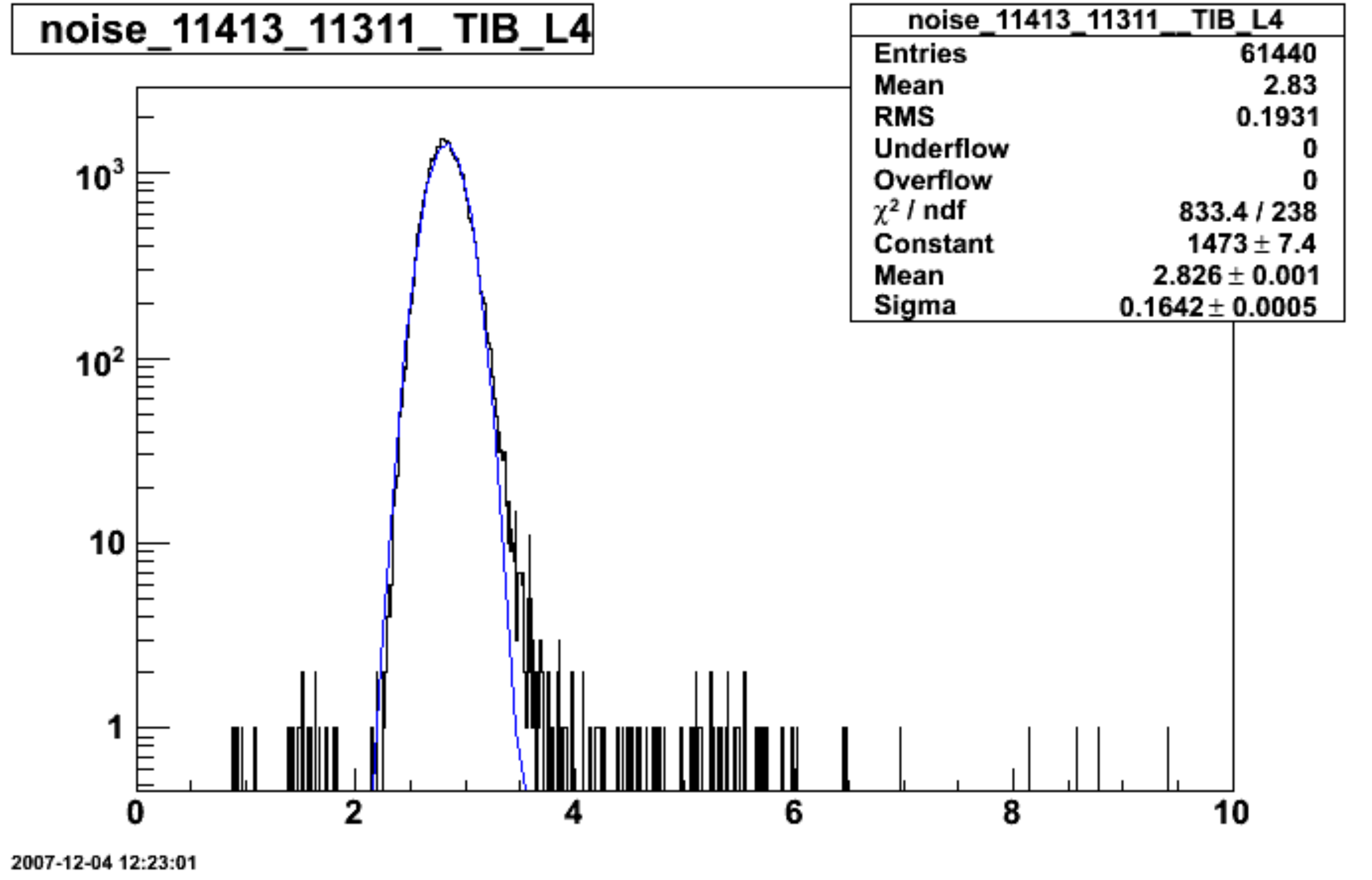} \\
    \end{tabular}
     \caption{\sl TIB noise profile for T= $-10\,^{\circ}\mathrm{C}$ for layers 1, 2, 3, and 4. Gaussian fit is shown. X axis is ADC value.}
    \label{fig:tib-noise}
  \end{center}
\end{figure}
\begin{figure}[!ht]
  \begin{center}
    \includegraphics[width=0.39\textwidth]{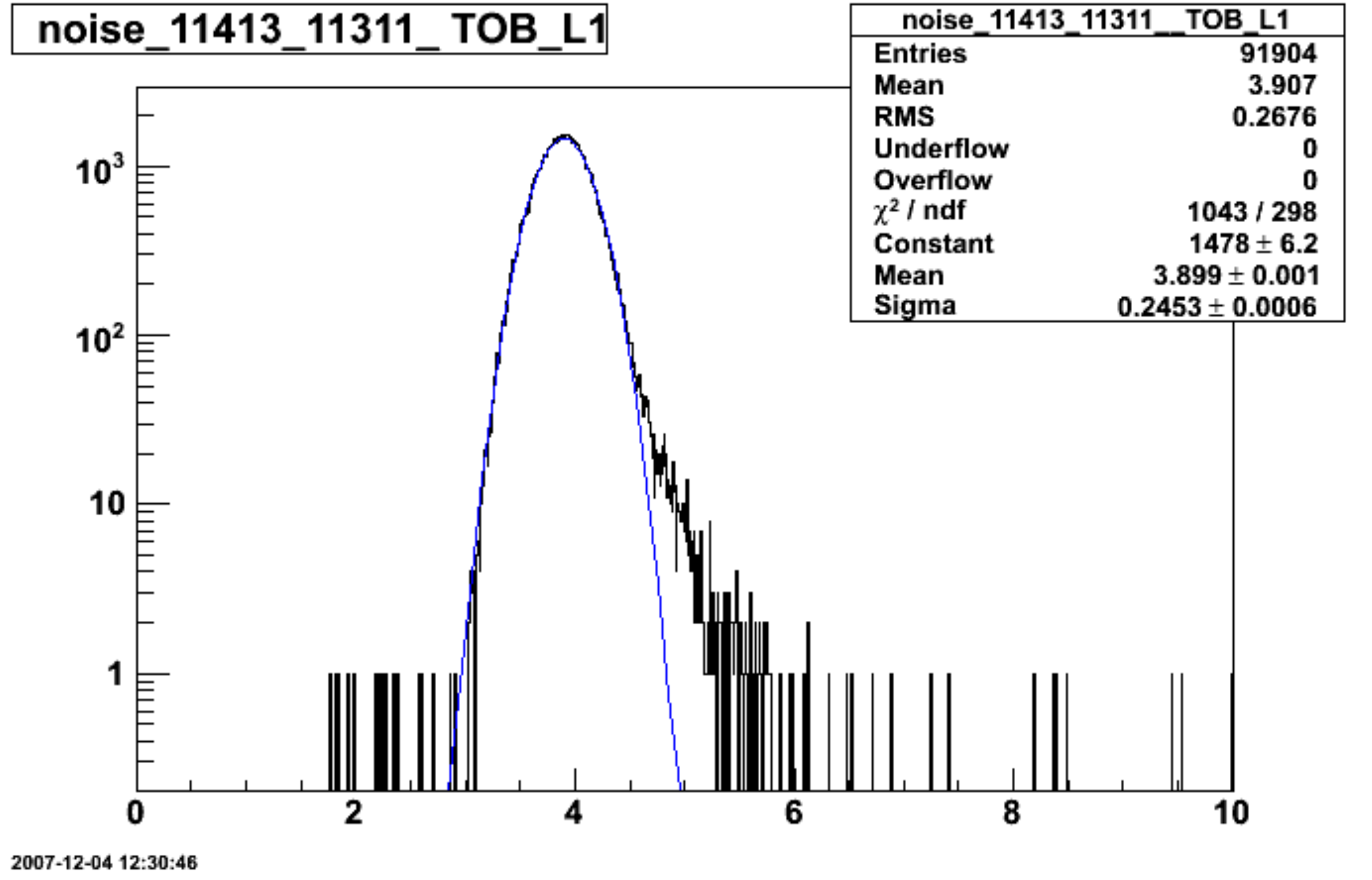}
    \includegraphics[width=0.39\textwidth]{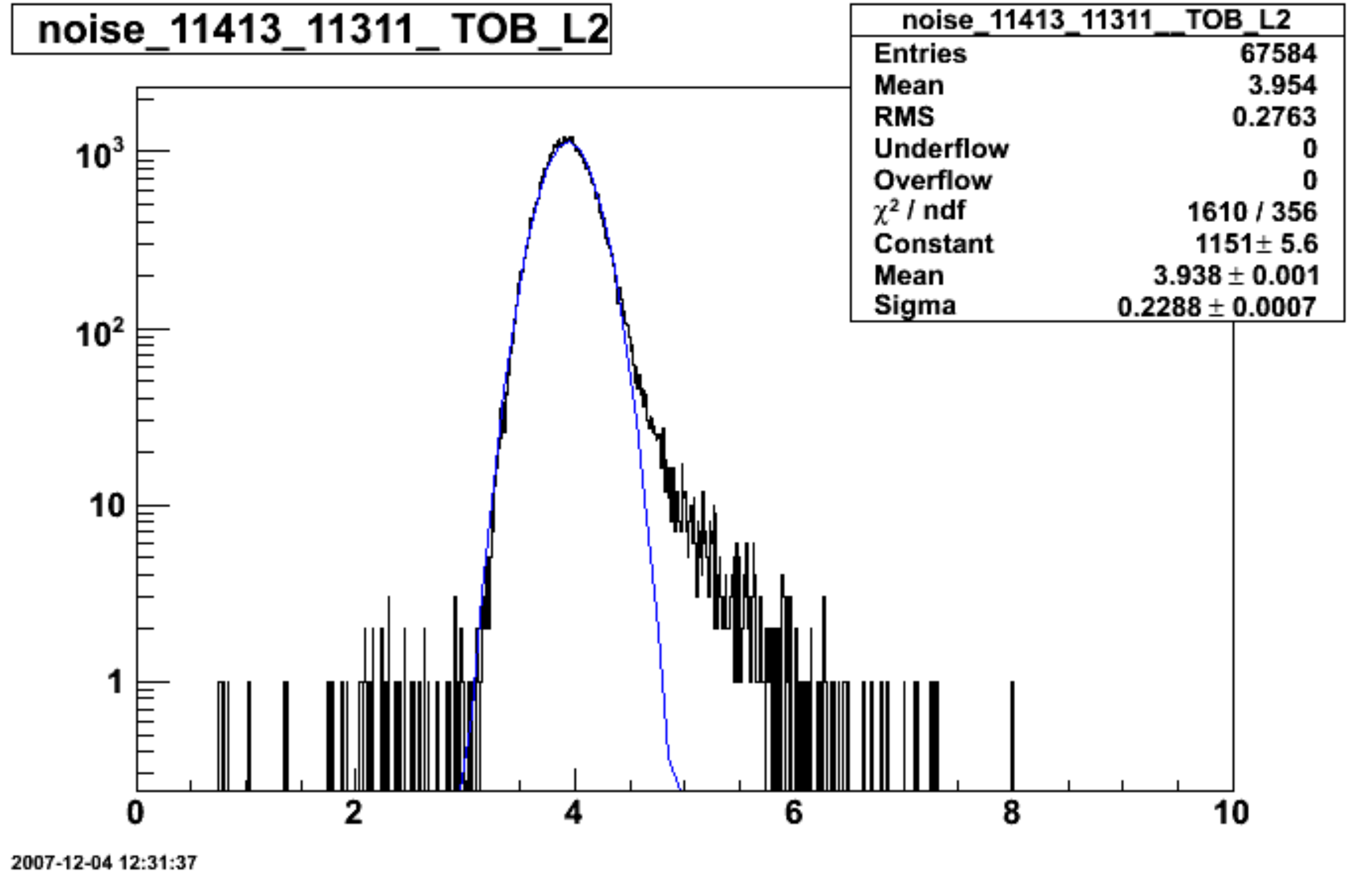}
    
    \includegraphics[width=0.39\textwidth]{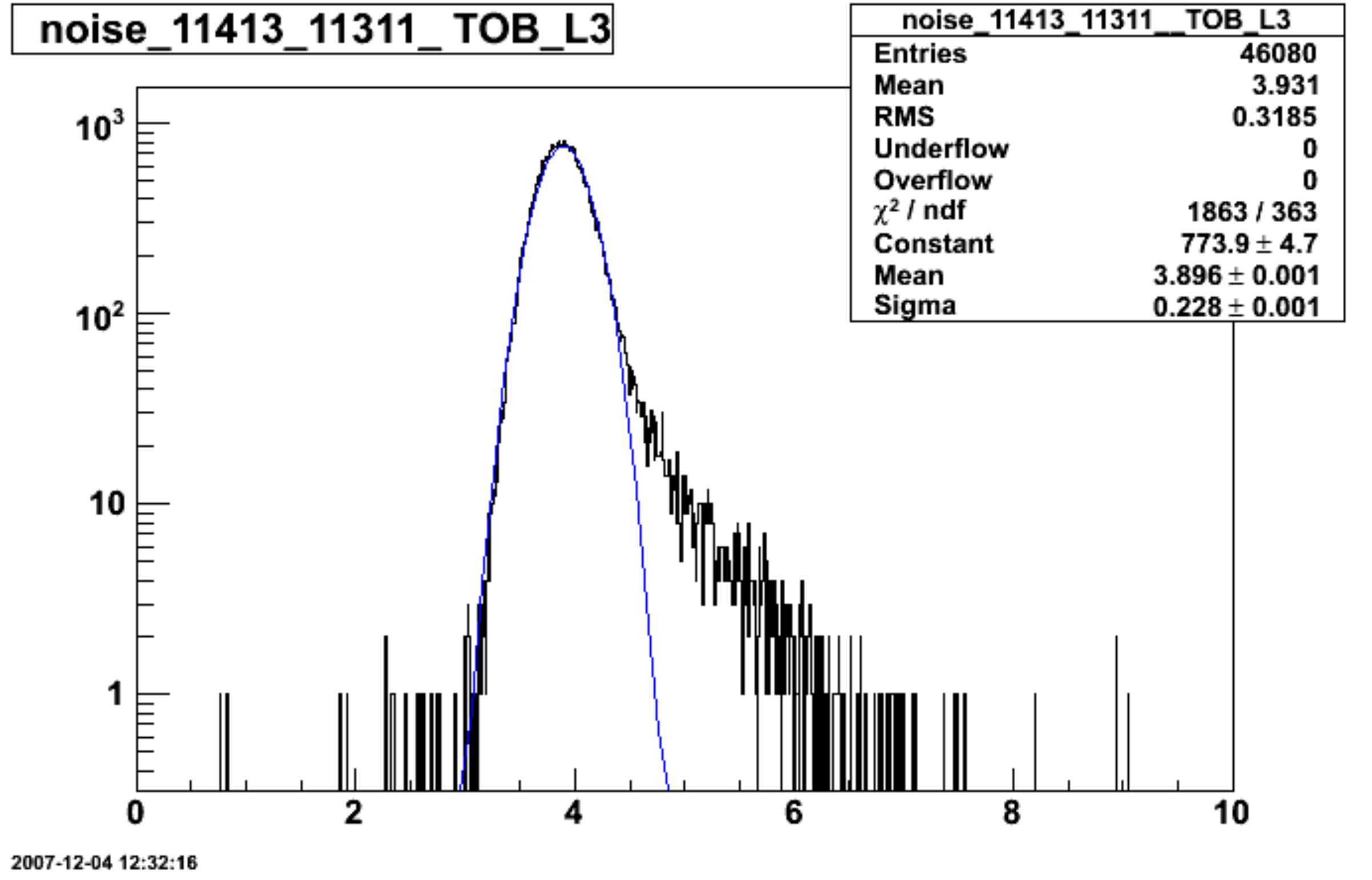}
    \includegraphics[width=0.39\textwidth]{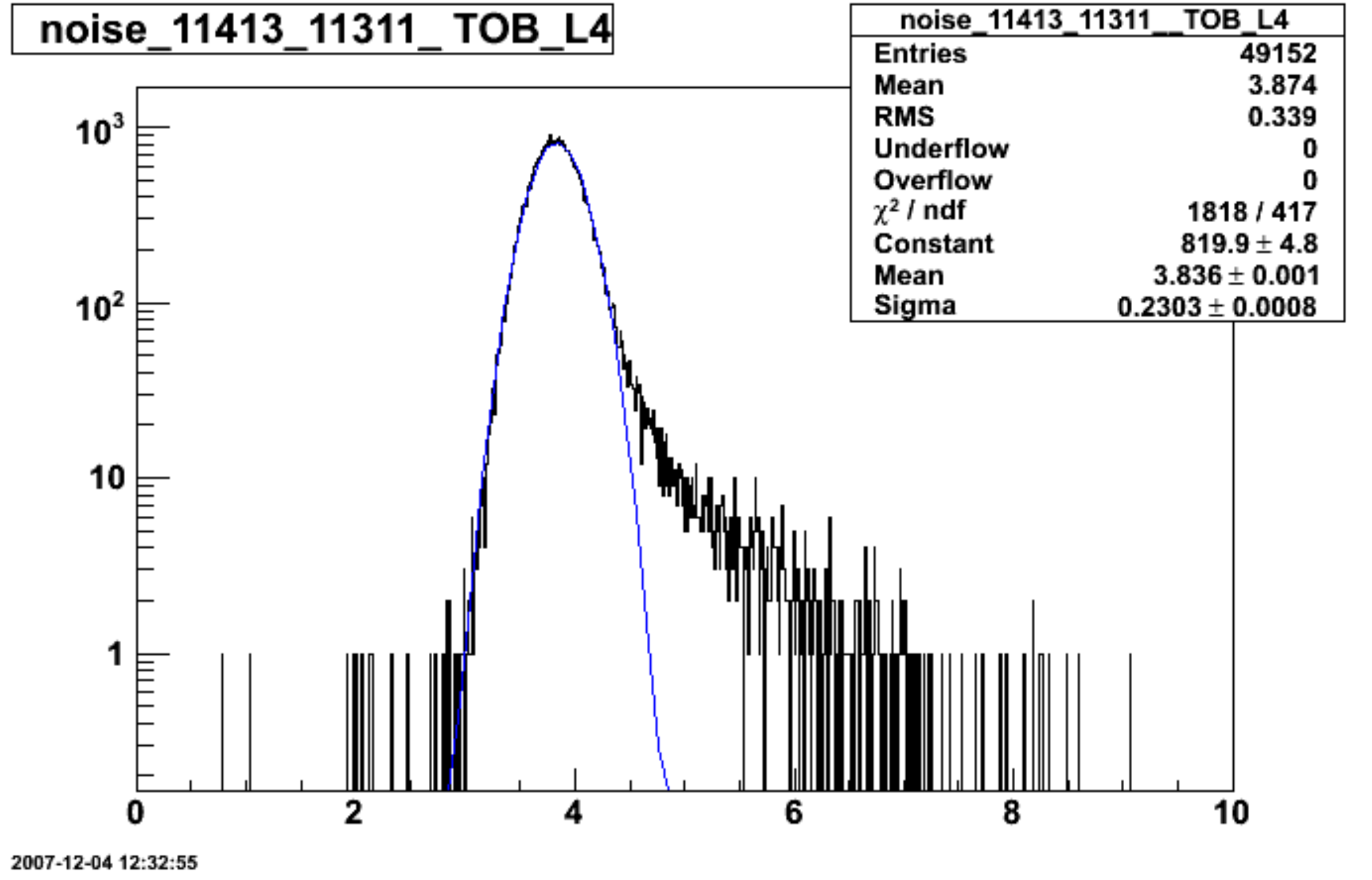}
    
    \includegraphics[width=0.39\textwidth]{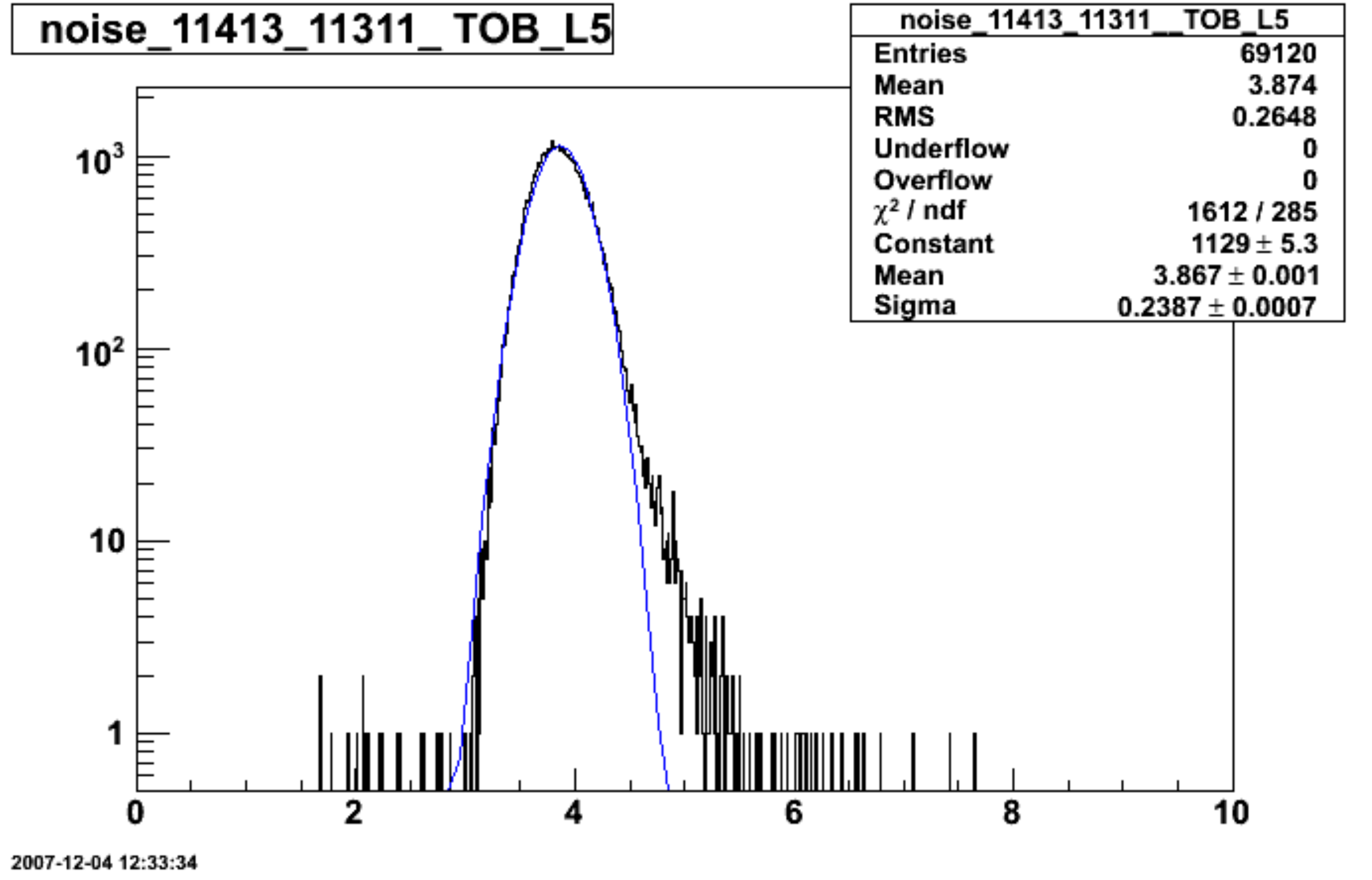}
    \includegraphics[width=0.39\textwidth]{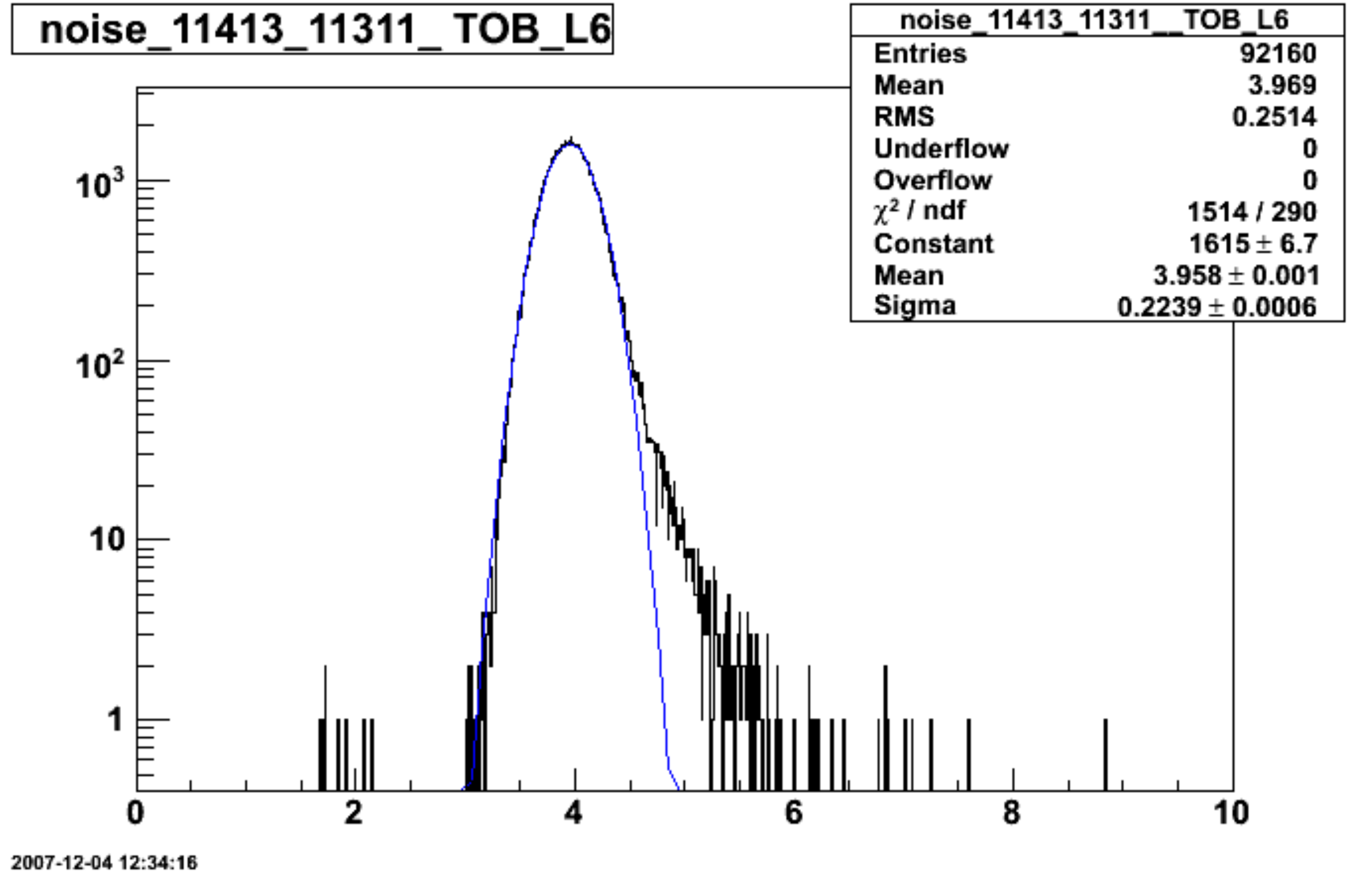}
    \caption{\sl TOB noise profile for T= $-10\,^{\circ}\mathrm{C}$ for layers 1, 2, 3, 4, 5 and 6. Gaussian fit is shown. X axis is ADC value }
    \label{fig:tob-noise}
  \end{center} 
\end{figure}
\begin{figure}[!htb]
  \begin{center}
    \includegraphics[width=0.39\textwidth]{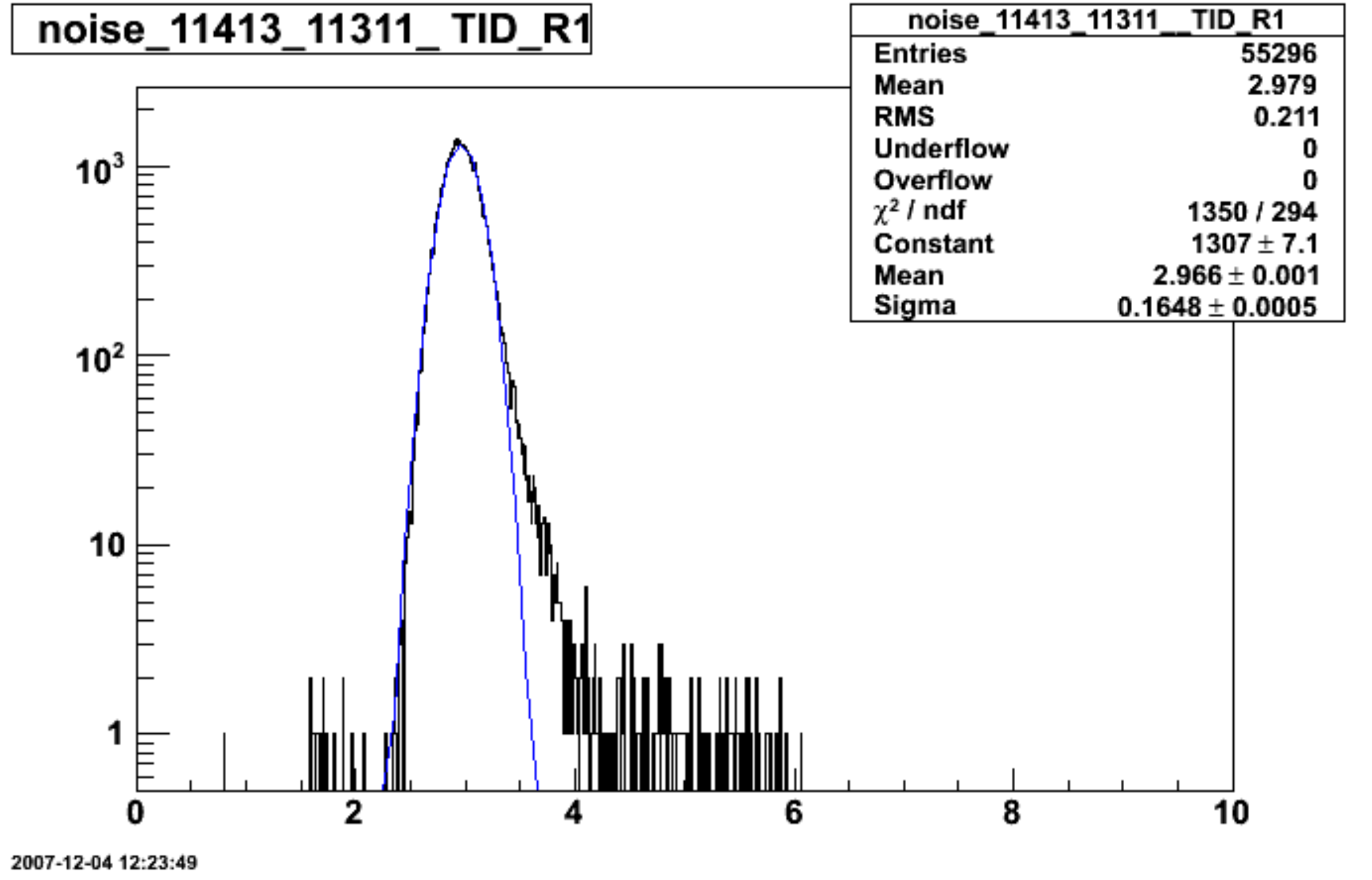}
    \includegraphics[width=0.39\textwidth]{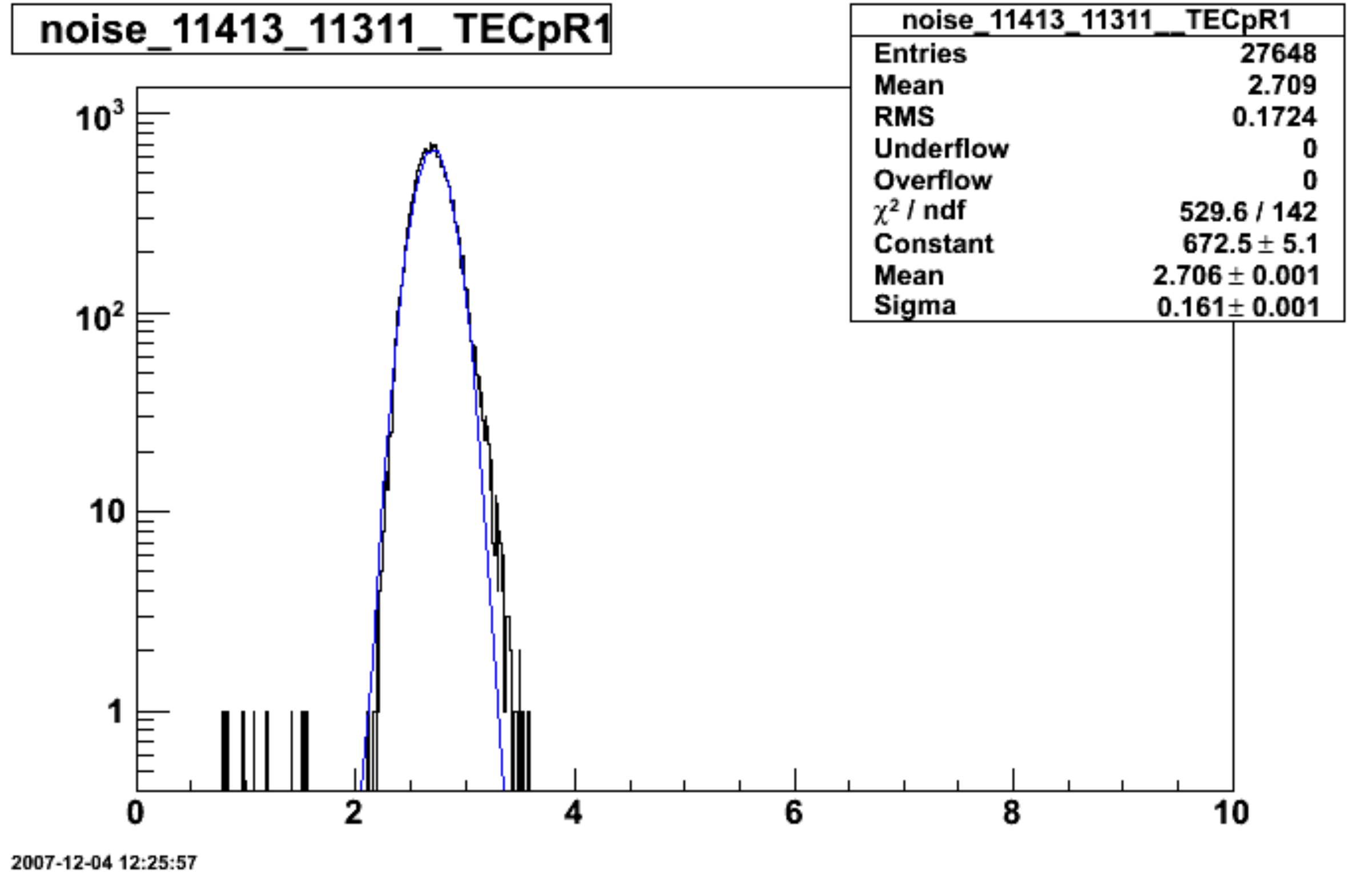}
    
    \includegraphics[width=0.39\textwidth]{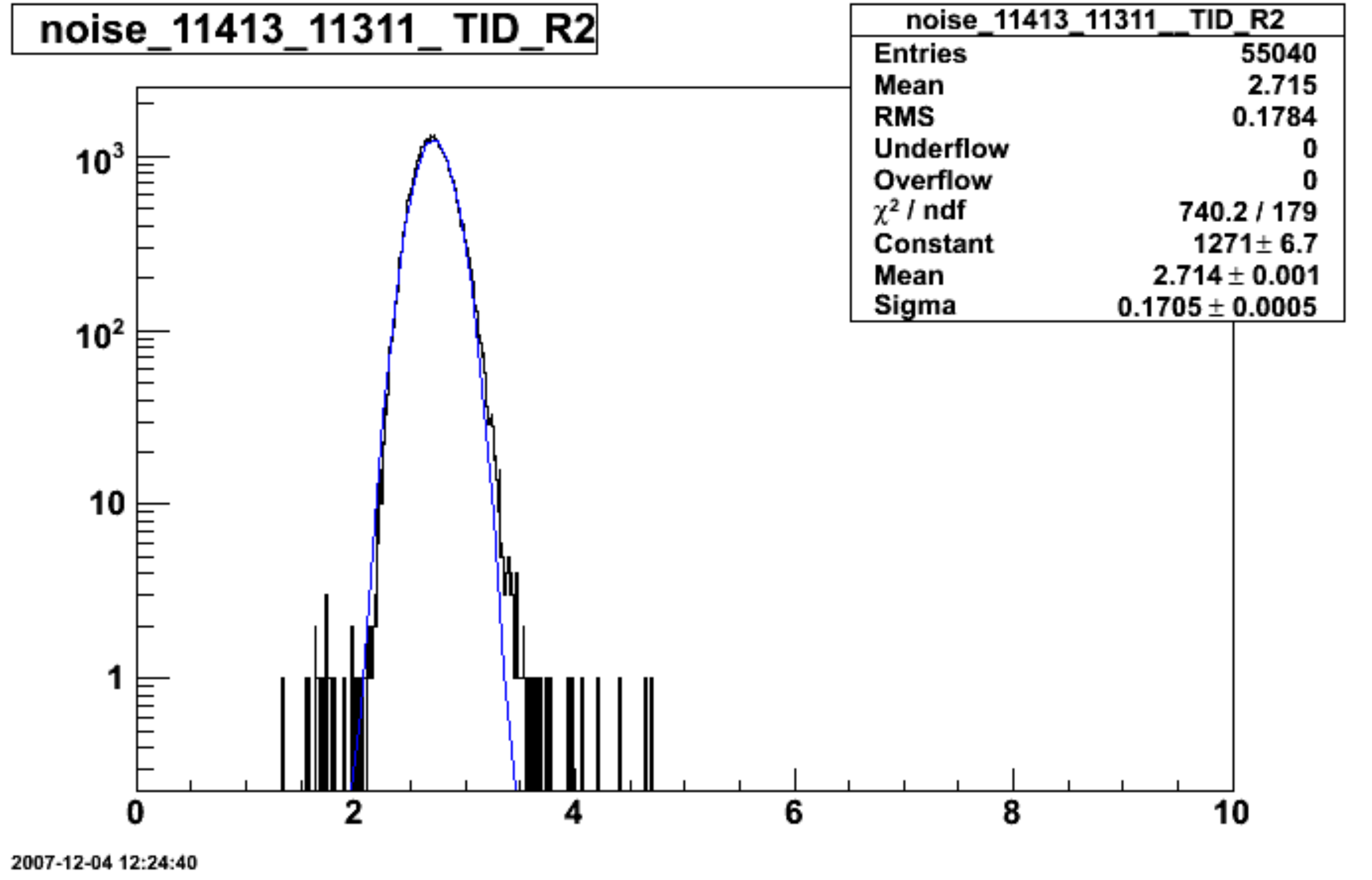}
    \includegraphics[width=0.39\textwidth]{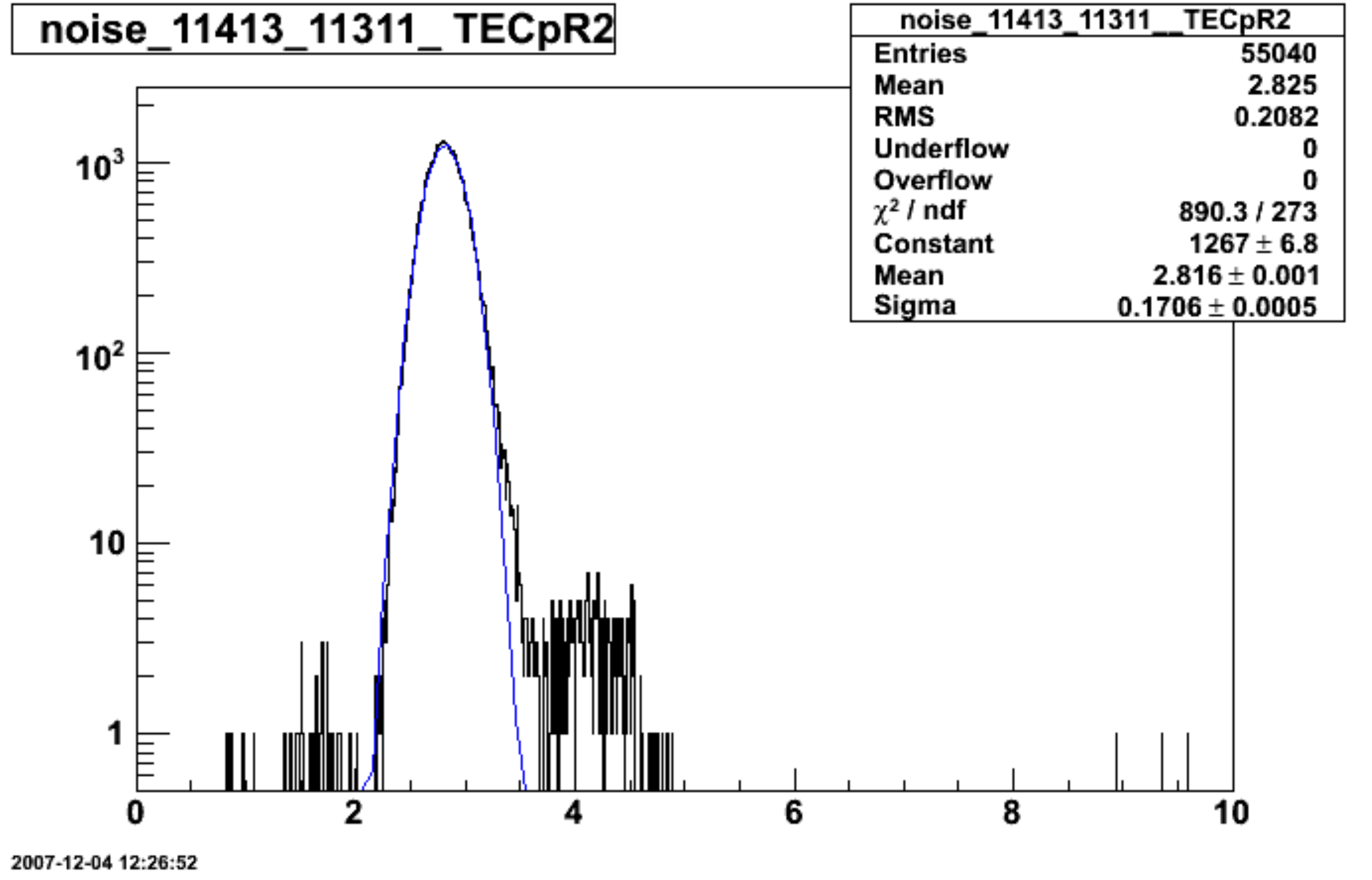}
    
    \includegraphics[width=0.39\textwidth]{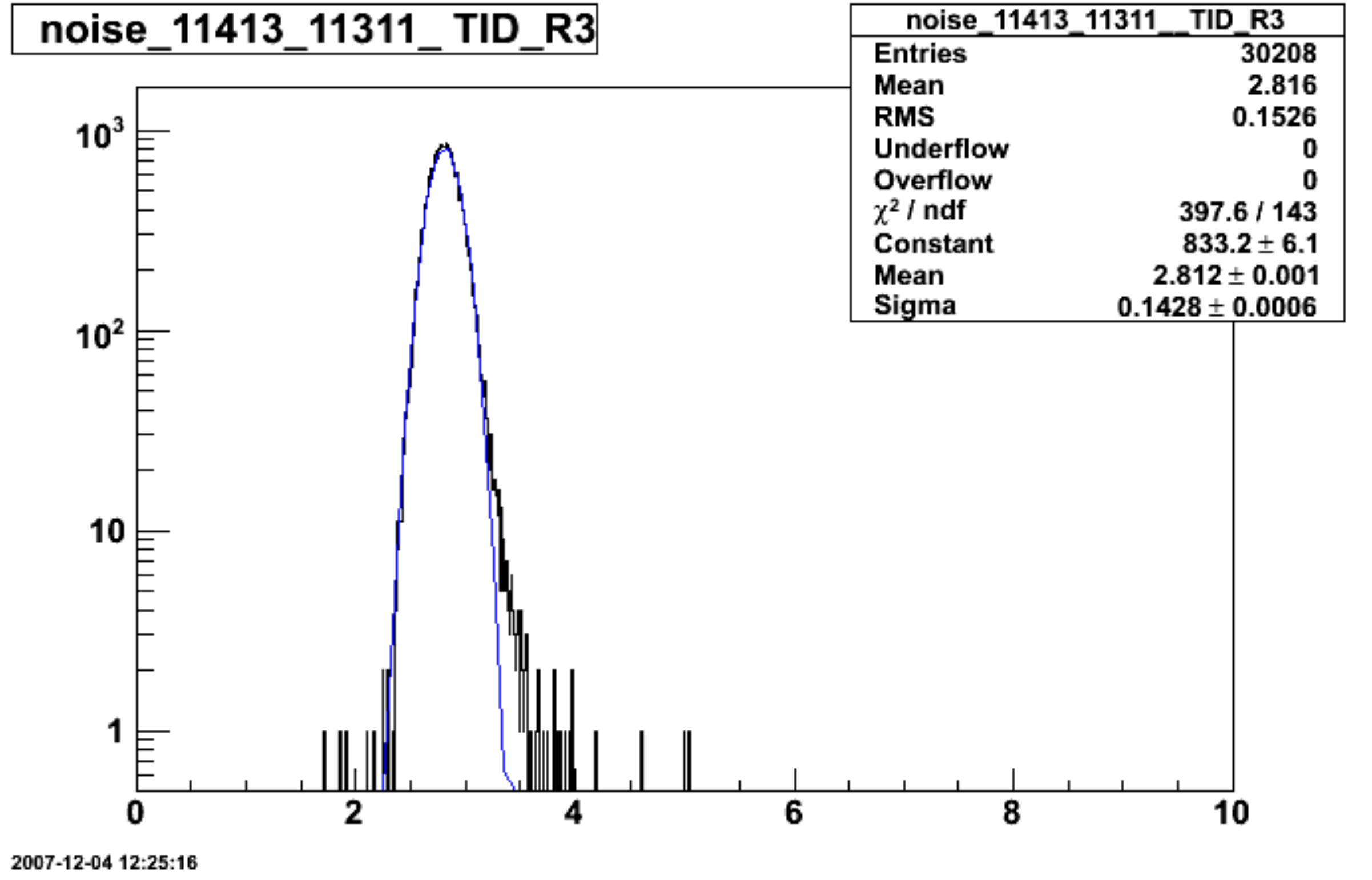}
    \includegraphics[width=0.39\textwidth]{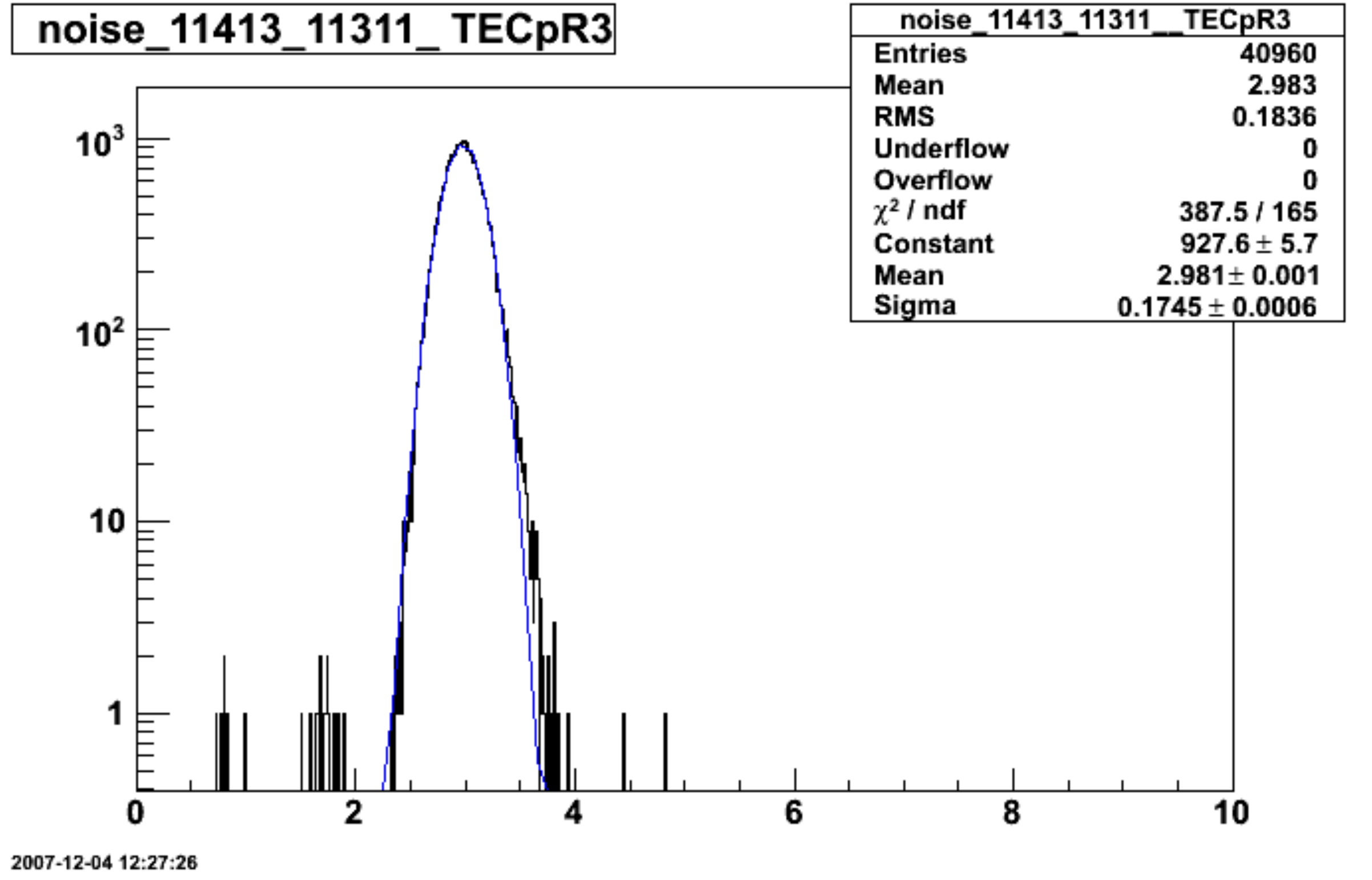}
    
    \includegraphics[width=0.39\textwidth]{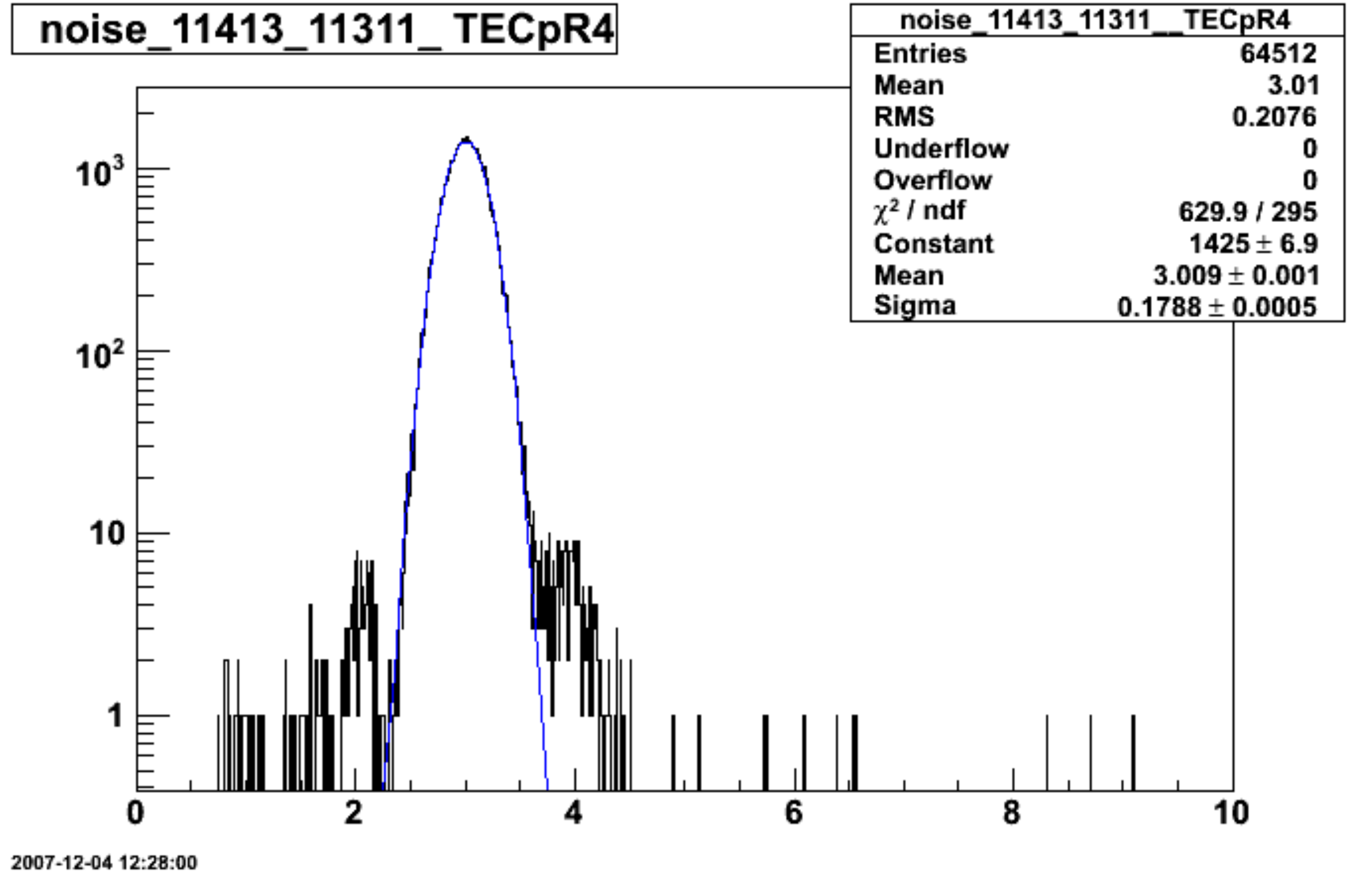}
    \includegraphics[width=0.39\textwidth]{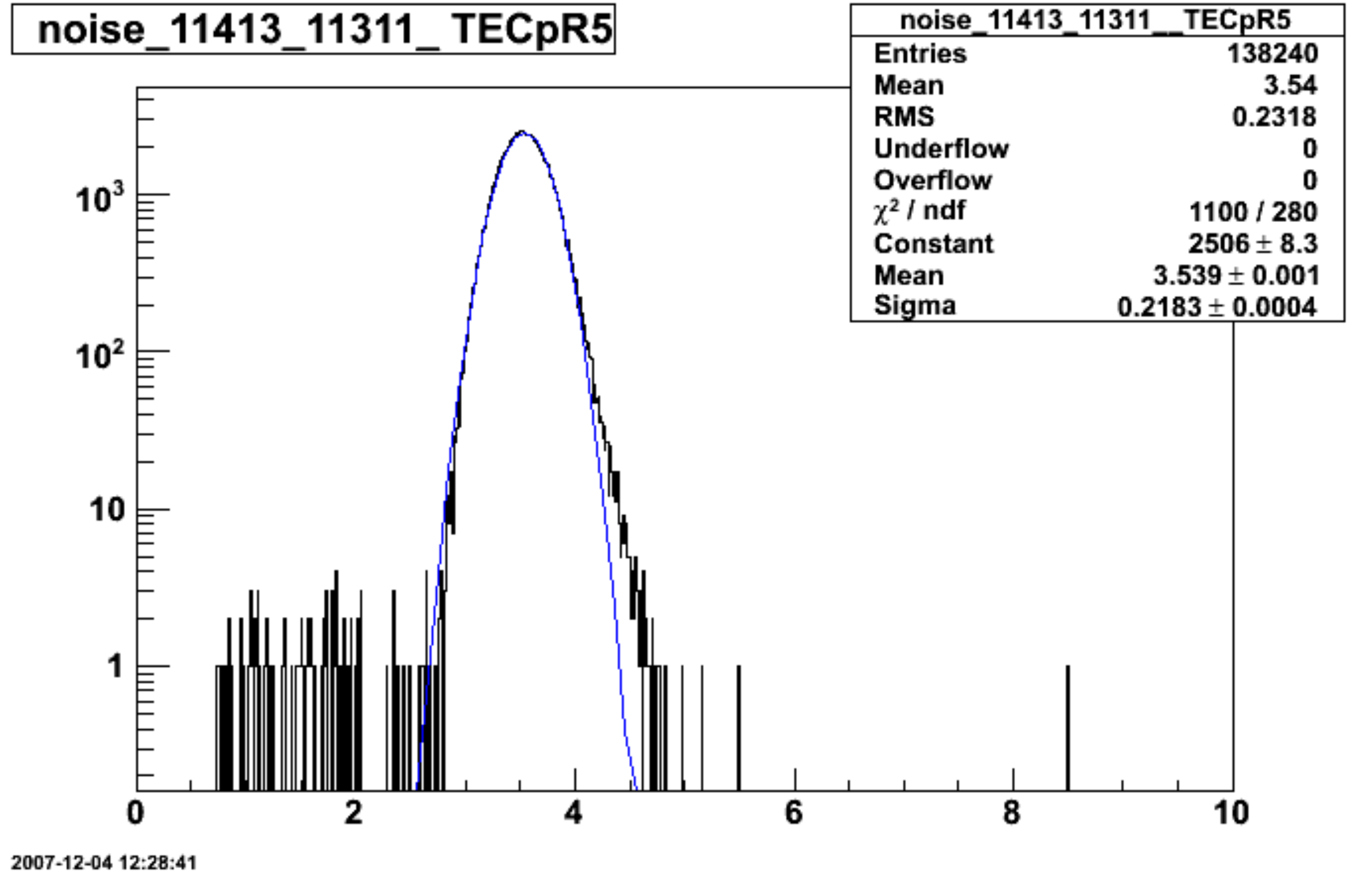}
    
    \includegraphics[width=0.39\textwidth]{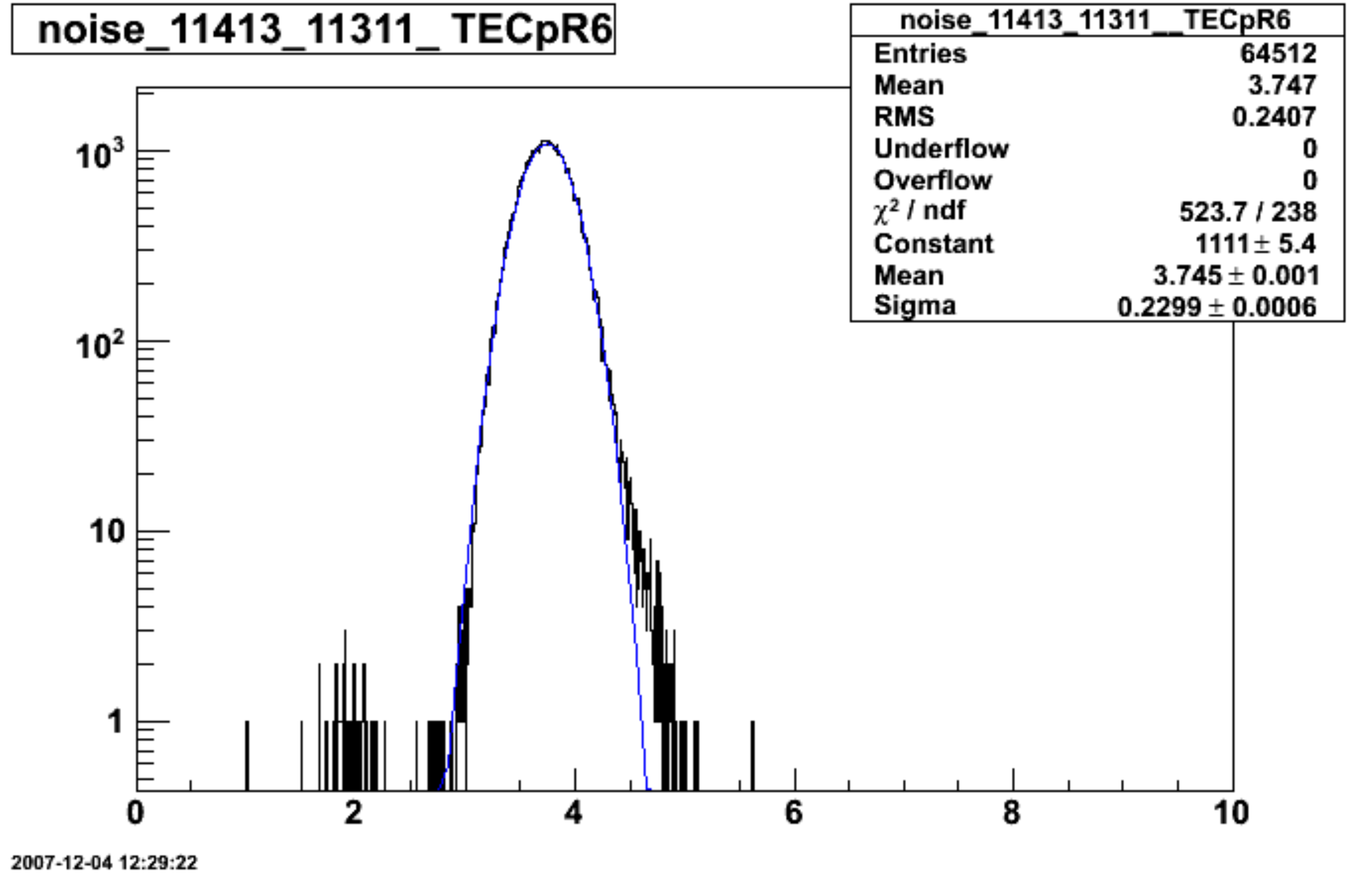}
    \includegraphics[width=0.39\textwidth]{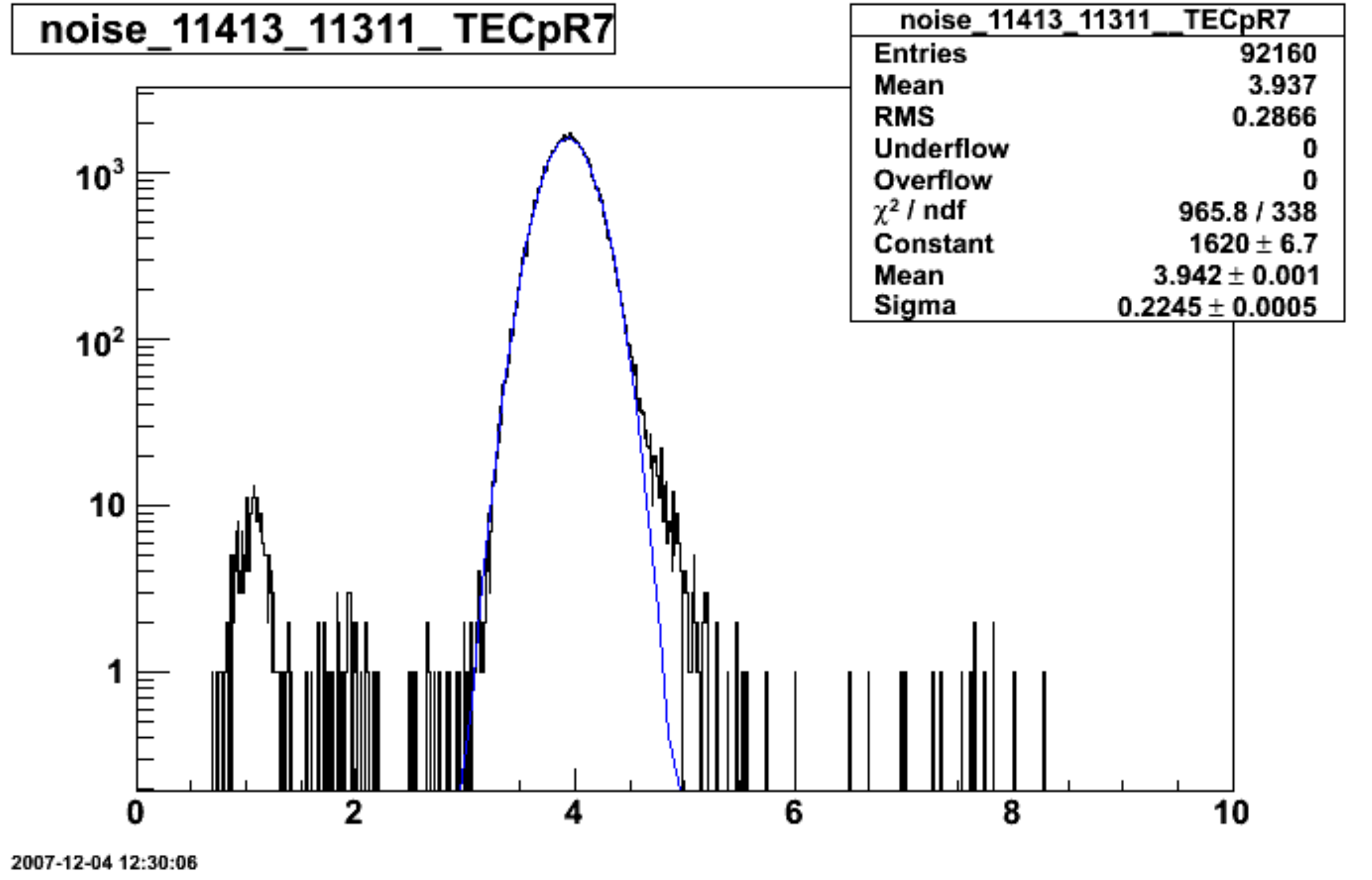} 
   
    \caption{\sl Noise profiles for T= $-10\,^{\circ}\mathrm{C}$ for different rings of TEC and TID. Gaussian
    fits are shown.  R2 and R3 silicon geometries are the same for TID
    and TEC.  X axis is ADC value}
    \label{fig:tectid-noise}
  \end{center}
\end{figure}

\subsection{Noise Performance Studies}
Prior to irradiation, the noise of a module is almost completely determined by the
input capacitance load at  the APV25, which in turn is dominated by the
 silicon strips. Thus, within 8\%, one expects a linear  dependence of the noise on the length of the silicon strip for all modules.

In the Sector Test, 
modules were mounted on the final support structures and therefore were in
close proximity to each other, which could have exposed  APV25
inputs to other sources of noise. In particular grounding loops, cross-talk from neighboring modules or other
noise sources, (digital noise, cables, power supply) could have affected the
final noise performance. Moreover, a study of the
temperature dependence is important. The commissioning procedure plays
an important role, since it optimizes many system parameters  and consequences are observed on the noise
performance. In the following the noise has been renormalized for each
pair of adjacent APV25, belonging to the same AOH laser,  in order to take 
into account the different electronic gains, by multiplying the noise
for the correction factor specified by equation~\ref{form:corr}.

 Modules with known problems that affect the noise
distribution calculation have been removed from the analysis at least
for the period when the problem was present.


Strip noise distributions for runs taken at $-10\,^{\circ}\mathrm{C}$
are shown for each layer for TIB and TOB in Fig.~\ref{fig:tib-noise} and  Fig.~\ref{fig:tob-noise}
respectively, while for TEC and TID are shown in Fig.~\ref{fig:tectid-noise}. 
For each distribution, a fit to a Gaussian has been performed and results are shown. 
For most of the layers the noise distribution is very well represented by a Gaussian, and fitted values and sigmas are almost identical for identical layers, showing  extra noise sources do not affect the detector performance significantly.
 

The only  relevant non-Gaussian tails at high values are visible for TOB layers 2, 3 and 4.  They are mainly due to channels close to APV25 edges and only for modules at specific positions within a rod, closer to a clock distribution board and a power cable. This noise pickup does not affect the TOB performance,  given its high signal to noise ratio, as shown in the next section. Nevertheless, during the Sector Test many possible grounding and filtering schemes were investigated in order to minimize this extra noise.  
A solution which was found to be very effective consisted in grounding  the TOB power cable shields at the patch panel close to the Tracker. This grounding implementation was possible on the Sector Test setup only for a fraction of the TOB, therefore the tails remained for most of the module rods. 
This grounding scheme has subsequently  been implemented during  installation of the Tracker inside CMS.



\begin{figure}[htb]
  \begin{center}
    \includegraphics[width=0.49\textwidth,angle=0]{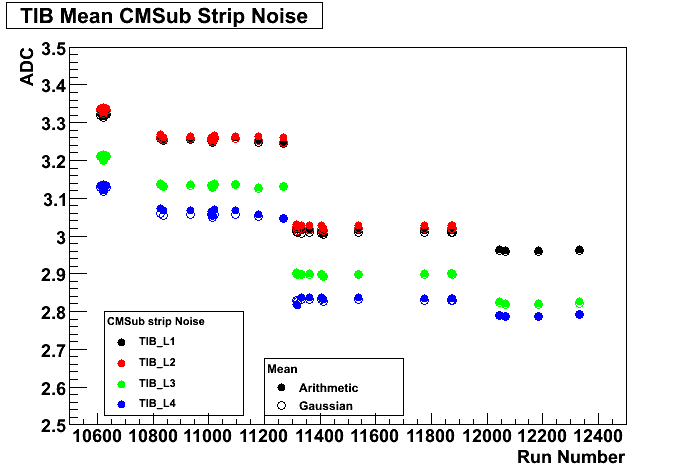}
    \includegraphics[width=0.49\textwidth,angle=0]{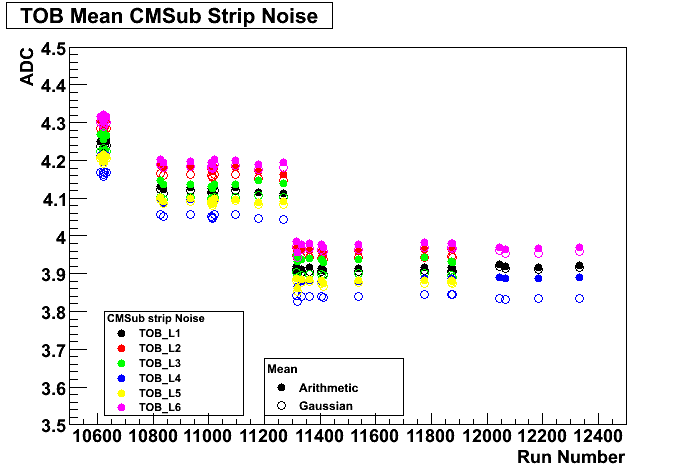}
    \caption{\sl  Noise vs run number for TIB on the left and for TOB on the right. 
   Four periods at different temperatures are visible: $10\,^{\circ}\mathrm{C}$, $0\,^{\circ}\mathrm{C}$,  $-10\,^{\circ}\mathrm{C}$ and  $-15\,^{\circ}\mathrm{C}$}.
    \label{fig:noivsrun}
  \end{center}
\end{figure}

Stability of the noise performance was studied  by taking 
pedestal and noise runs at different times when the Tracker was running in a stable conditions
 with fixed  electronics configuration settings. Results are shown in  Fig.~\ref{fig:noivsrun} displaying  noise of all layers/rings versus the run number for TIB and TOB: the steps in values represent the different 
temperatures considered,  $10\,^{\circ}\mathrm{C}$,$0\,^{\circ}\mathrm{C}$,  $-10\,^{\circ}\mathrm{C}$ and  $-15\,^{\circ}\mathrm{C}$. The data average and mean value of a Gaussian fit are
displayed as solid and open symbols respectively.  For constant coolant temperature the noise is stable to better than $\pm 0.5 \%$. Most importantly, the noise decreases with
decreasing temperature as  expected by the laboratory studies made 
on the APV25 performance.

\begin{figure}[htb]
  \begin{center}
      \includegraphics[height=0.45\textwidth,angle=0]{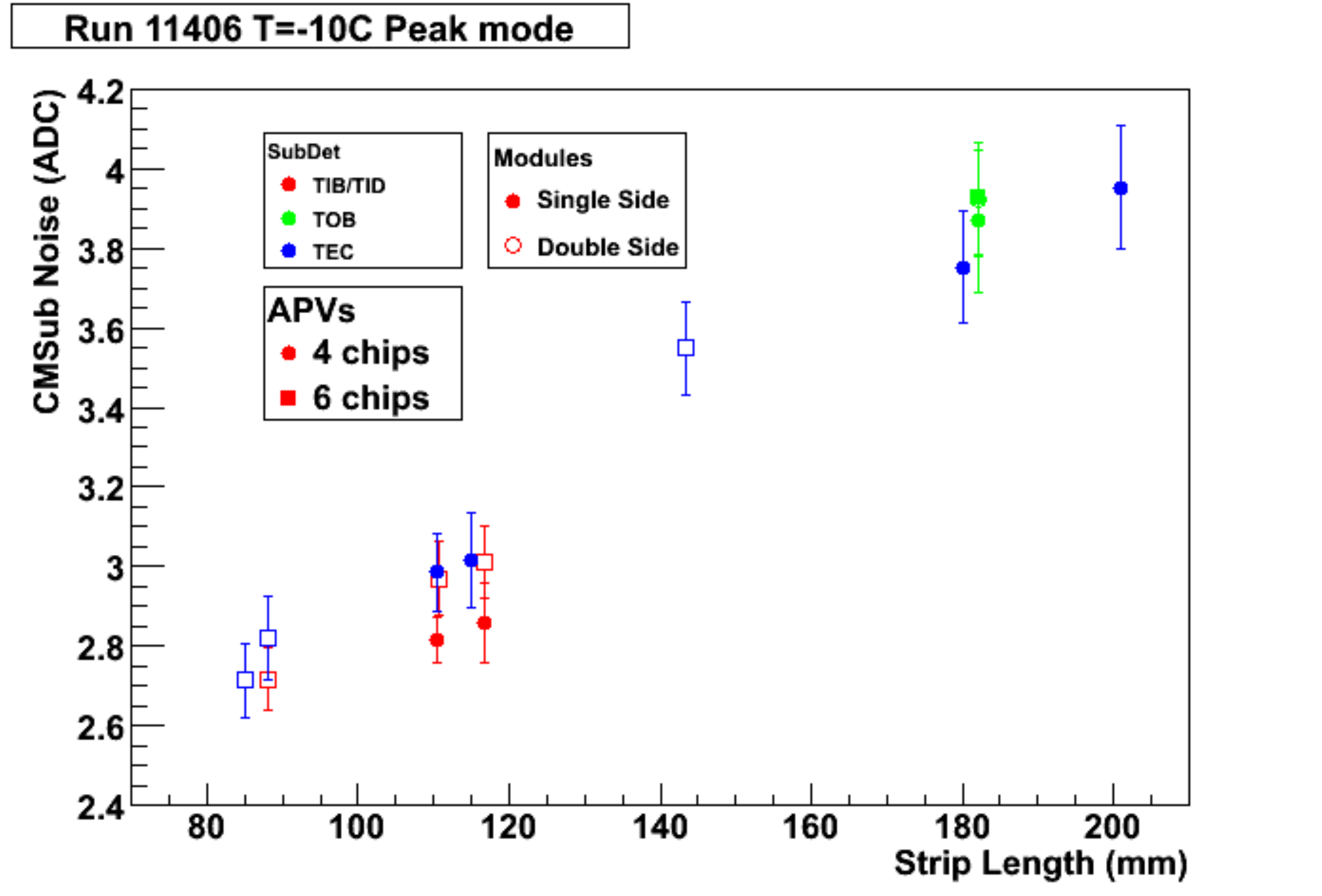}
      \caption{\sl Noise vs strip length for T= $-10\,^{\circ}\mathrm{C}$.}
      \label{fig:noise-vs-l}
  \end{center}
\end{figure}

The (Gaussian fit) mean strip  noise obtained for each different
type of module has been correlated with the module strip
length. Results are shown in Fig.~\ref{fig:noise-vs-l} for a run 
taken at $-10\,^{\circ}\mathrm{C}$, but similar results are obtained at the other temperatures.
  Error bars represent the
spread of the mean module noise for each module type. As expected the
behavior is well represented by a linear behavior, within 
statistical fluctuation, for modules from the same layer, although the average
values show  differences for the same strip length but different
module geometries. This can be explained by the difference in supply  voltage
and by the different APV25 settings used for different sub-detectors.

\subsection{Detector quality}\label{sec:quality}




Faulty channels have been studied both from the perspective of  badly behaving modules or missing fibers and of individual bad channels.

Modules with known problems or that were badly behaving were removed either from DAQ or from the data analyses. The resulting fraction of missing modules was at the 0.5\% level. Dead fibers were identified during timing runs based on low tick-mark heights. They correspond to the broken fibers whose channels showed problems during the timing runs. The number of missing fibers in the Tracker was at the  0.1\% level.

Remaining isolated bad channels were identified having 
higher than five sigmas, referred as noisy, or lower than five sigmas, referred as dead, noise compared to the average noise per module, and this was done for each module geometry of the  Tracker. 

\begin{figure}[!h]
  \begin{center}

    \includegraphics[width=0.4\textwidth,angle=0]{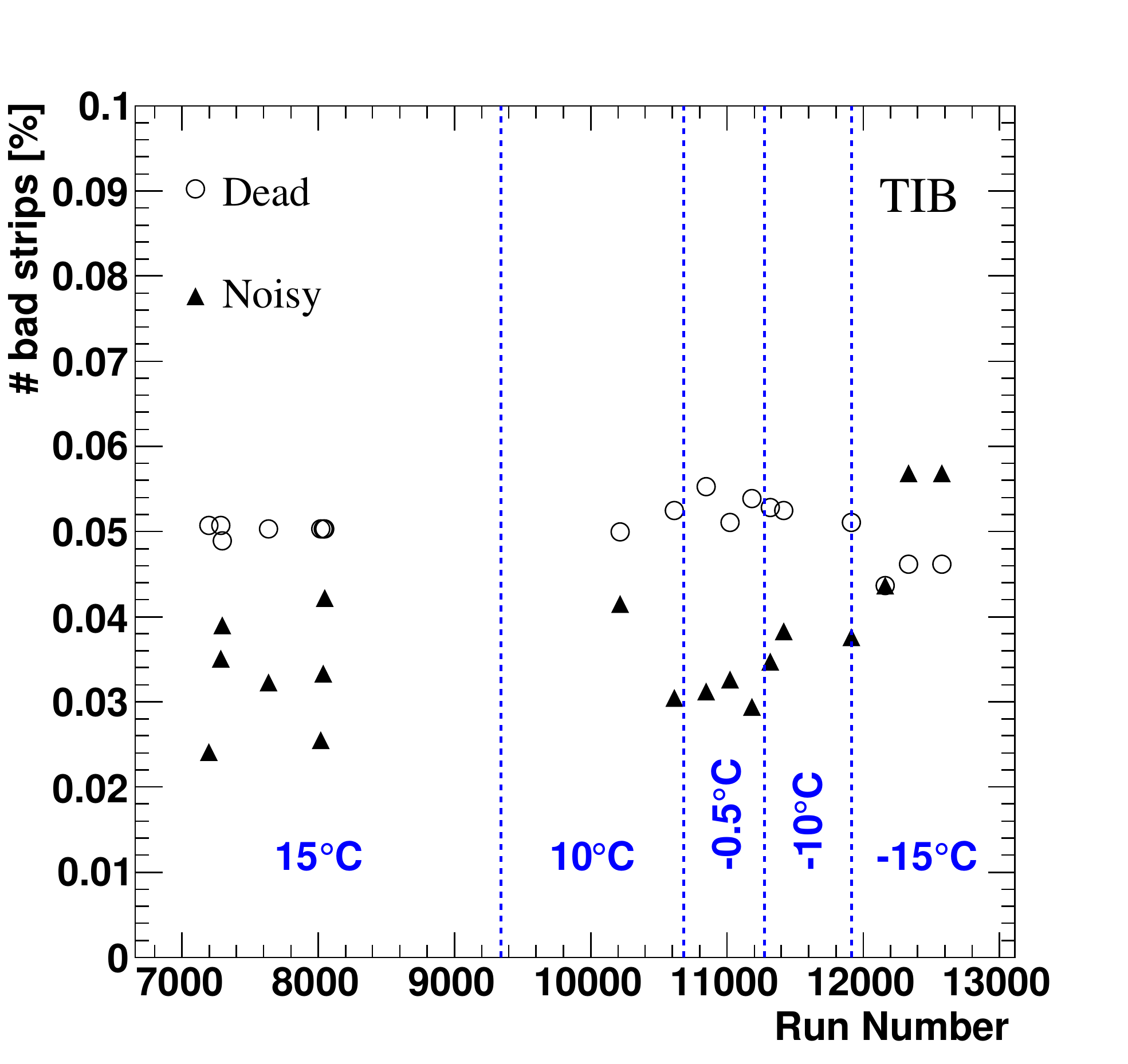}
    \includegraphics[width=0.4\textwidth,angle=0]{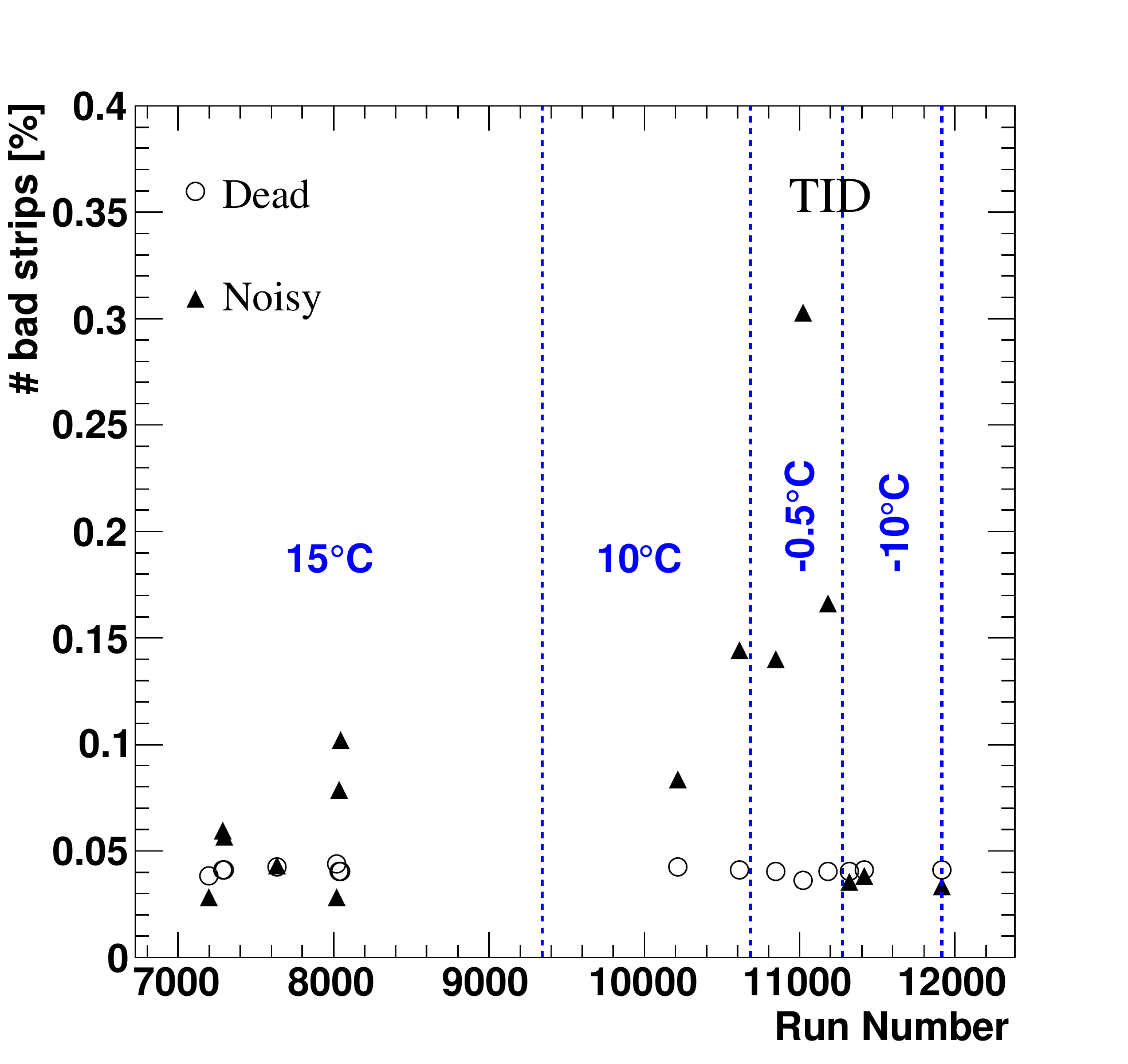}

    \includegraphics[width=0.4\textwidth,angle=0]{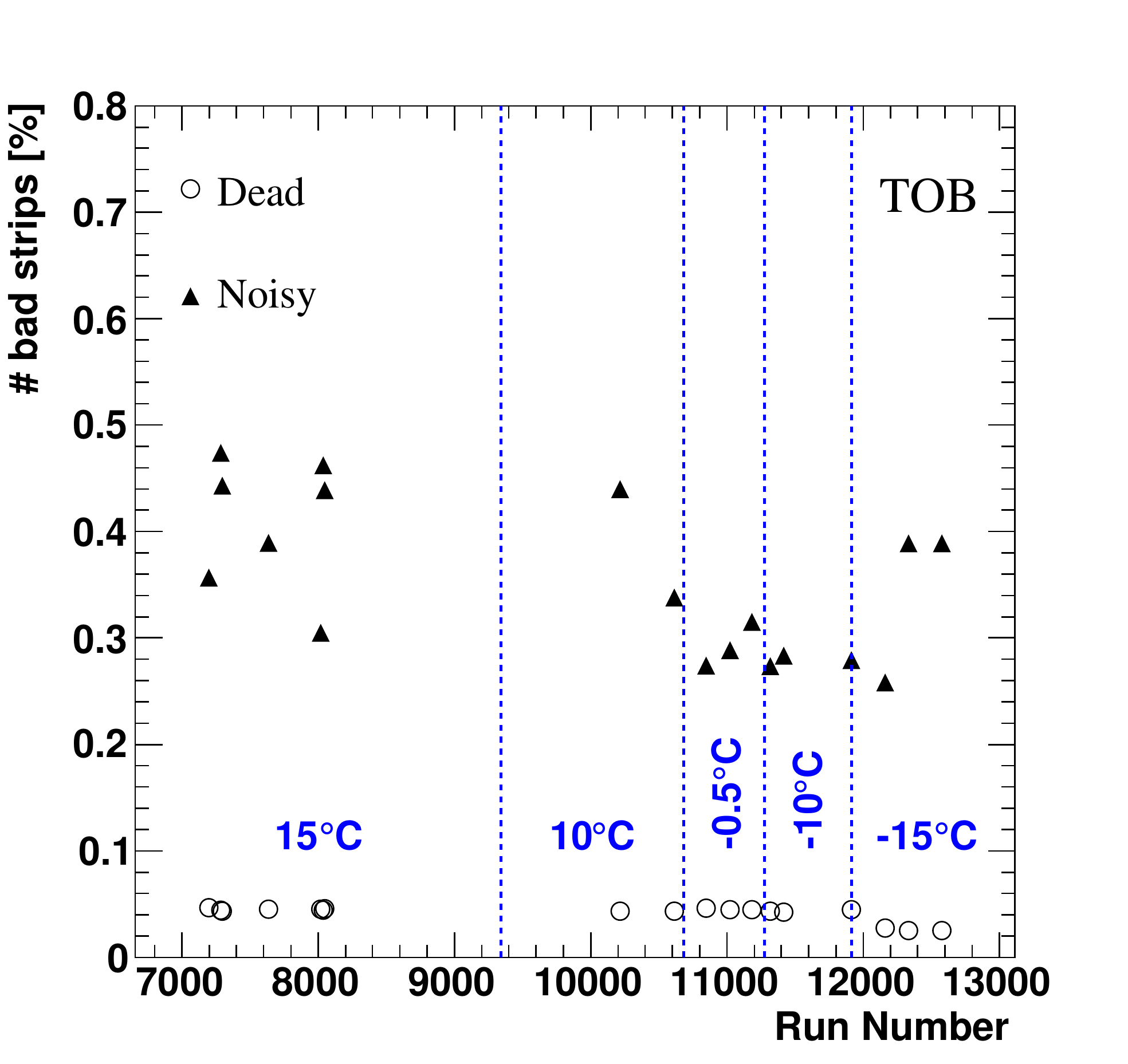}
    \includegraphics[width=0.4\textwidth,angle=0]{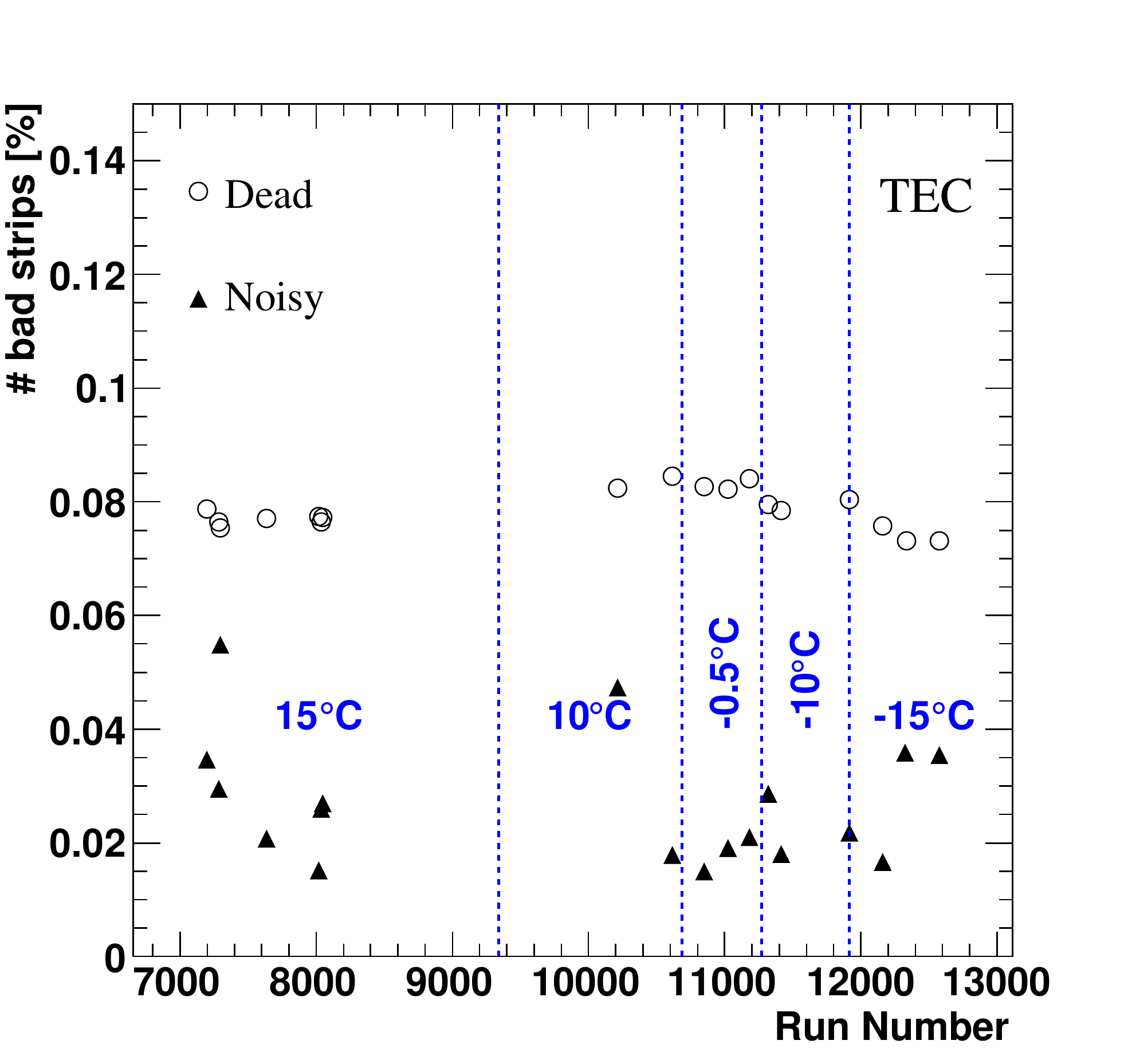}
      \caption{\sl Noisy and dead strips vs run for TIB,
      TID, TOB, and TEC. }
      \label{fig:noisy-dead-gordon}
  \end{center}
\end{figure}

The results of this analysis as a function of the run number are shown in Fig.~\ref{fig:noisy-dead-gordon}. 
The number of dead channels is almost constant among several runs for all sub-detectors, showing that the identification of dead channels is clear and stable. The noisy components are instead subject to  fluctuations, in particular  for TOB and TID.
 The fraction of dead (noisy) strips is 0.05\% (0.04\%) for TIB, 
0.04\% (0.15\%) for TID, 0.04\% (0.3\%) for TOB, and 0.08\% (0.02\%) for
TEC.

Since the analysis of defective strips is made on a per run basis, it
is important to understand the number of runs in which a strip was
identified to be bad.  Dead strip
identification is stable: the majority of dead strips (70\%) were flagged 
in all runs. About 30\% of the classified dead strips appear only in a single run which likely had a timing issue or other  unusual problem.
On the contrary, only a small fraction of the noisy strips  were noisy throughout the Sector Test. In most cases anomalous noise persists for one or two
 runs at the most. These runs may reflect special operating conditions or non-optimal system configuration.

Finally, a comparison of the identified faulty channels with data from the 
Tracker construction database shows
that 90\% of the dead channels and 40\% of the noisy channels had been
flagged as such by the end of the construction period.

\section{Detector Performance based on Cosmic Ray Data}
The signal performance of the Tracker is very important;
it depends on several factors: charge collection of the silicon sensors,
 performance of the APV25 with well defined parameters, performance of the full electronic chain.
 
A study of cosmic trigger timing is presented in the first subsection,
to verify that the maximum signal was taken for all the
runs. As tracking efficiency relies on a sufficiently large signal to noise ratio (S/N) on all
Tracker modules, it is important to measure the S/N on a
layer by layer basis. Moreover, 
S/N analysis does not depend on the gain
calibration. Lastly, analysis of the signal size allows a
straightforward comparison of the performance of different layers as a
means of determining the absolute gain calibration and 
measurements of temperature dependence.

In order to obtain a high purity signal, only those hits that are
associated with a reconstructed track have been used. The CTF track
algorithm was used and only tracks with 
$\chi^2 /d.o.f. < 30$ 
were considered. Only events with low track multiplicity (less than
three) and with a low hit multiplicity (less than 100) were considered.

\begin{figure}[ht]
  \begin{center}
    \includegraphics[width=0.3\textwidth,height=4cm]{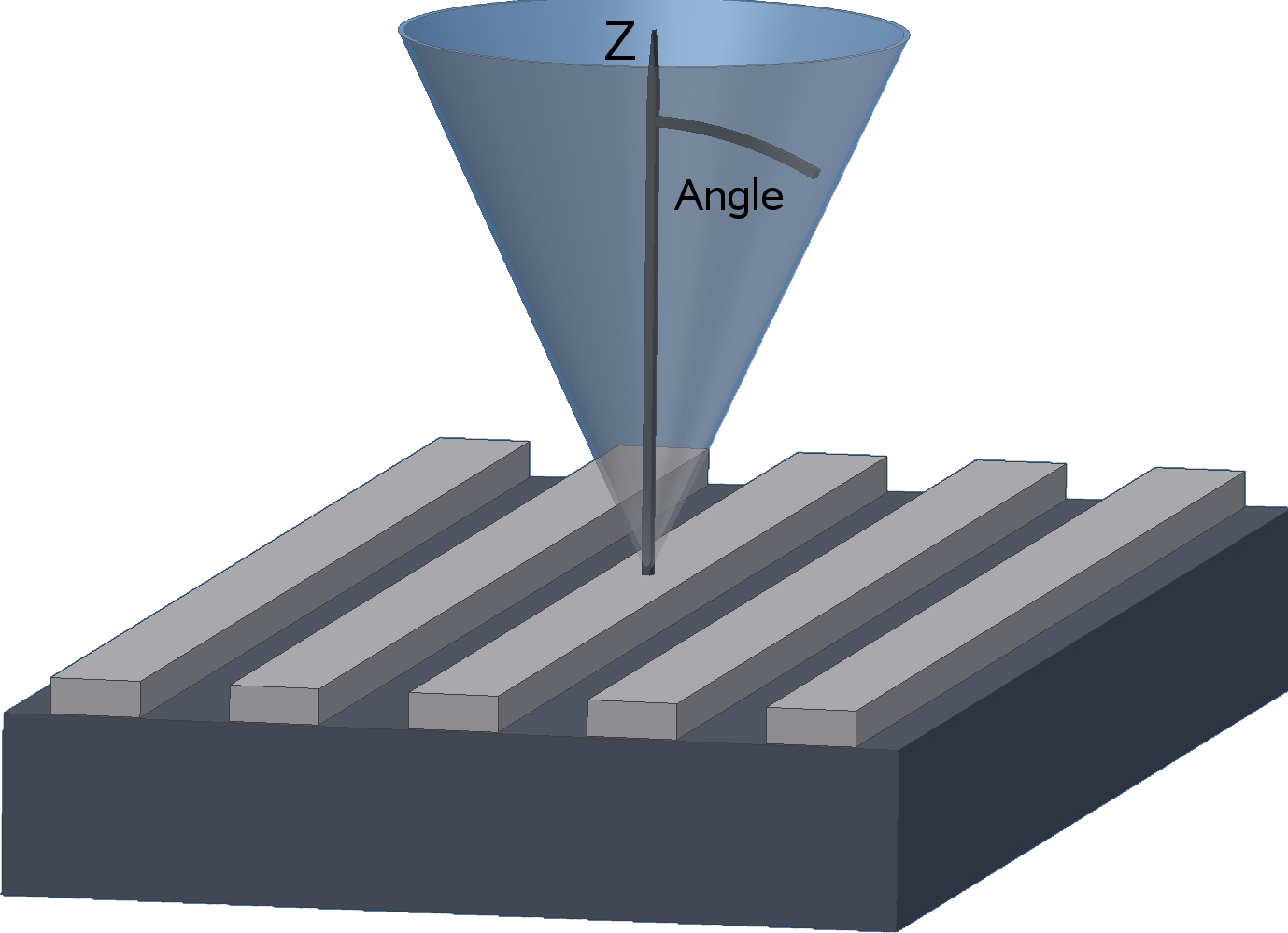}
    \includegraphics[width=0.3\textwidth,height=4cm]{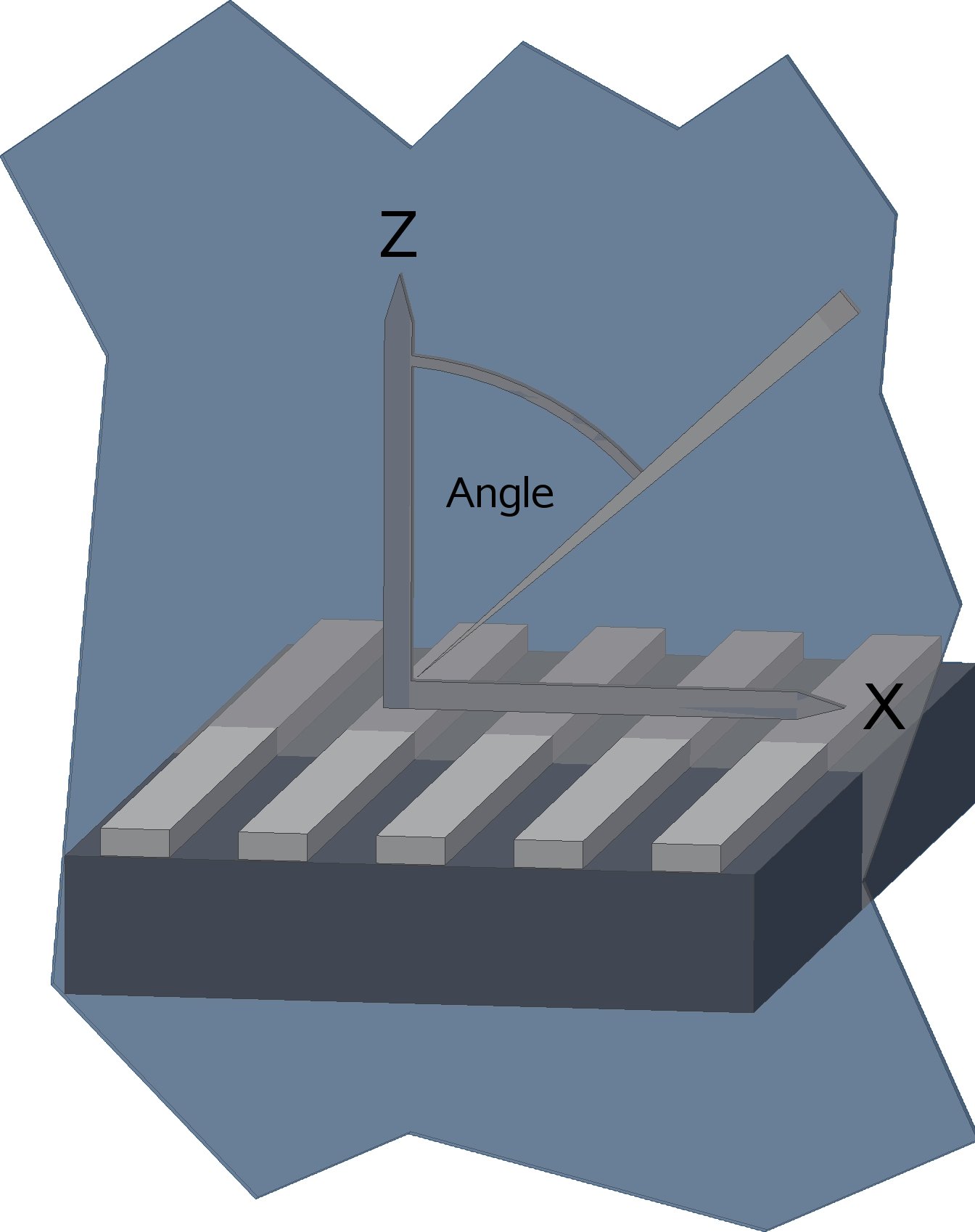}
    \caption{\sl Angle definitions: $\theta_{3D}$ on the left;
    $\theta_{XZ}$ on the right.}
    \label{fig:local_angles}
  \end{center}
\end{figure}

The energy ($S_{tot}$) deposited in Tracker modules can be parameterized:

\begin{equation}
 S_{tot}  =   \frac{dE}{dx} t K ;  K = \frac{1}{cos( \theta_{3D} )}
\end{equation}
 
where $\theta_{3D}$ is the angle of incidence of the track with
respect to the silicon detector normal, see Fig.~\ref{fig:local_angles};
$t$ is the silicon thickness. 

In the following  analyses signal values
are normalized to the thickness of the silicon detectors.

\begin{equation}
 S_{ren}  =   \frac{S_{tot}}{K} = \frac{dE}{dx} t; 
     \label{form:SK}
\end{equation}

Fig.~\ref{fig:local_angles} illustrates the definitions of two of the three
 angles referred to in this section. The $XZ$ angle is angle made by
the track in a plane orthogonal to the sensor surface and whose $X$
axis is transverse to the strip direction. There is a direct
correlation between the $XZ$ angle and cluster size. The $YZ$ angle is
defined in a plane perpendicular to the surface and oriented along the
strip direction.

\subsection{Latency Scan}

The response of CMS silicon modules to signals is detailed
in~\cite{timingNote}, and only some key points are highlighted in this
section. 
 Ideally, the analytical form of the pulse shape in peak mode
is the transfer function in the time domain of a CR-RC circuit:

\begin{equation}
S_{\mathrm{peak}}(t) \propto \frac{t}{\tau} e^{-t/\tau},
\label{eq:peak}
\end{equation}

where $\tau$ is the rise time, and the time $t$ is positive. The pulse from the APV25 amplifier lasts about 300\,ns, which is large
with respect to the 25\,ns time which separates two bunch crossings.

\begin{figure}[!hbt]
  \begin{center}
    \includegraphics[width=0.7\textwidth]{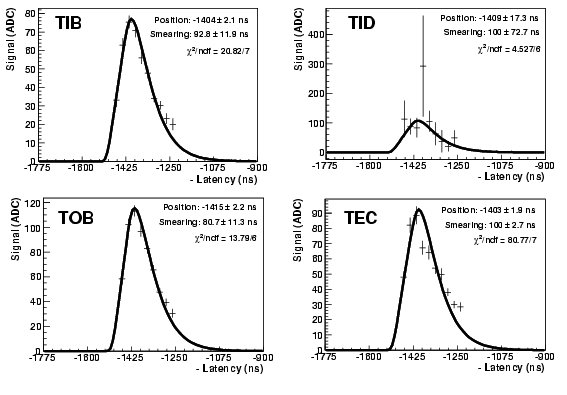}
    \vskip -2ex
    \caption{\sl. 
     Fit of the signal as a function of the latency, for each of the
     four sub-detectors. Data are from run 10947 and were used to set
     the latency for the following period.  
    }
    \label{fig:lat-fitsub}
  \end{center}
\end{figure}
\begin{figure}[!hbt]
  \begin{center}
    \includegraphics[width=0.4\textwidth]{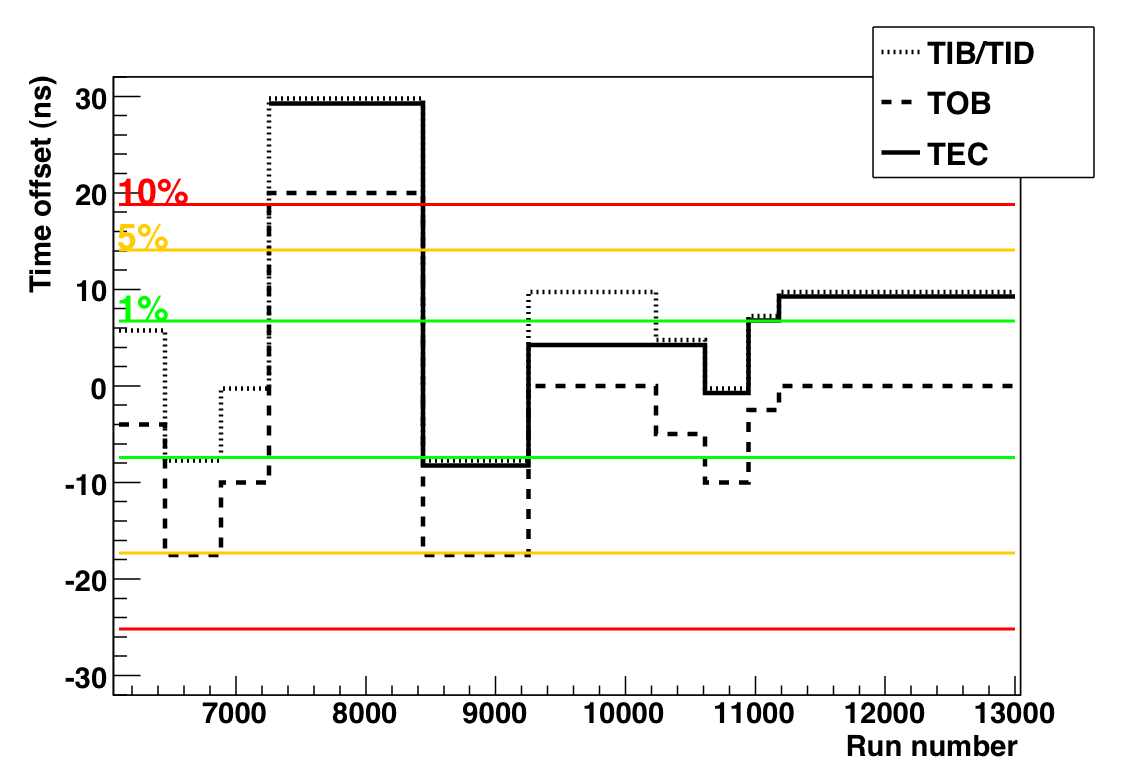}
\includegraphics[width=0.4\textwidth]{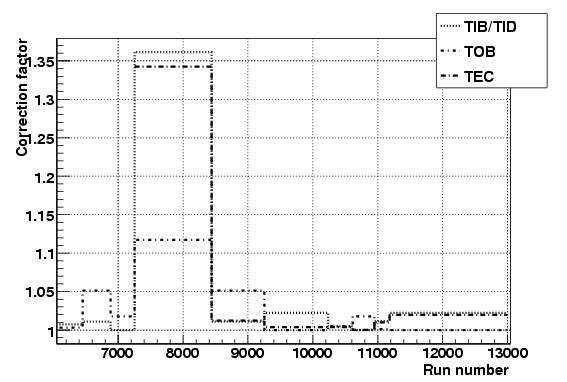}
    \caption{\sl Offset from the on-peak timing (left) and signal
    correction factor (right) as function of the run number for the
    three sub-detectors TIB/TID, TOB, and TEC.}
    \label{fig:lat-offpeak}
  \end{center}
\end{figure}

The APV25 stores a voltage proportional to the input charge  in an internal memory pipeline every 25 ns. The latency
of the trigger determines the pipeline cell containing the maximum
charge from the traversing particle. This pipeline cell number is the same
for all the Tracker modules, since they have been synchronized by the
commissioning ({\em timing}) runs, where the absolute timing of
each module is set in order to accommodate the delays introduced by
the hardware configuration (fiber lengths, CCU configurations, FEDs,
etc.) This is obtained with a precision of 1.04\,ns, the time step of
the dedicated programmable delay available in the Phase-Locked Loop
(PLL), mounted on  each module. The PLL allows the independent shifting of the clock and
trigger signals.

During the Sector Test, latency scans were done after almost every
significant change of APV25 settings or temperature to determine the new
 latency parameters.  Typically about 50 events were taken for each latency step during the
scans. About ten steps in latency were performed for a total
of about 500 triggers on average.

Results from one of the latency scans are shown in
Fig.~\ref{fig:lat-fitsub}. The latency that maximizes the signal peak
for each single subdetector is obtained from fits to these plots.
There are small differences in optimal latency among the four
sub-detectors due to the different lengths of readout fibers (from the
front-end hybrid to the FED). Since no tuning of the pulse shape
was performed, deviations from an ideal RC-CR shape are present and
therefore a  smearing term has been introduced in the fit, to take into account
 the non-nominal behavior of the pulse shape. The smearing obtained from the fit is larger
for TEC, since the pulse shape of individual APV25s was not optimized
 individually for the many different TEC module geometries.

The estimated difference between the optimal latency and the latency
used in the various phases of the TIF Sector Test is shown in
Fig.~\ref{fig:lat-offpeak}.  During the many different operating
periods of the Sector Test, the commissioning process generally
resulted in timing well within a 25\,ns window. In one week of running,
though, an incorrect latency value was determined by mistake whose 
error in timing was about 30\,ns. A systematic
relative difference of about 10\,ns between the optimal timing for TOB
with respect to TIB/TID and TEC is also visible. The statistical error
of these measurements has been estimated to be about $\pm$1 ns.

The non-optimal sampling determines the loss in the signal performance
of the various sub-detectors. On the right side of
Fig.~\ref{fig:lat-offpeak}, the horizontal lines mark the offset
corresponding to a loss of signal of 1\%, 5\%, and 10\% respectively.
The correction factor to compensate for the loss of signal is plotted
for each run in Fig.~\ref{fig:lat-offpeak}. The anomalous week
where a wrong latency was chosen, resulted in large
correction factors: 10\% for TOB and 35\% for TIB/TID and TEC.
Otherwise, the correction is less than 5\% and this sets a scale for
the expected accuracy of the absolute calibration. That is, in
comparing results between sub-detectors or for the same sub-detector at
different temperature conditions, this level of uncertainty is larger
than, for example, the contributions from time-of-flight differences.

\begin{figure}[!htb]
  \begin{center}
    \includegraphics[width=0.99\textwidth]{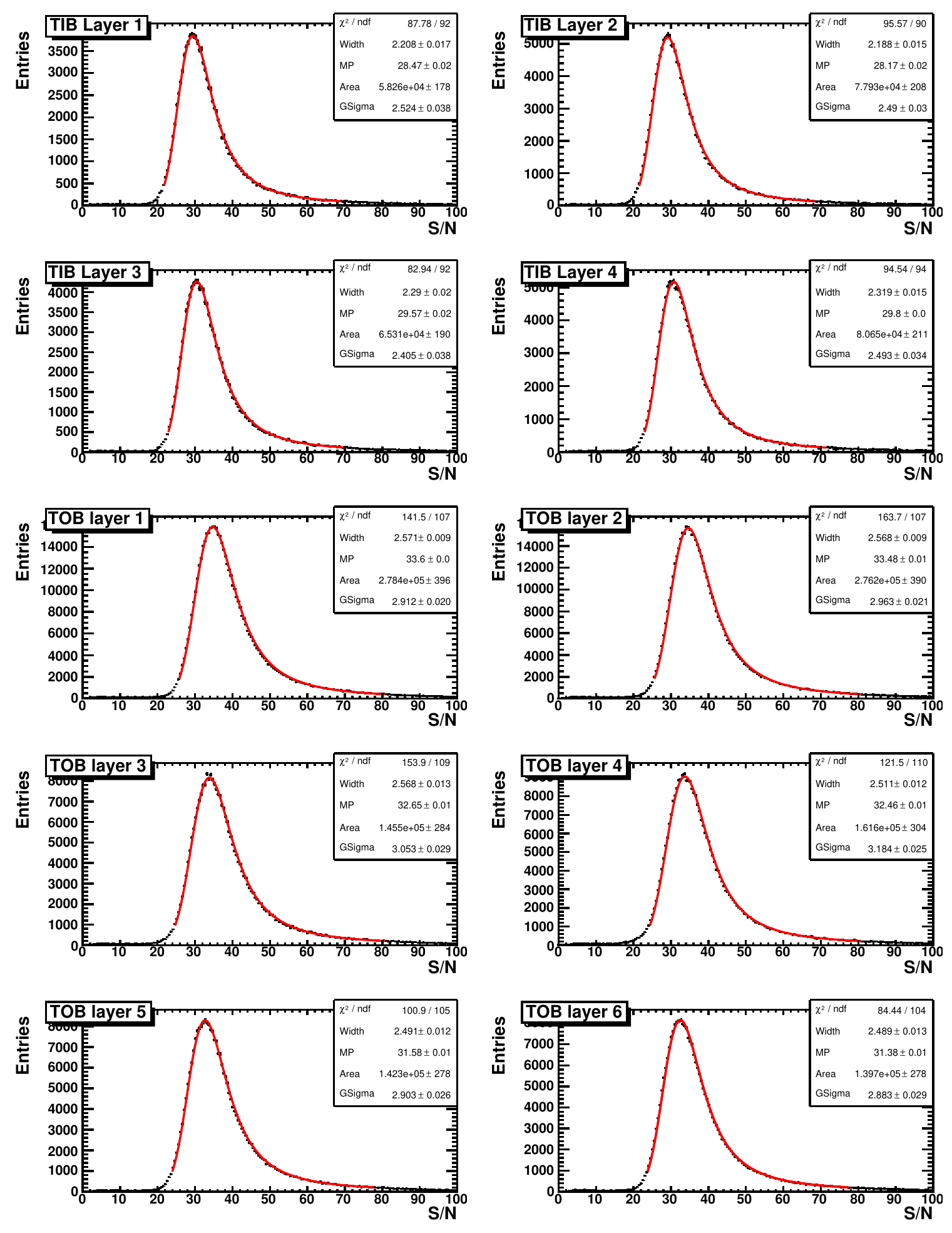}

    \caption{\sl The signal-to-noise corrected for the track angle for TIB
    and TOB layers at $T=-10\,^{\circ}\mathrm{C}$.}
    \label{fig:ston-tibtoblayers}
  \end{center}
\end{figure}
\begin{figure}[!htb]
  \begin{center}
    \includegraphics[width=0.99\textwidth]{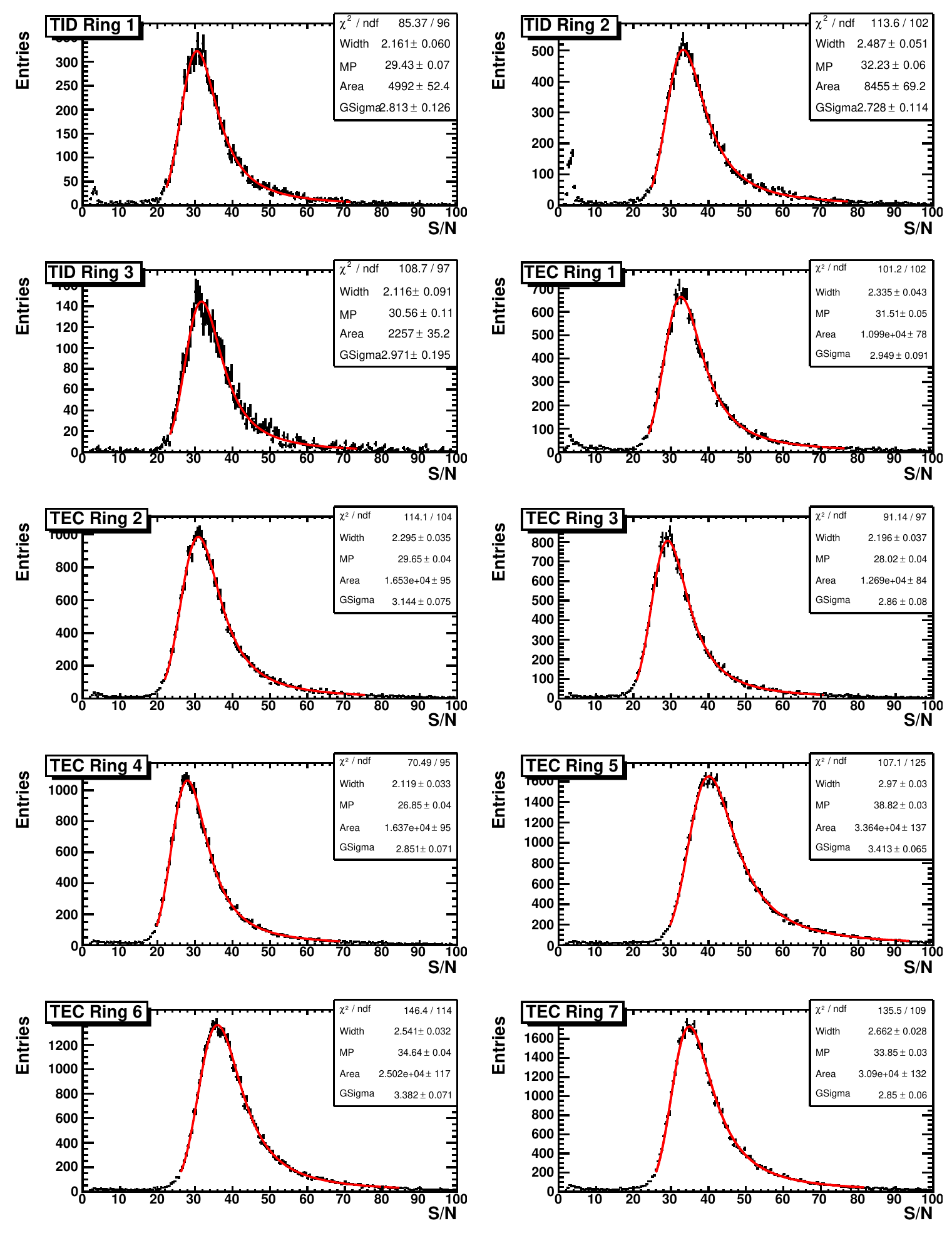}
    \caption{\sl The signal-to-noise corrected for the track angle for TID and TEC rings at $T=-10\,^{\circ}\mathrm{C}$}
    \label{fig:ston-tidteclayers}
  \end{center}
\end{figure}

\subsection{The signal-to-noise}

The signal-to-noise quantity normalized to the sensor thickness ($S_{ren}$/N) 
should be largely independent of gain corrections and therefore allows
a more accurate comparison of results from modules in the same layer
or at different temperatures or from run to run. For this reason $S_{ren}$/N
is an ideal parameter for measuring the stability of the Tracker
during the Sector Test.

$S_{ren}$ is the signal as defined in eq.\ref{form:SK}; $N$ is the cluster noise, defined as $N =  \sqrt{ \sum_i{ {N_i}^2 } /n_{strips} } $ where $N_i$ is the noise
of the i-th strip of the cluster and $n_{strips}$ is the number of
strips of the cluster.  It should be emphasized that if the noise is
constant for all strips in the cluster, then the cluster noise is
independent of the cluster size and equal to the strip noise.  $N_i$
is determined in each pedestal run and written to the offline
database.  It was not remeasured continuously during the cosmic data taking.

Results of $S_{ren}$/N performance at $T=-10\,^{\circ}\mathrm{C}$ are shown in
Fig.~\ref{fig:ston-tibtoblayers} for TIB, TOB and in Fig.\ref{fig:ston-tidteclayers} for
TID, TEC.  The fit to the $S_{ren}$/N is based on a Landau function convoluted with a Gaussian, and therefore has four parameters.  The
range of the fit starts from a minimum value equal to the 10\% of
the peak, goes to a maximum equal
to peak position plus three times the FWHM. The $\chi^2 / ndf $ for the fit
is excellent. For TEC and TID the statistics are lower than for TIB
and TOB, since the rate of large angle cosmic tracks is much reduced
with respect to vertical ones. All Tracker layers show a large  $S_{ren}$/N, in all cases higher than 26.

\begin{figure}[!htb]
   \begin{center}
\includegraphics[width=0.45\textwidth,height=5.9cm]{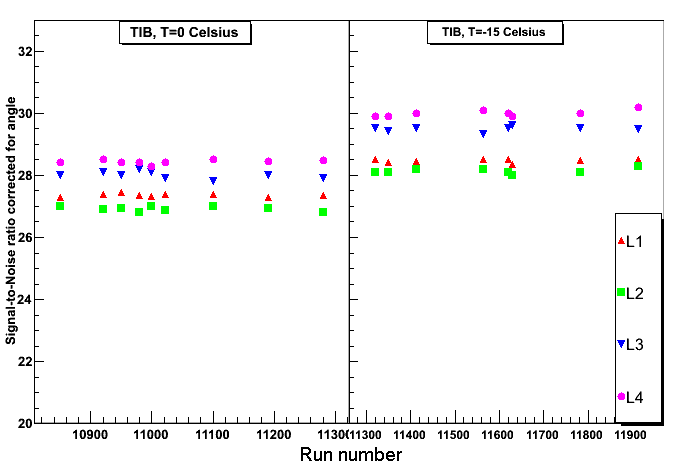}
 \includegraphics[width=0.45\textwidth,height=5.9cm]{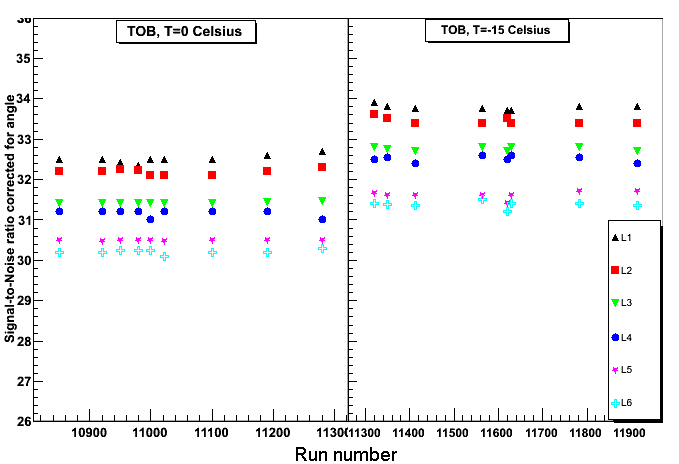}

     \caption{\sl Signal over noise corrected for the track angle: from the left to the right for TIB and TOB. }
     \label{fig:ston-stab}
   \end{center}
 \end{figure}

A study of the stability of $S_{ren}$/N performance was done, examining  measurements for every single run  during long periods where the Tracker was running in stable conditions: the best periods are when the Tracker was running cold.  Fig.~\ref{fig:ston-stab} shows the results for TIB and TOB, for all  the layers, versus the run number when Tracker was running at $0\,^{\circ}\mathrm{C}$ and $-10\,^{\circ}\mathrm{C}$: TIB and TOB show a very stable behavior with variations less than 0.3\%. Similar results are obtained for TEC and TID, but the lower statistics per each run gives rise to a higher statistical error on the run by run measurement.

$S_{ren}$/N increases with decreasing temperature, as expected from the results on the temperature dependence of the noise. A more quantitative analysis of the temperature dependence was not possible since it requires an optimization of the APV25 parameters for each module geometry at each temperature.

After normalizing the signal to the sensor thickness, $S_{ren}$/N should be identical for identical modules, regardless to their position.  Nevertheless, the segmentation of the silicon in microstrips, combined with the effect of the
clustering algorithm, which uses thresholds to determine strip inclusion in a cluster, might introduce
some charge loss. 
The fractional charge loss due to clustering  threshold will be more significant when the charge is small. 
An analysis has been done based on the angle of incidence of the tracks, 
as charge tends to be shared among more
strips at higher angles. Therefore the dependence on the
track angle has been studied first with respect to the 3D angle. In
order to separate the effect of illuminating more strips from that of
increasing the charge the study has
been done for: the dependence on the XZ local angle for tracks
perpendicular to the silicon strips (YZ local angle less than 6
degrees), referred as transverse tracks, where the sharing among strips is maximized; the dependence
on the YZ local angle for tracks that come along the direction of the
silicon strip (XZ local angle less than 6 degrees), referred as longitudinal tracks, where the sharing
among strips is almost independent of the angle. These studies have
 been performed by looking at the dependence in $S_{ren}$/N on track path (K) within the silicon, where the path is expressed in units of silicon thickness. 
$S_{ren}$/N should be independent of $K$, see equation~\ref{form:SK}.

\begin{figure}[h]
  \begin{center}
    \includegraphics[height=0.45\textwidth,angle=90]{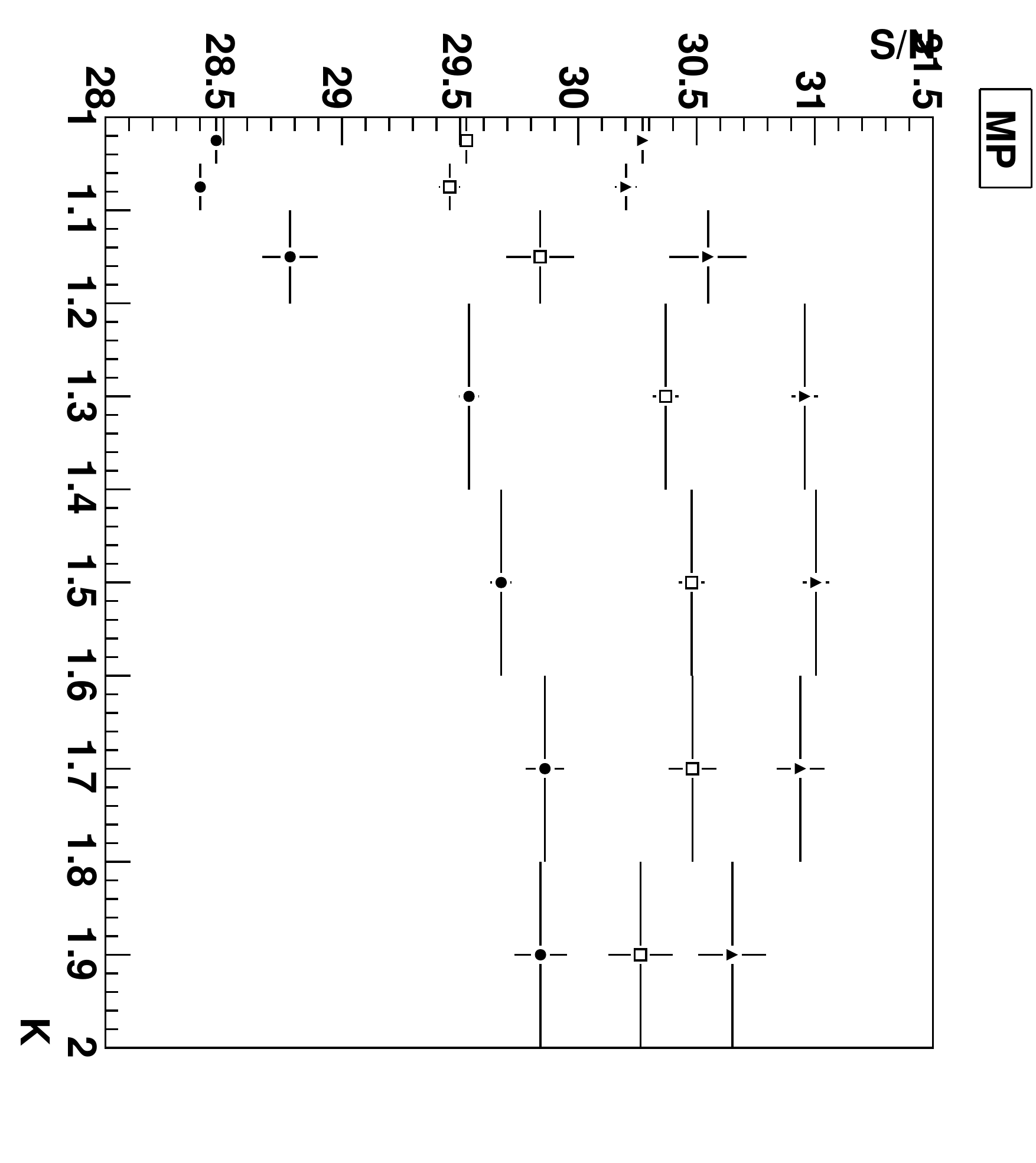}
    \includegraphics[height=0.45\textwidth,angle=90]{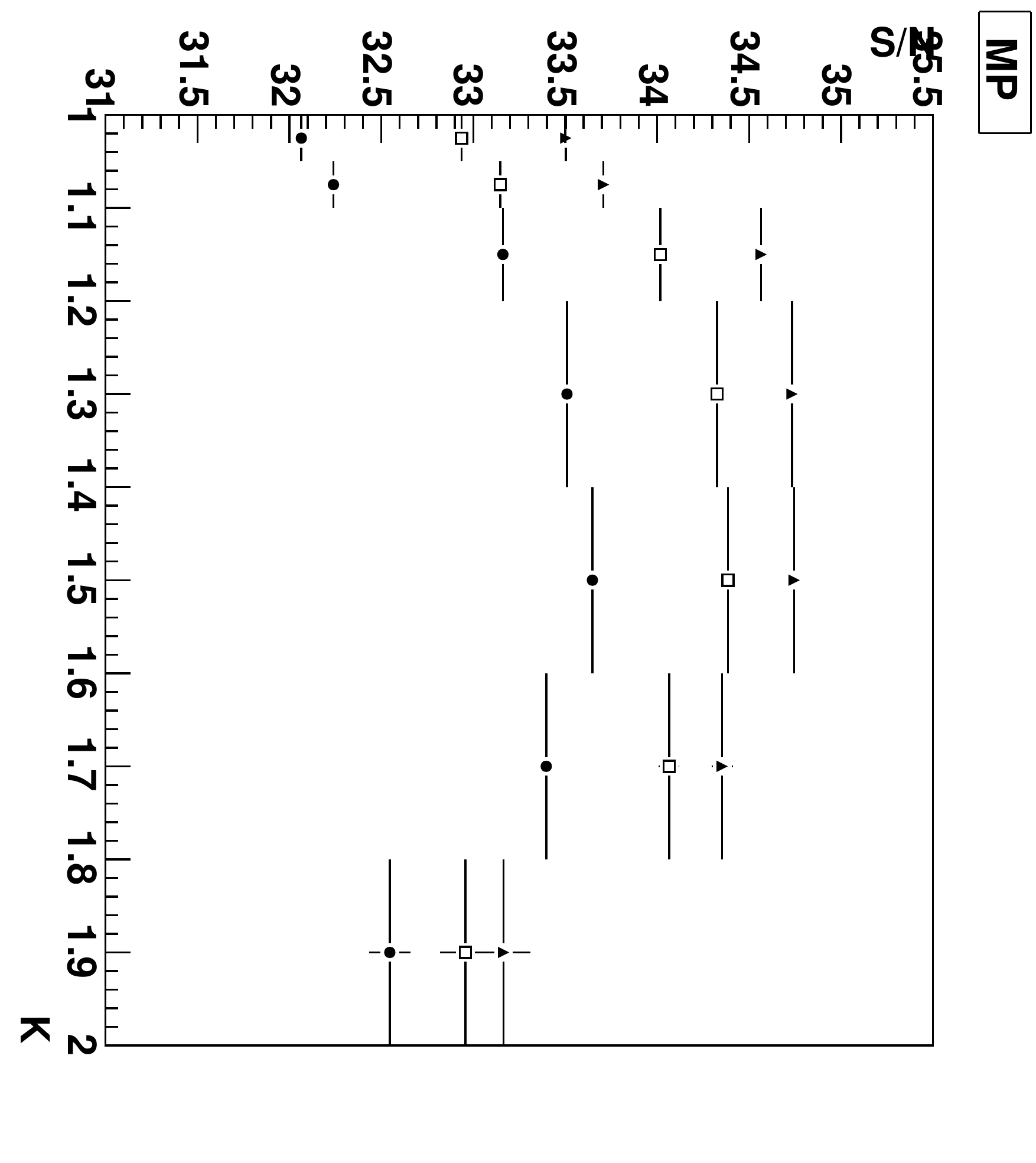}
    \caption{\sl $S_{ren}$/N dependence on K for tracks
    perpendicular to the strip direction (open triangles) , i.e. the XZ local direction,  and   for tracks
    parallel to the strip direction (black circles) , i.e. the YZ local direction, for TIB and TOB. }
    \label{fig:SN-vs-xz}
  \end{center}
\end{figure}

Results are shown in  Fig.~\ref{fig:SN-vs-xz} for TIB and TOB: with black circles  for the YZ angle and with open triangle  for the XZ angle dependence. TEC and TID results have not been included since the variation of $S_{ren}$/N for the different rings is large and there is a correlation between K and the statistics for each ring.

It is clear in Figs.~\ref{fig:SN-vs-xz}  that, for small K, there is a nearly 5\% loss of $S_{ren}$/N (and presumably signal) for TIB.  For longitudinal tracks  most of the charge is concentrated in one strip, signal in neighbouring strips increases with K: once above the clustering threshold, they are included in the cluster and therefore almost all the charge delivered by the track is taken into account already for $K=1.3$ where a sudden increase is visible. For transverse tracks the charge is shared among several strips and the charge loss at the edges of the cluster,  will be less significant as K increases. For $K=2$ the transverse and parallel tracks give the same signal-to-noise value.

For TOB a similar effect is seen, with a estimated loss of charge for small K of the order of nearly 6\%: the charge of transverse tracks increases almost linearly with K and parallel tracks show a sudden increase at $K=1.2$. Contrary to the TIB, for TOB the parallel tracks do not reach a constant value, but $S_{ren}/N$ decreases starting for $K=1.6$. This can be explained by the effect of departure from linearity of the APV25 for charges higher than 3 mips with gradual fall off beyond, that influences the tail of the Landau distribution for large charge release to single strips. 



\subsection{Signal calibration}
\begin{figure}[!hbt]
  \begin{center}
    \includegraphics[height=0.4\textwidth,angle=90]{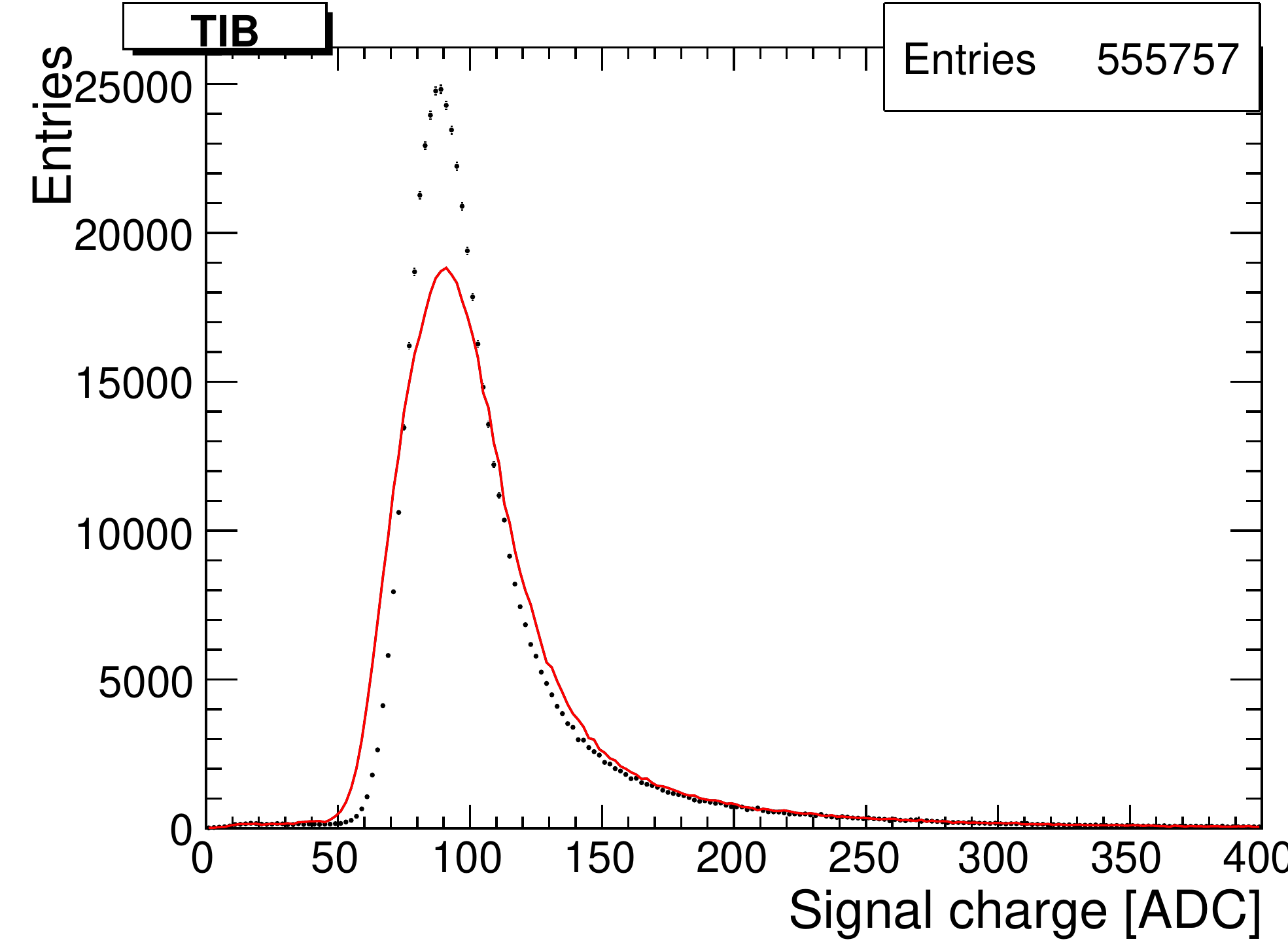}
    \includegraphics[height=0.4\textwidth,angle=90]{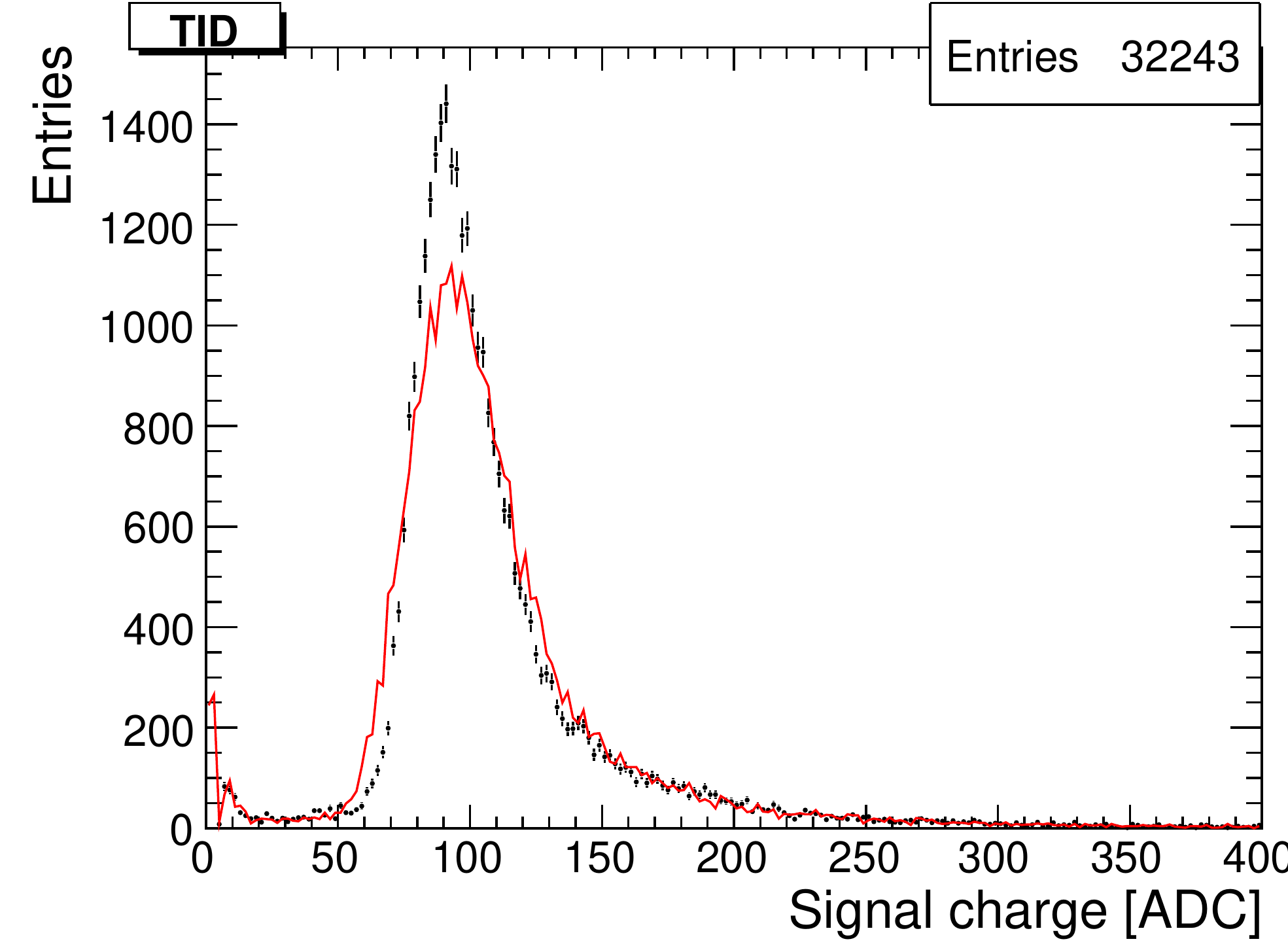}

    \includegraphics[height=0.4\textwidth,angle=90]{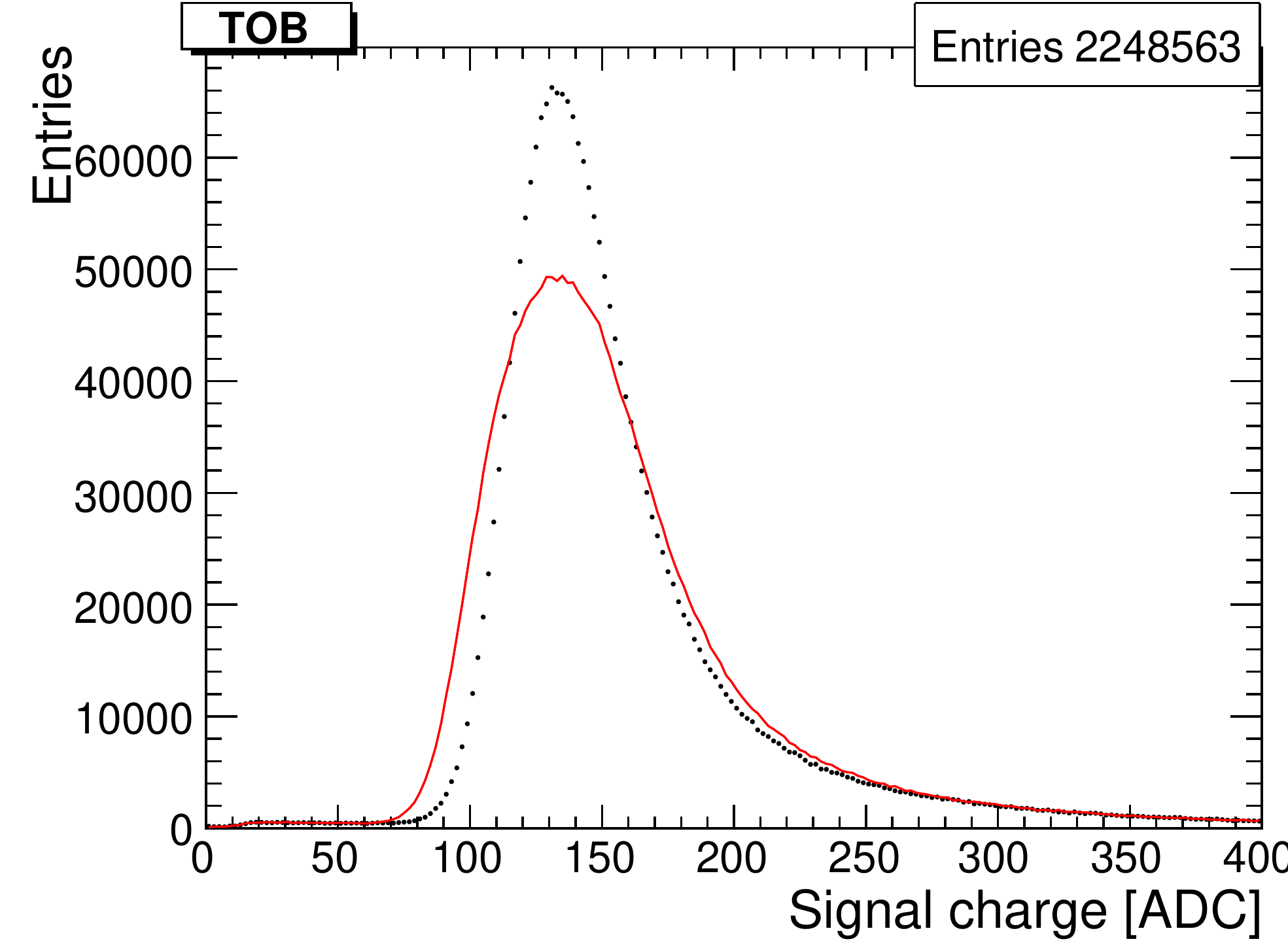}
    \includegraphics[height=0.4\textwidth,angle=90]{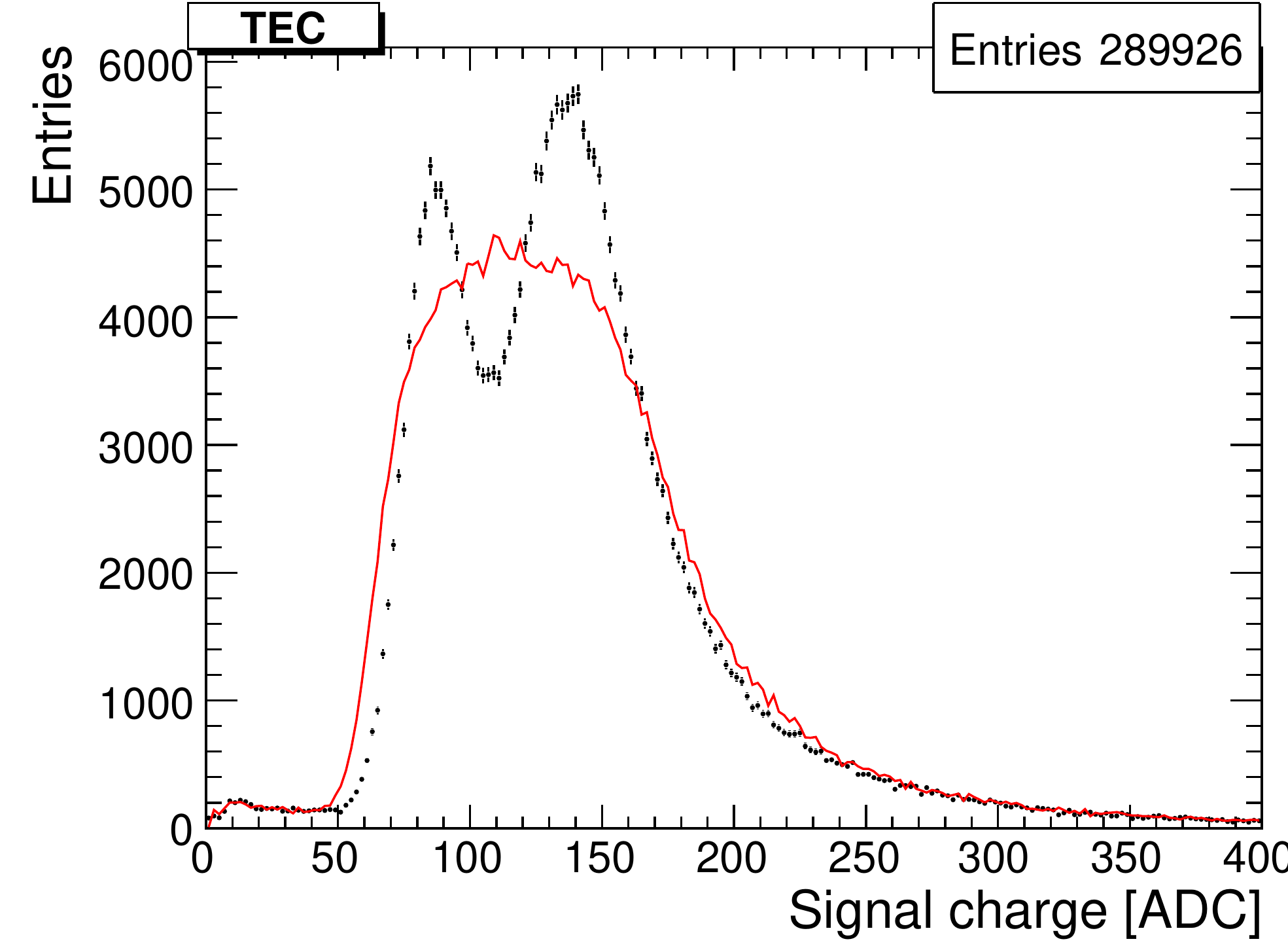}

    \caption{\sl Signal corrected for the track angle ($S_{ren}$), in full curve without
    electronic gain calibration and in dots with it.}
    \label{fig:signal-layer}
  \end{center}
\end{figure}

The correction for the electronic gain, see equation~\ref{form:corr}, can be used to improve the resolution on the signal performance.  Fig~\ref{fig:signal-layer}  shows the uncorrected and the corrected signal. For all sub-detectors the calibration results in a decrease of the FWHM; it is clear that the electronic gain calibration inproves the FWHM. In the case of the TEC, after the electronic calibration, two peaks are visible, which are the consequences of the use of thin and thick silicon sensors.

\begin{figure}[!h]

  \begin{center}
    \includegraphics[height=0.4\textwidth,angle=90]{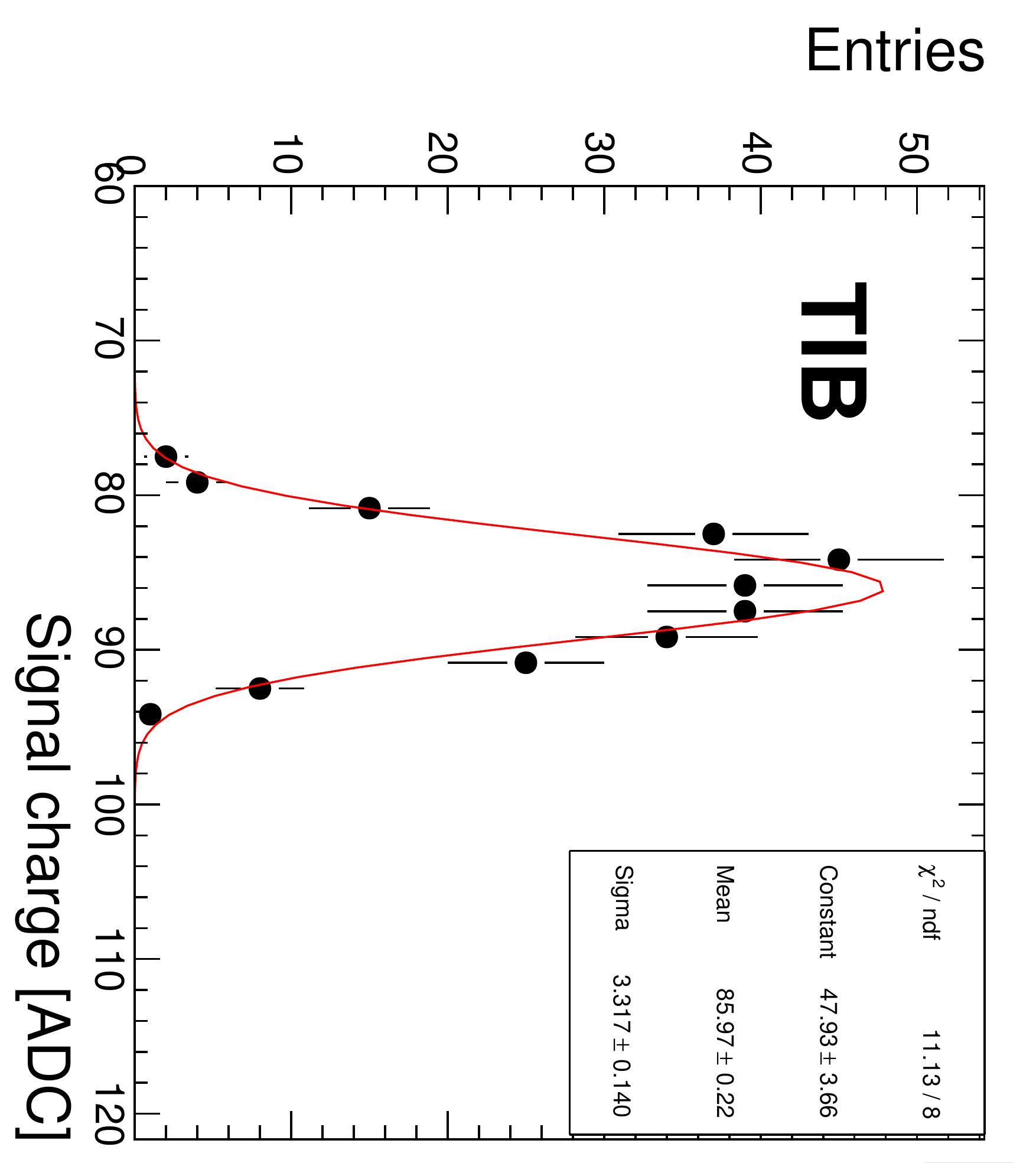}
 \includegraphics[height=0.4\textwidth,angle=90]{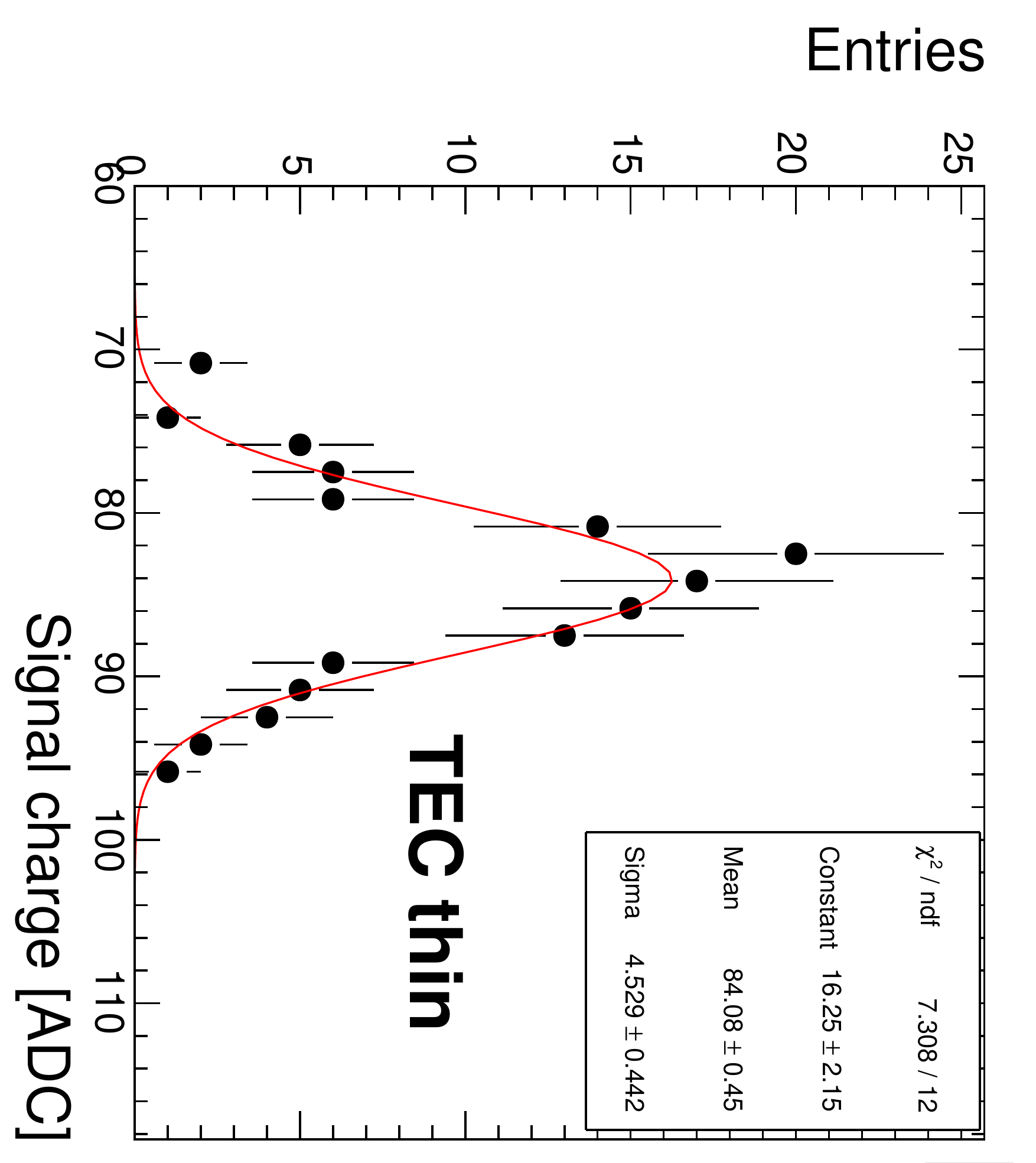}

 \includegraphics[height=0.4\textwidth,angle=90]{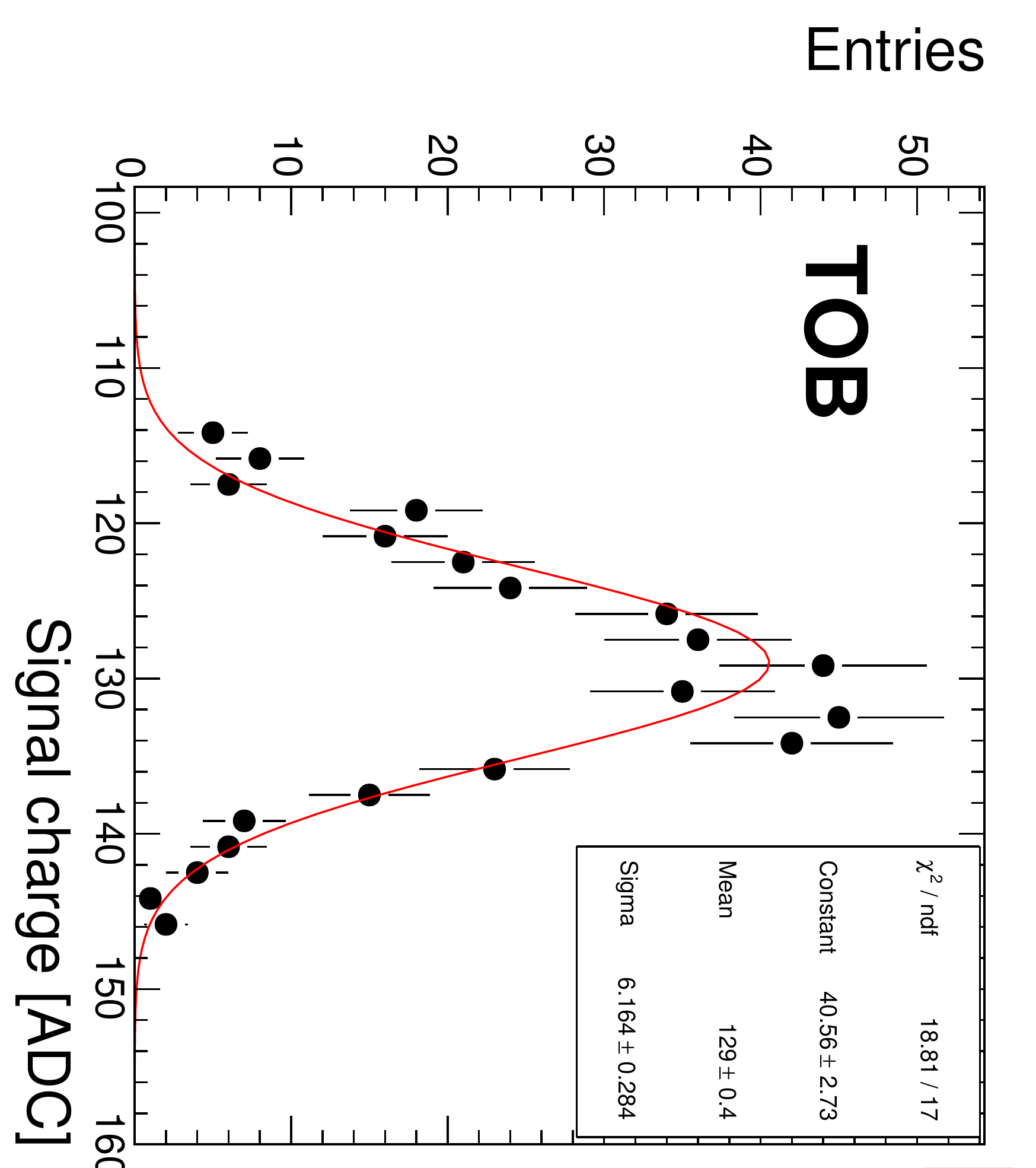}
 \includegraphics[height=0.4\textwidth,angle=90]{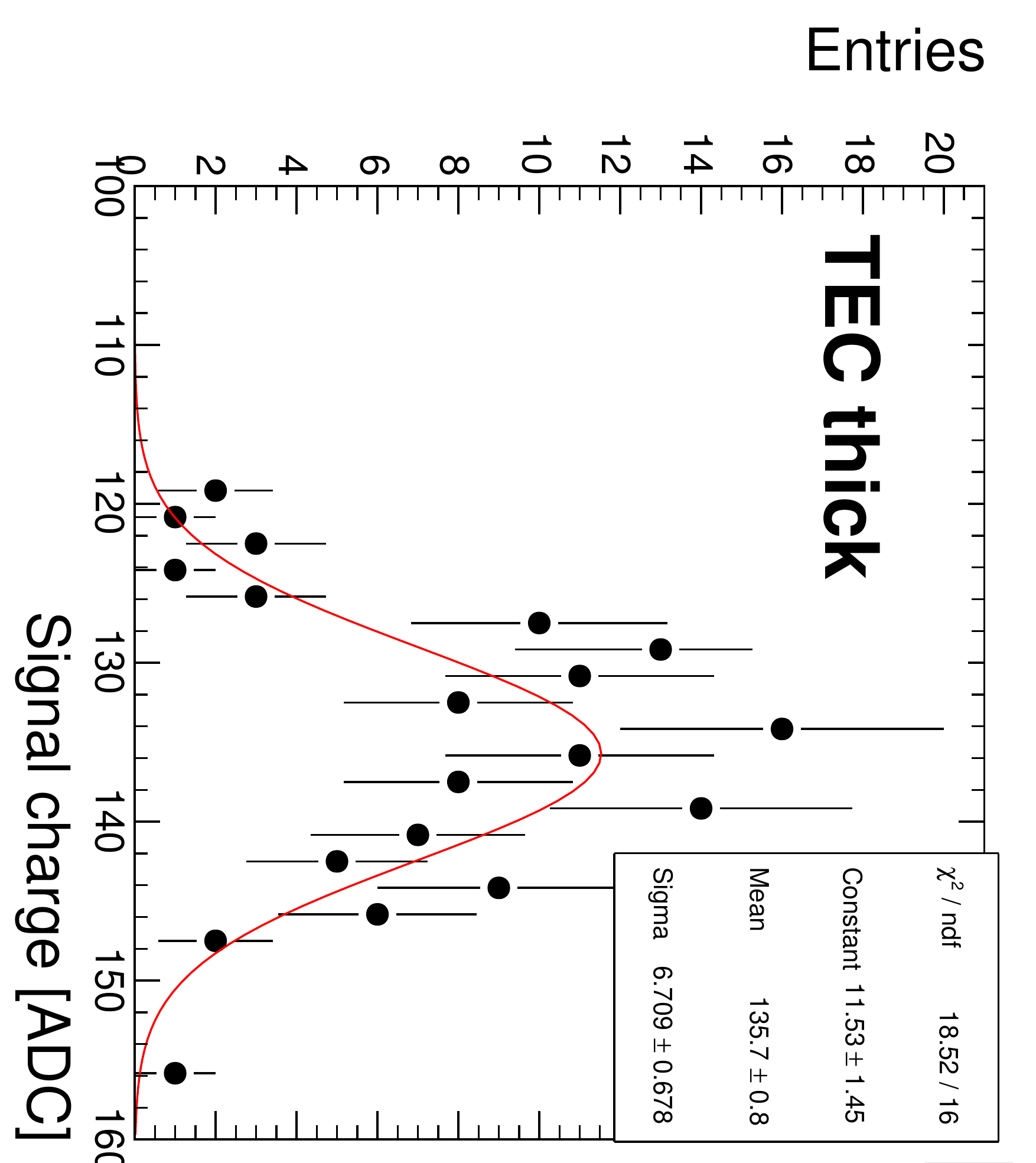}

    \caption{\sl Distribution of signal ($S_{ren}$) most probable value, module
    per module for TIB (top left), TEC thin sensors (top right), TOB
    (bottom left), and TEC thick (bottom right). }
    \label{fig:gain-dis}
  \end{center}
\end{figure}

Having included this calibration it is interesting to check the level
to which the absolute calibration is understood. To do this, an
analysis at the module level has been made, looking at the
distribution of the most probable value of the signal, expressed in
ADC counts, for the thin detectors (TIB and TEC rings 1 to 4), and for the
thick detectors ( TOB and TEC rings 5 to 7). Results are shown in
Fig.~\ref{fig:gain-dis} for the $-10\,^{\circ}\mathrm{C}$ period.

Modules of the same sub-detector are well represented by Gaussian
distributions with fairly large sigmas. These are at the level of 3.8\% and 5.3\%
for TIB and TEC thin sensor modules; of 4.6\% for TOB and TEC thick
sensor modules. Therefore, single module signal performance can be
understood to these levels of precision within the same sub-detector, to be compared with the spread of the tickmark height of about 13\% before taking into account the electronic gain, see Fig~\ref{fig:elgain-dis}. Further improvements in the evaluation of the electronic gain can come, for example, by taking into account the supply voltage for individual modules.  

It should be noted that modules of the same geometry, but placed in different  position, are differently illuminated by the cosmic rays and this varies $S_{ren}$ as it can be seen in Figs.\ref{fig:SN-vs-xz}. This introduces another level of complexity to this study.

Some lessons can also be learned by comparing the average results obtained by 
modules with silicon of the same thickness: for the TIB and thin sensor TEC modules, the average values 
are within 2.2\%, while TOB and thick sensor TEC modules have average values
that differ by 5\%. It should be borne in mind that signal
performance can be affected by changes in APV25 parameters. Typically,
these are adjusted for reasons other than equalizing the peak amplitude of the output signal.

TEC has the interesting feature that all modules, both thin and thick,
have been run with the same APV25 parameters. The ratio of the mean
values of thick over thin sensor module signal is: 
$\frac{<Signal(TEC_{thick})>} {<Signal(TEC_{thin})>} = 1.614 \pm 0.008$, which is compatible
with the ratio $1.62$ of thicknesses of $470\,\mu$m and $290\,\mu$m.
 This is not so precisely the case when  TIB and TOB are compared.

In order to better understand the signal performance, this analysis has
been performed at two different temperatures:
$10\,^{\circ}\mathrm{C}$ and $-10\,^{\circ}\mathrm{C}$.
The signal changes of 2-3\% within a given layer for the 20 degrees change
in temperature. It can be concluded that the conversion factor between ADC counts and
electrons must be done at individual layer level for each temperature
if a 5\% level has to be reached. 
Further refinements to the calibration would require a module by module study based on track information.




\subsection{Validation of Online Zero Suppression}
The correct functioning of the FED clustering algorithm and its
simulation was validated in the Sector Test through the
study of cosmic data runs and special configuration runs.

Special runs were taken in which the data from a single
TOB rod was optically split and sent simultaneously to two FEDs. One
FED passed the data in VR mode, and the other applied ZS.  Offline
analysis of the output from each FED allowed a direct comparison of
the FED ZS output with a simulation of the ZS algorithm applied to the
VR data.  Only small differences, consistent with gain variation
between the two data paths, were observed.

Cosmic data was further used to validate the functionality of the
FED.  Several consecutive runs, one with ZS enabled and the second
without ZS enabled, were taken and the data from these to check for
global FED failures, to further validate the offline algorithm, and to
study the effect of varying the thresholds. 

The only significant difference, an excess at about 128 counts, is due
to a small number of channels that had pedestals outside the range
allowed by the FED.  


\subsection{Hit Occupancy}
Occupancy for a given strip is defined as the fraction of events in which it registers 
a signal exceeding the threshold required to be ``hit''.
When LHC is running at high luminosity, the occupancy of the innermost strip layer 
(TIB L1) is estimated to be a few percent~\cite{TK}.
As described in the reconstruction section,
seed strips and the cluster require a signal-to-noise value greater than 3
and 5 respectively.  If strip noise is purely Gaussian then the probability of a
statistical fluctuation to become a cluster is very low, less than
$5\cdot 10^{-5}$, and is therefore  negligible. It is interesting to
measure the subdetector occupancy during the Sector Test, by taking
 the mean value of the distribution of the probability of a strip to be in a cluster, that is, the strip occupancy distribution. 

\begin{figure}[!hbt]
  \begin{center}
\includegraphics[width=0.4\textwidth,angle=0]{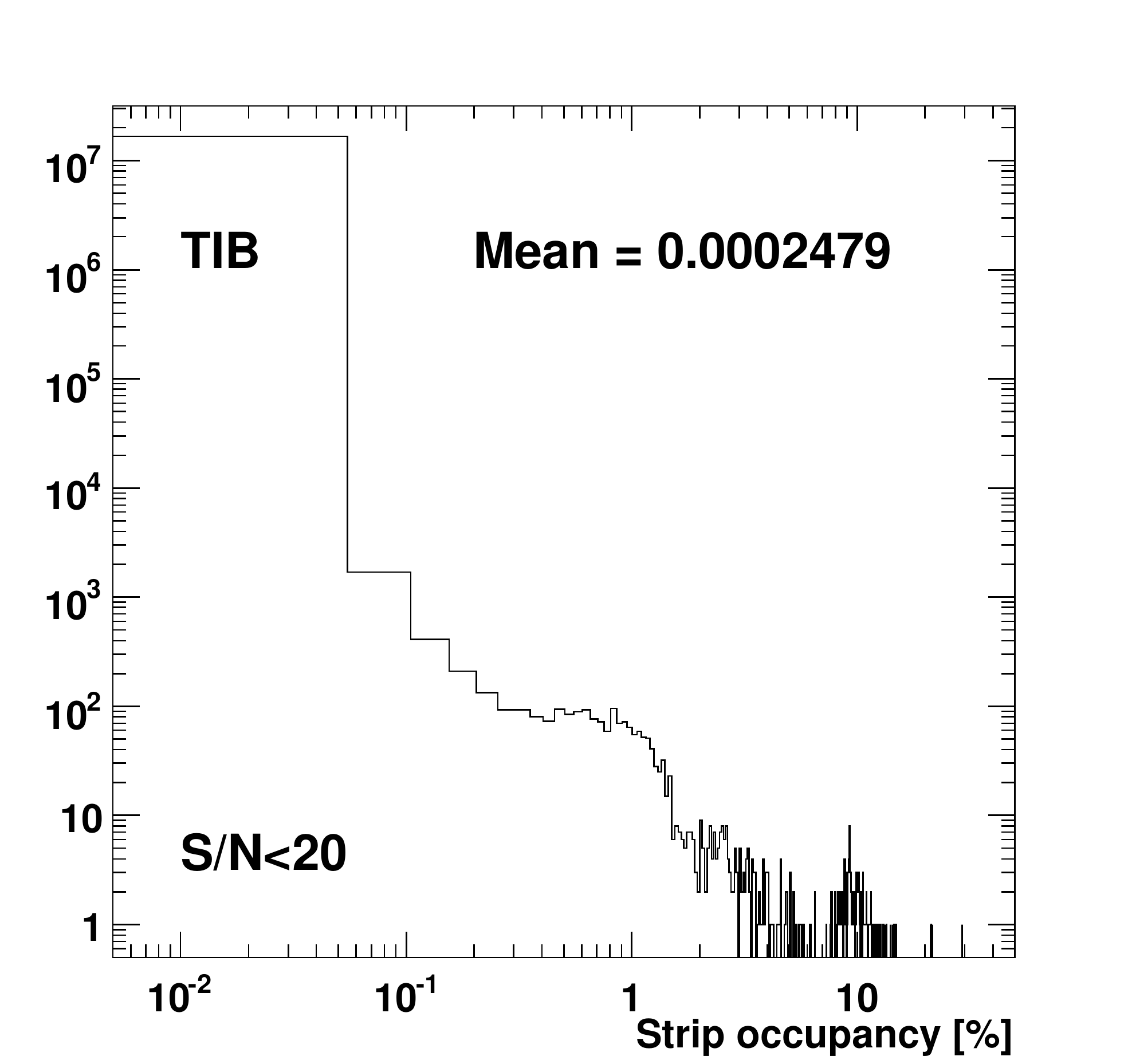}
\includegraphics[width=0.4\textwidth,angle=0]{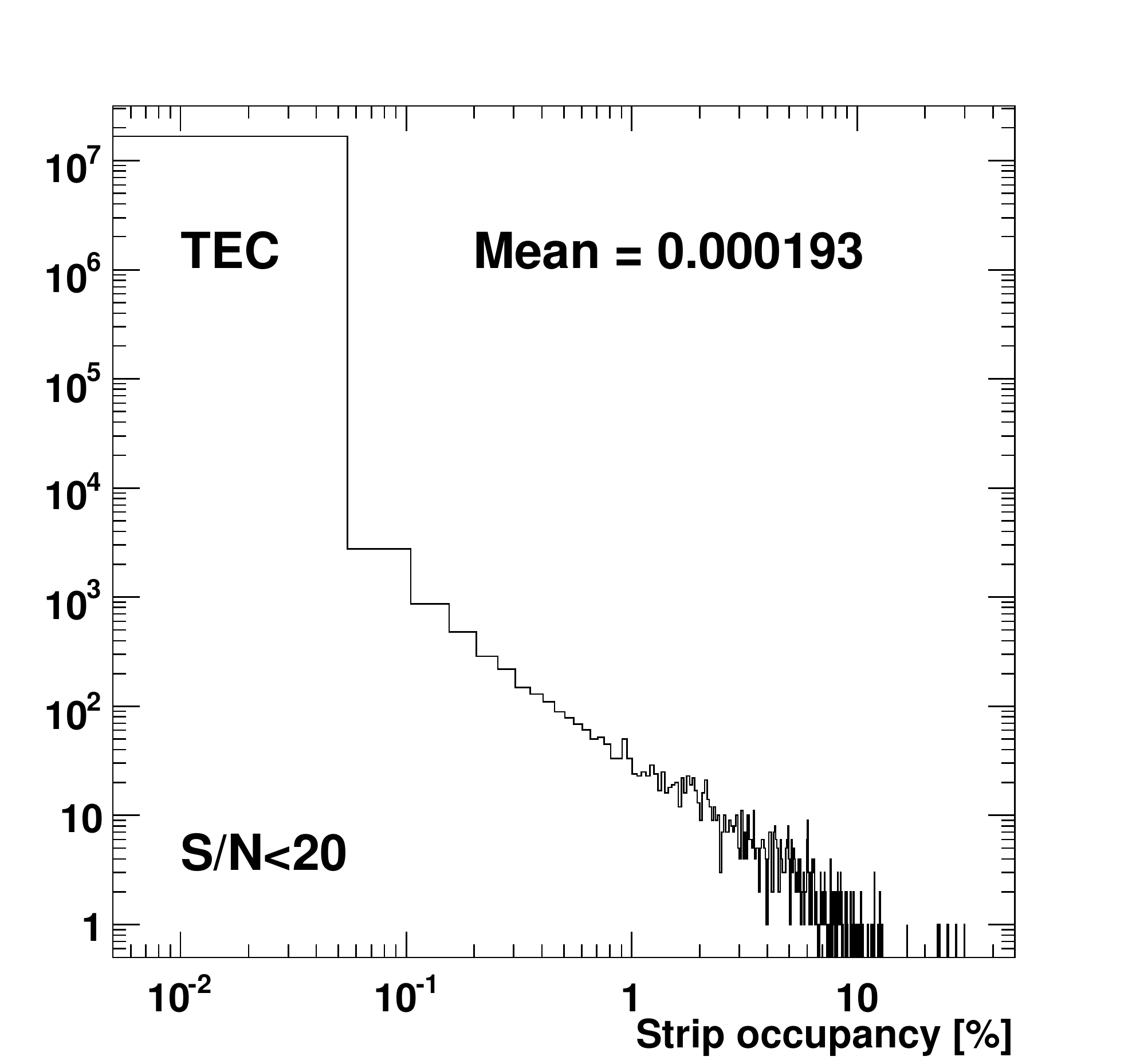}

\includegraphics[width=0.4\textwidth,angle=0]{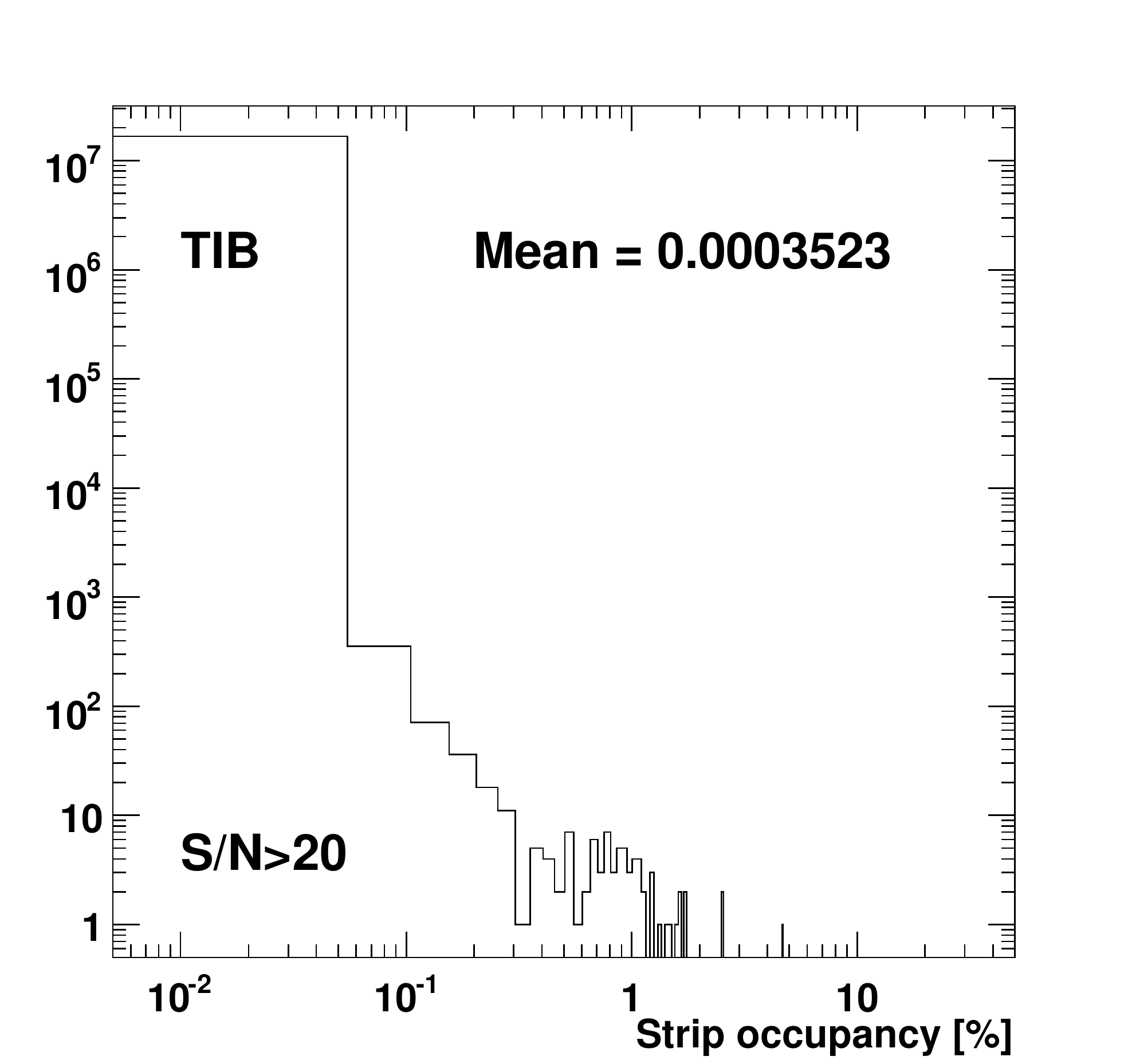}
\includegraphics[width=0.4\textwidth,angle=0]{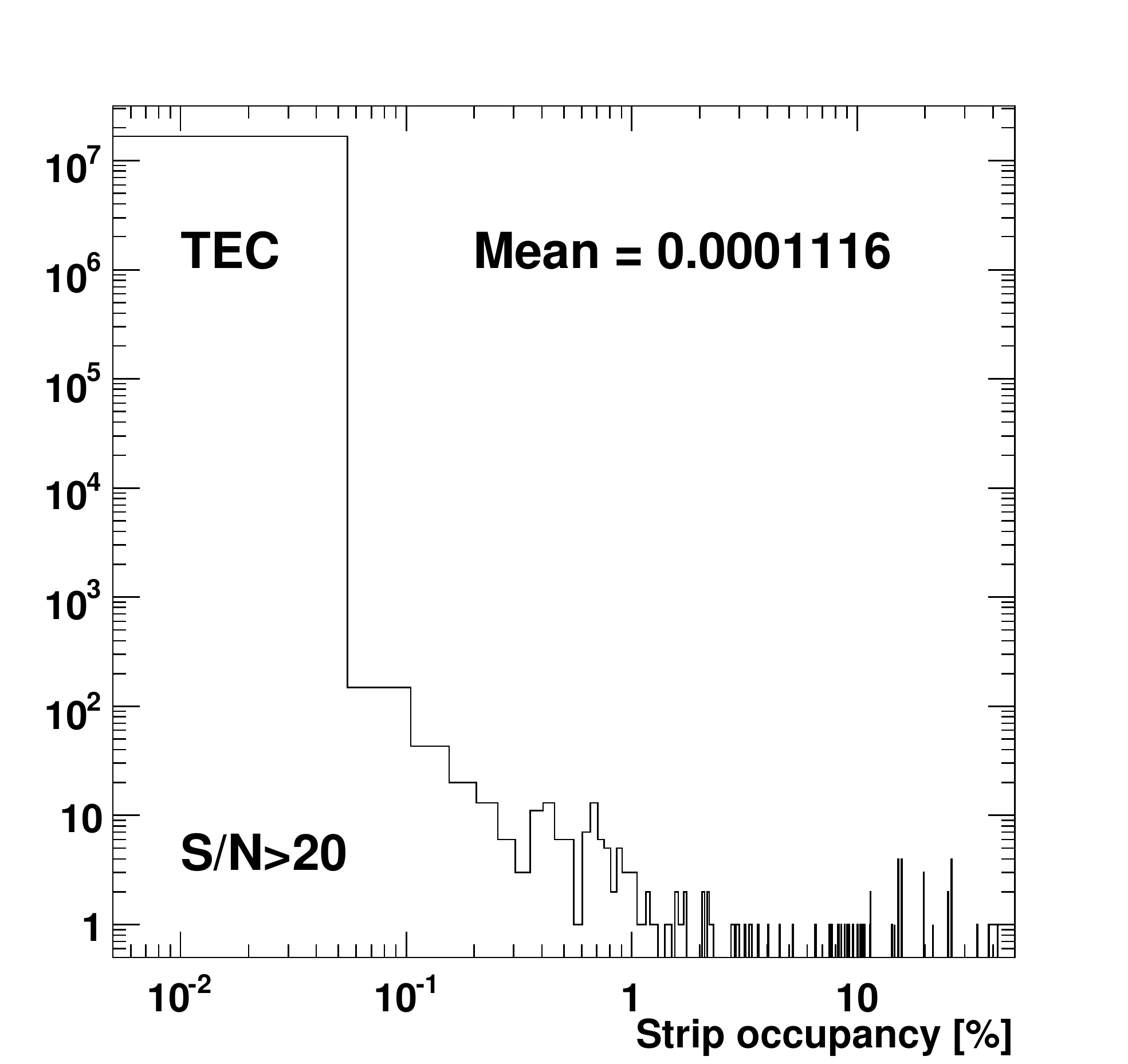}

      \caption{\sl Distribution of strip occupancy for TIB and TEC: upper (lower) plots for S/N below (above) 20.}
      \label{fig:occupancy}
  \end{center}
\end{figure}

In order to separate the contribution coming from real tracks, the strip occupancy has been studied for clusters with S/N higher or lower than 20.  Results are shown in Fig.~\ref{fig:occupancy} for TIB  and TEC. 
Cosmic rays account for an occupancy of $3.5\cdot 10^{-6}$ (TIB) and $1.1\cdot10^{-6}$ (TEC) with no tails in the 
distribution. Clusters from statistical fluctuations accounts instead for occupancy of $2.5\cdot 10^{-6}$ and $1.9\cdot10^{-6}$ 
and show much longer tails. It is apparent that there are strips that are more active than others, with strip occupancies up to 5\% or larger.

It seems clear that the  tails observed in some of the occupancy distributions are the result of active (``hot'')
strips. These strips have not to be confused with the noisy strips described in Sect~\ref{sec:quality} since these ``hot'' strips are characterized by non-Gaussian fluctuations, above five times the noise value. The number of the ``hot'' strips with strip occupancy above 1\% is below 0.05\% and therefore does not significantly affect module occupancy. 
Similar results have been obtained for TOB and TID sub-detectors.

\begin{figure}[!hbt]
\begin{center}
\begin{minipage}{0.49\textwidth}
\includegraphics[width=0.8\textwidth]{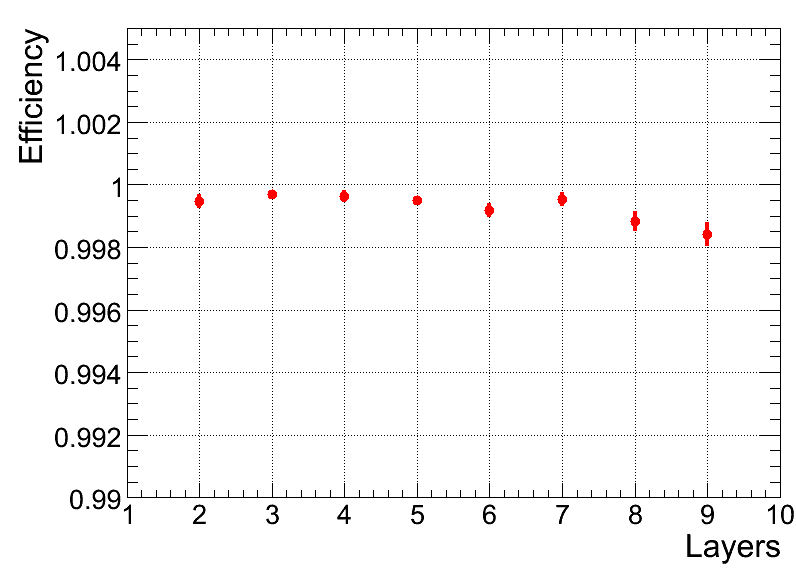}
\end{minipage}
\begin{minipage}{0.49\textwidth}
\includegraphics[width=0.8\textwidth]{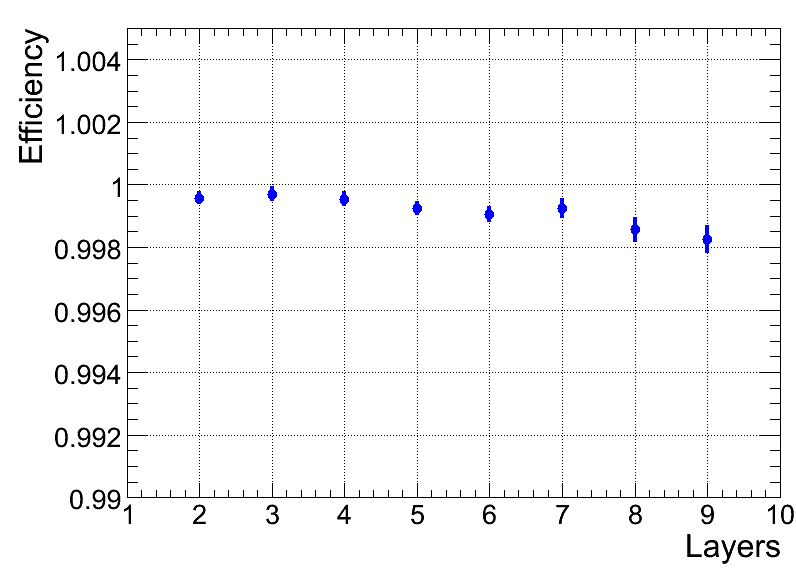}
\end{minipage}
\caption{\sl Summary of layer efficiency at room temperature (left) and at
$T=-10\,^{\circ}\mathrm{C}$ (right).}
\label{fig:hiteff}
\end{center}
\end{figure}

\subsection{Hit Reconstruction Efficiency}
The efficiency of a Tracker module to observe a hit when traversed by a particle,
is one of the most important characteristics of
the detector performance as this can  be affected by losses in the
silicon, electronic chain, or data acquisition and analysis. Efficiency has been measured by modifying the track
reconstruction algorithm to skip the layer under study during the
pattern recognition phase. A sample of high quality events was
selected by requiring only one track reconstructed by the CTF
algorithm, one hit in the first TIB layer, one hit in the two
outermost TOB layers, and at least four reconstructed hits (of which
at least three matched from stereo layers). Tracks were required  to have no more than five lost hits during the pattern recognition phase and not more than three
consecutive ones; finally an upper cut of $30,^{\circ}$ on $\theta_{3D}$
was applied to select tracks almost perpendicular to the modules.

In order to avoid genuine inefficiencies present at
the edge of the sensor and in the bonded region between two sensors,
additional cuts have been applied to restrict the regions in which
efficiency is measured.
 With this event selection criteria and
fiducial area restrictions, the efficiency exceeds 99.8\% for all measured
layers as is shown in Fig.~\ref{fig:hiteff}.
 Additional details on the tracking and
analysis procedures can be found in Ref.~\cite{TIFTrackingNote}.


\section{Simulation}
Cosmic ray muons are simulated with the CMSCGEN generator\cite{CMSGEN}
and the parameterizations of the energy dependence and incident angle 
are based on a generator developed for the
L3+Cosmics experiment.  
The muon energy spectrum at the surface level TIF contains very soft
muons and simulation is needed down to 0.2\,GeV as this is the cutoff imposed by the 5\,cm of lead. 
The uncertainties in the cosmic ray spectrum in this low energy range are very high and
the CMSCGEN parameterizations do not permit an accurate simulation
of muons below 2\,GeV. 
For the TIF data simulation, 
the events with cosmic muons below 2\,GeV are produced according to 
the energy spectrum measured by the CAPRICE\cite{Caprice} experiment that describes cosmic muons down to 0.3~GeV. 
The angular dependencies in this simulation are taken to be the same as for muons
at 2 GeV.  Using the CAPRICE energy spectrum, the remaining
uncertainties at this low energy are largely due to solar modulation,
angular correlations, latitude dependencies, altitude effects, the
earth's magnetic field, and specifics of the building ceiling
structure where the tests take place. 
However, this approximation of the energy spectrum seems satisfactory for Tracker commissioning
purposes.  

\begin{figure}[hbt]
   \centering
     \includegraphics[width=0.6\textwidth]{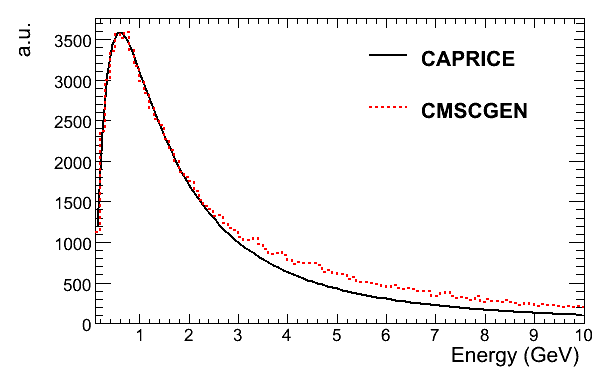}
  \caption{
  \label{fig:Spectrum}
  {\em The CMSCGEN cosmic muons energy spectrum. Below 2 GeV the events are
  reweighted according to the CAPRICE spectrum.} }
\end{figure}



\begin{figure}[hbt]
  \centering
\begin{minipage}{0.3\textwidth}
    \includegraphics[width=0.9\textwidth]{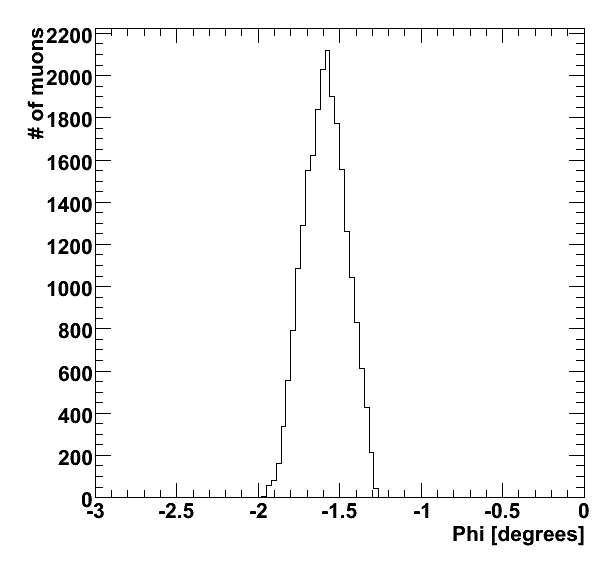}
\end{minipage}
\begin{minipage}{0.3\textwidth}
    \includegraphics[width=0.9\textwidth]{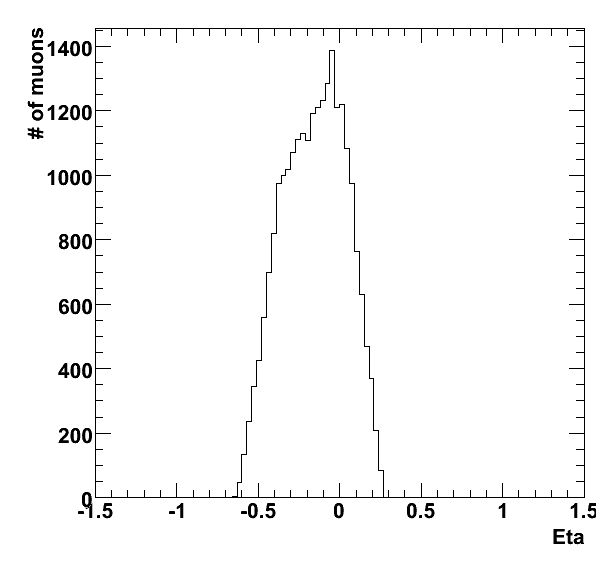} 
\end{minipage}
\begin{minipage}{0.3\textwidth}
    \includegraphics[width=0.9\textwidth]{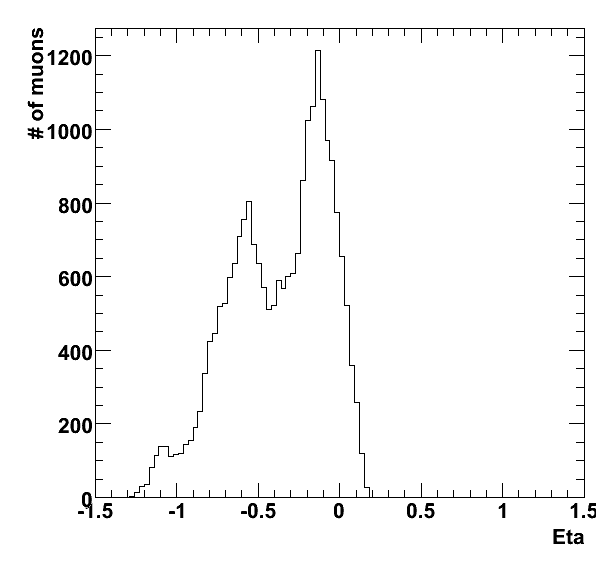} 
\end{minipage}
\caption{ {\em Cosmic muon track parameters: $\phi$(left), $\eta$ for scintillator Conf(TA)(middle) and Conf(TC)(right)}}
  \label{fig:simu_muon_param}
\end{figure}
\begin{figure}[!t]
  \centering
\begin{minipage}{0.3\textwidth}
    \includegraphics[height=5cm]{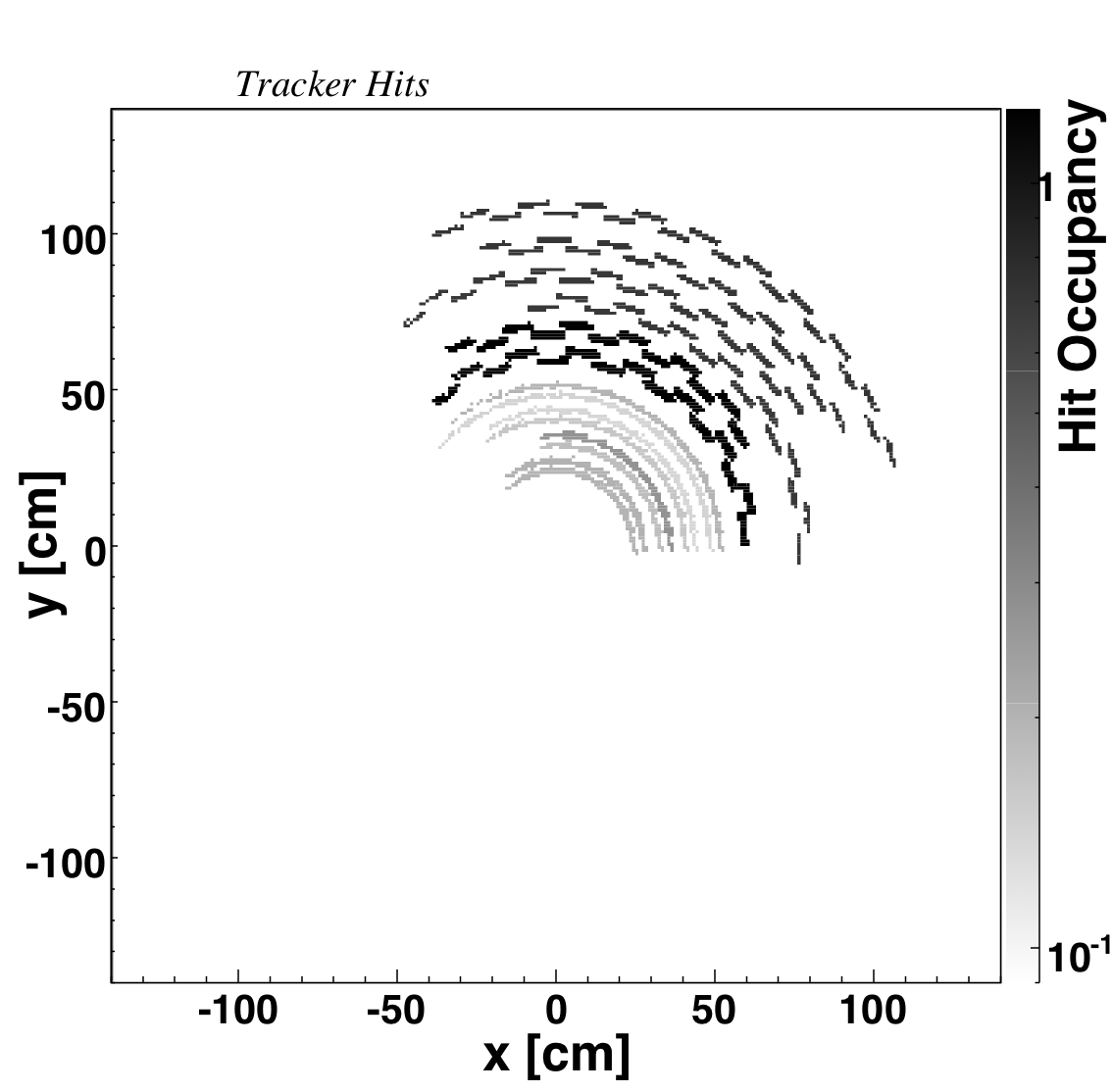}
\end{minipage}
\hspace{5mm}
\begin{minipage}{0.6\textwidth}
    \includegraphics[height=5cm]{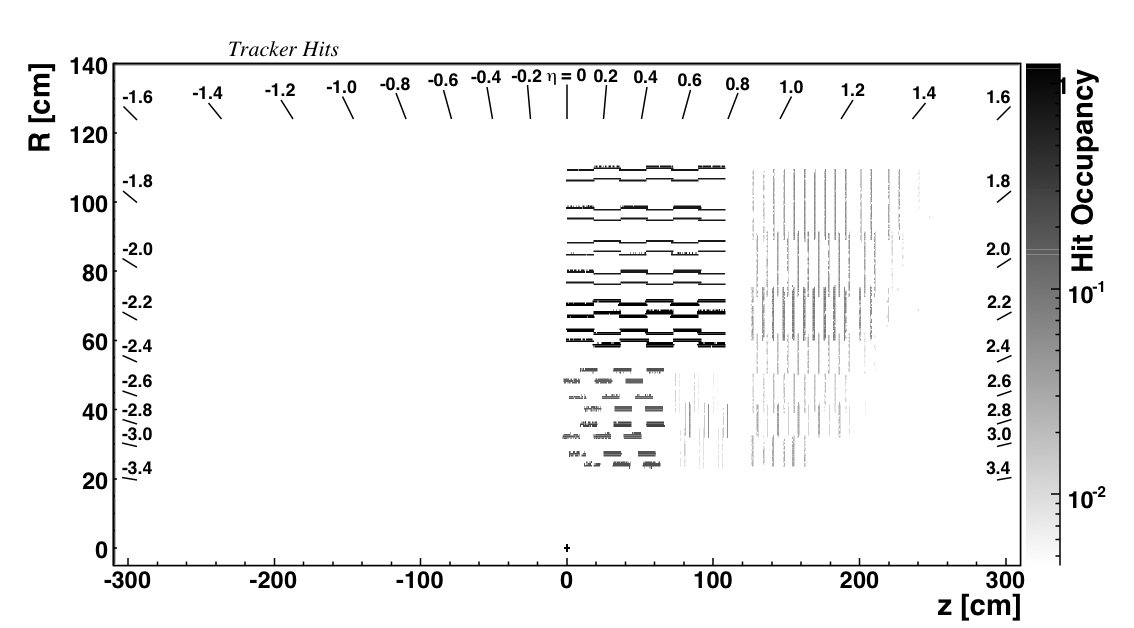}
\end{minipage}
  \caption{
  {\em Hit distribution for the cosmic muons in the x-y(left) and y-z(right) plane in Configuration TC} }
  \label{fig:simu_cablemap_confC}
\end{figure}

%
Before the simulation stage, the generated cosmic muons are
filtered in order to mimic the experimental trigger setup described in
Section 2.6, where triggering was based on the coincidence of separated
scintillation counter arrays.
In order to simulate the effect of the trigger, the scintillators,
three or four depending on the configuration, have been modeled as
simple $1\times 1$\,m$^2$ surfaces. Muons are initially extrapolated to
the outer radius of the Tracker, the intersection points with the
scintillator surfaces calculated, and the trigger logic
applied. 
No materials other than those known to be present in the Tracker were
taken into account in the simulation process. 
The $\phi$ and the $\eta$ distributions of the simulated cosmic tracks for the two scintillator 
configurations are shown in Figure~\ref{fig:simu_muon_param}.  
The simulation includes the information of the Tracker region actually connected for 
the readout, this can be seen in Fig.~\ref{fig:simu_cablemap_confC} for the trigger Configuration TC.


Details of the signal and noise simulation implementation in the CMSSW
software are described in \cite{PTDR1} and will be quickly summarized here.  
GEANT4 propagates each track through  the detector
volume. The particle entrance and exit point is recorded for each
sensitive layer, together with the deposited energy. 
During this step, cuts for $\delta$-ray production
are applied (120 keV for the strip Tracker sensitive volumes) tuned to optimize the 
timing of the simulation while minimizing the effects on the description of the
the charge distribution. To digitize the signal collected by each
channel, the charge is multiplied by a gain factor and rounded to the
nearest integer, thus simulating ADC digitization. Signals exceeding
the 8-bit ADC range are assigned the maximum allowed ADC value (256
ADC counts).
The simulation of the response of the detector and the readout chain happens during this 
digitization phase. 
FED zero-suppression can also optionally be simulated.
The code has been developed trying to balance the accuracy of the simulation, the timing
performance, and the quality of the simulation in order to be able to
reproduce position resolution, charge collection, detector efficiency
and occupancy. 
The resulting one-dimensional charge distribution is mapped to the strip
readout geometry and the fraction of charge collected by each readout
channel is determined. 
After this step the simulated data can be treated by the reconstruction software 
in the same way as actual data.

\section{Simulation Tuning}

One of the main goals of the Tracker simulation is to be able to reproduce the detector response 
to real data from p-p collisions as accurately as possible. It is important to identify the quantities
 that are most sensitive to disagreements between real data and
Monte Carlo and can have an impact on the performance of the simulation for 
physics analysis (for instance cluster position resolution, occupancy, track reconstruction 
efficiency and fake rate).  The Sector Test was an opportunity to 
perform a first tuning of the simulation using cosmic muons, keeping in mind the limitations 
due to the fact that these muons do not have the same timing as collision particles and their momentum 
is unknown. 
However, the large statistics of the cosmic muons from the Sector test allow one to verify 
if the granularity available in the simulation parameters is adequate for achieving the 
tuning with real data. 
The quantities that received the most
attention in this study relate to noise and signal clusters: noise is
considered both at the individual channel level and for its effect on
occupancy; whilst typical parameters in studying signal clusters include
cluster width, absolute calibration (electrons to ADC counts), and charge sharing.

\subsection{Signal and Noise}

The noise simulation in the CMS Tracker, as explained in \cite{PTDR1} 
is based on a parametrization of the equivalent noise charge (ENC) as a function of the 
strip length $L$ and corresponds to: 
\begin{equation}
ENC(Peak) = (38.8\pm 2.1)e^-{\rm cm}^{-1} \times L + (414\pm 29)e^-
\end{equation}

\begin{figure}[!b]
\centering
    \includegraphics[height=9cm,angle=90]{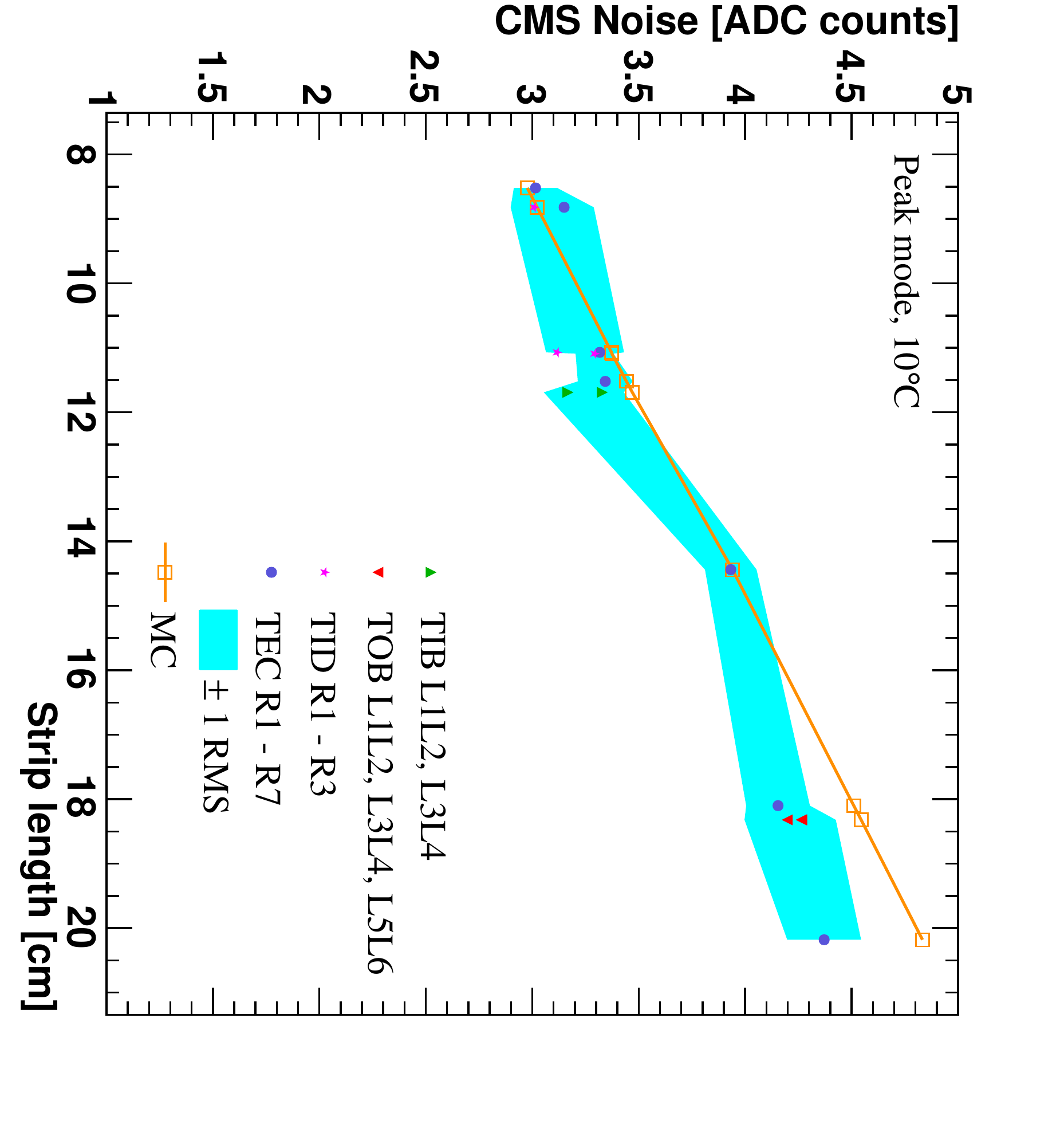}
    \caption{\sl Measured noise in the various sub-detectors as a function
    of strip length compared with the Monte Carlo parameterization. The
    band corresponds to the spread of the mean noise in the
    layer.}  
 \label{fig:noise_data_mc}
\end{figure}

The noise measured with the data, as explained in Section\,4, is compared with the 
parametrization in the simulation in Fig.\,\ref{fig:noise_data_mc}. 
In the real data, effects such as temperature changes, can give rise to significant 
differences of the noise even for sensors of the same length. Other effects can also come into play 
(such as different APV settings) and a slight disagreement is present 
between the real and the simulated noise for the long sensors. 
One learns that an exact parametrization of the noise is not possible. To achieve a perfect simulation of it, 
the rms noise values used in the simulation, will have to be taken strip-by-strip from those measured with real
data.
 
\begin{figure}[hbtp]
\centering
    \includegraphics[height=4.5cm,angle=0]{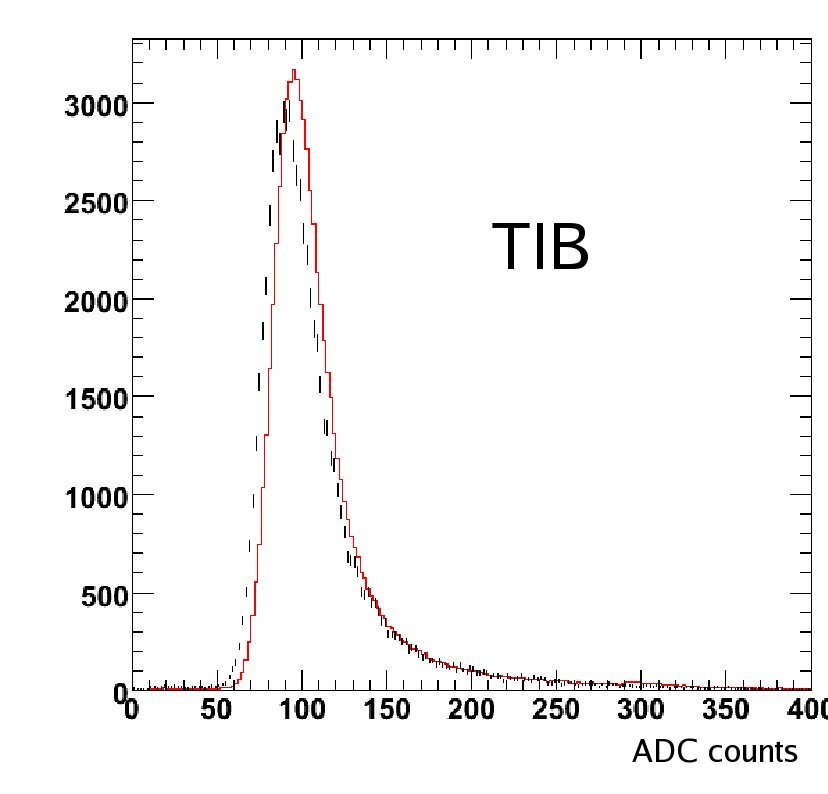}
    \includegraphics[height=4.5cm,angle=0]{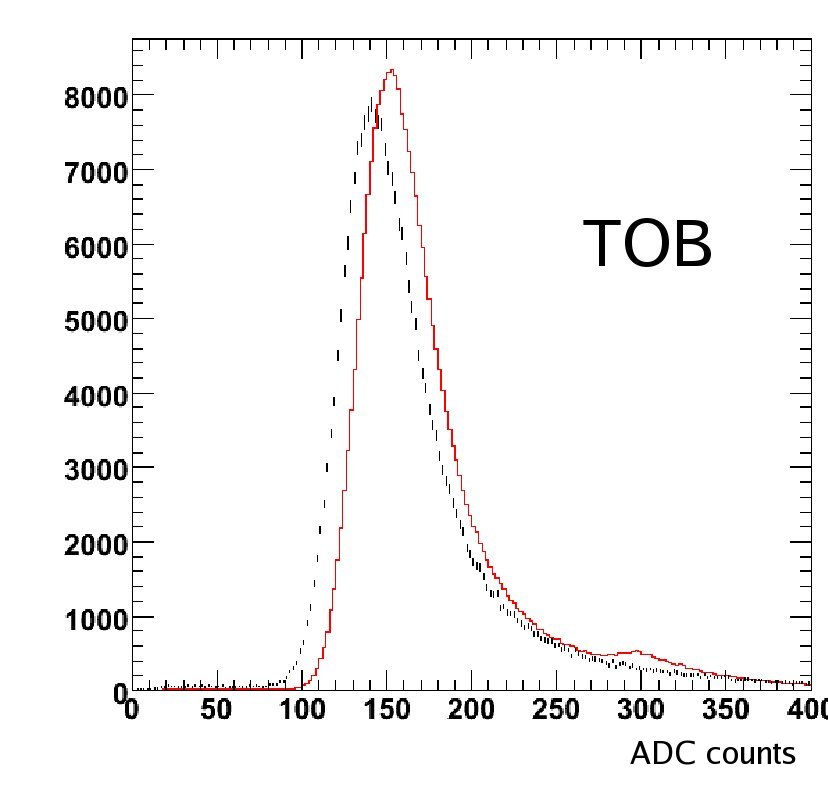}
    \includegraphics[height=4.5cm,angle=0]{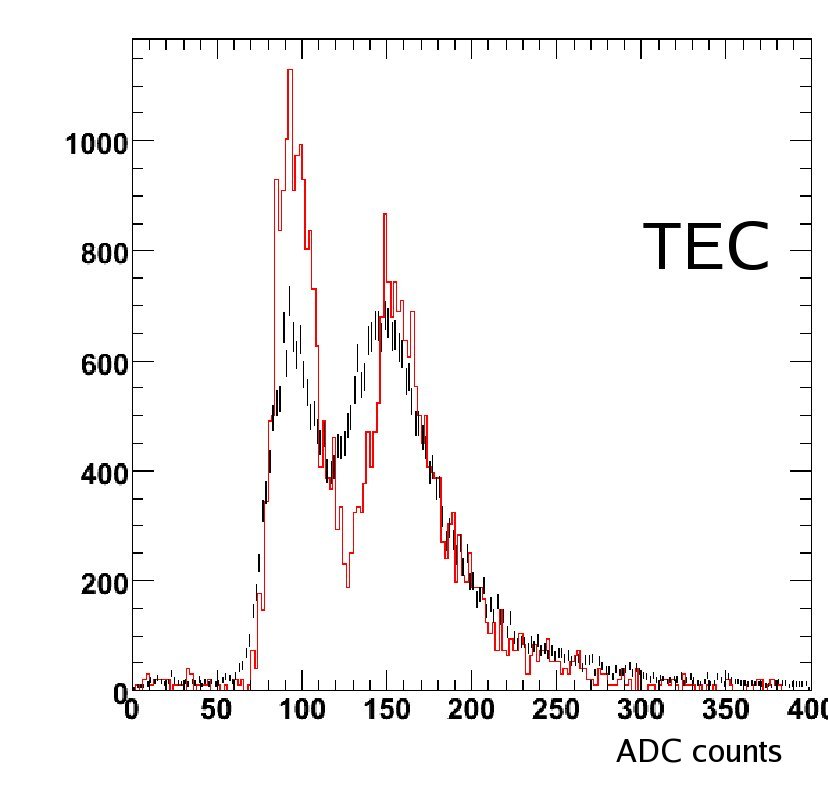}
    \caption{\sl Signal distribution for clusters on tracks in data (dashed line) and Monte Carlo (solid line). From left to right: TIB, TOB and TEC}  
 \label{fig:Landau_data_mc}
\end{figure}

\begin{figure}[hbtp]
\centering
    \includegraphics[height=4.5cm,angle=0]{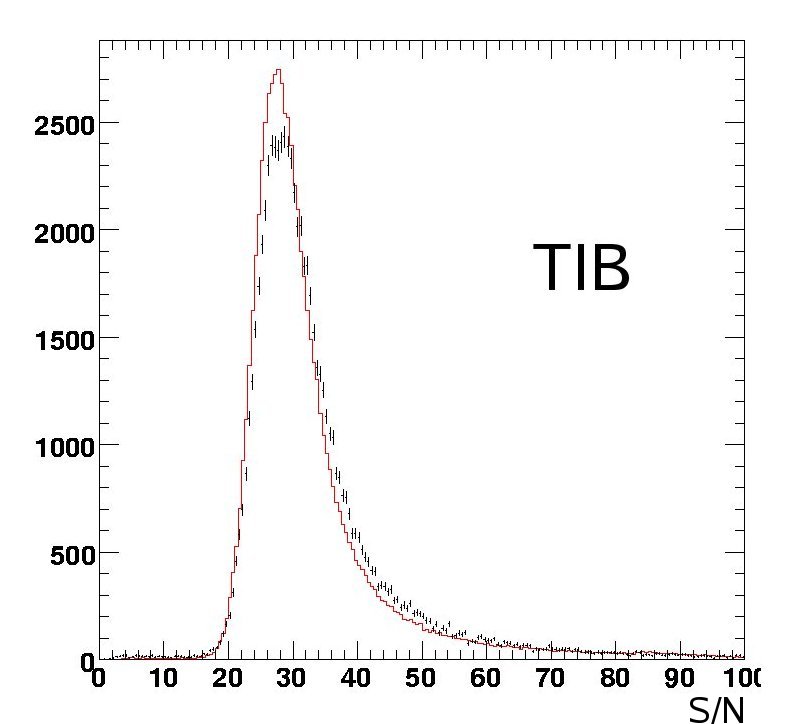}
    \includegraphics[height=4.55cm,angle=0]{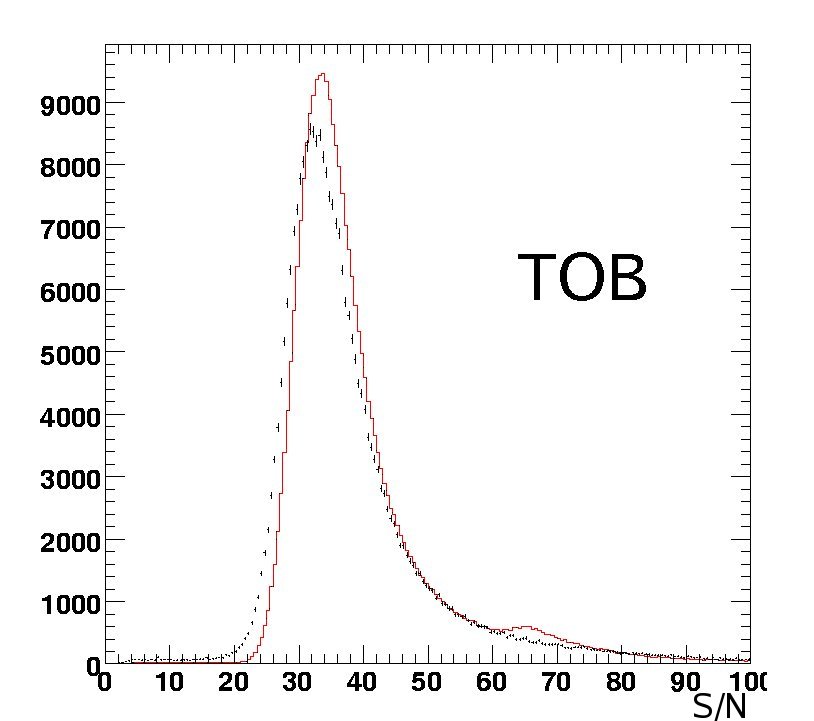}
    \includegraphics[height=4.5cm,angle=0]{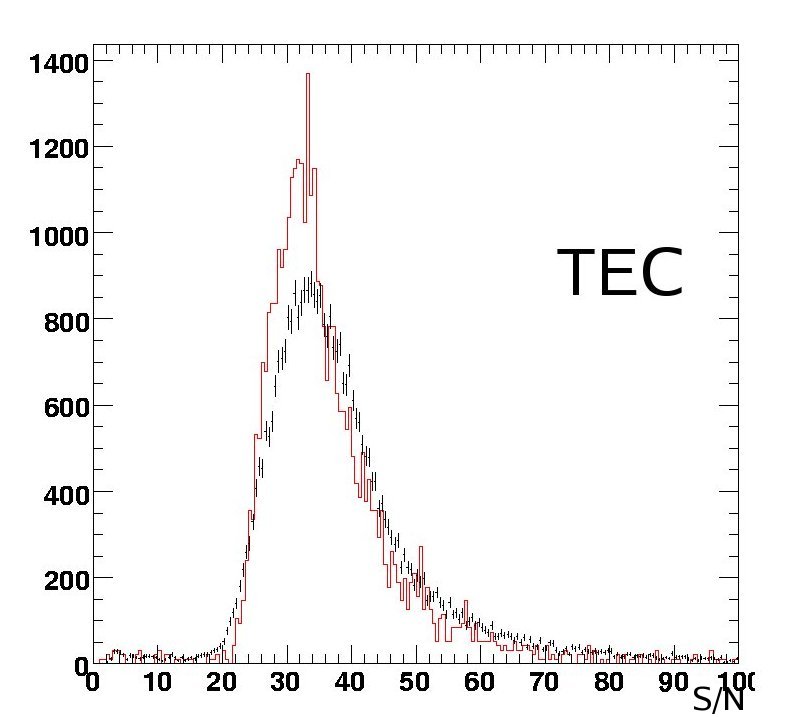}
    \caption{\sl Signal to noise distribution for clusters on tracks in data (dashed line) and Monte Carlo (solid line). From left to right: TIB, TOB and TEC}  
 \label{fig:SN_data_mc}
\end{figure}

For the comparison with the simulation, it should be noted that both the noise and the signal generation is performed in terms
of electrons and then converted in ADC counts using a factor that depends on the absolute calibration of the signal peak in data and 
MonteCarlo (for this analysis $e^-/ADC=250$). 
It is important to emphasize the fact that an absolute calibration is meaningful only for a specific configuration of 
the APV parameters, running temperature etc. From signal studies presented earlier we can expect that an agreement around 
5\%-10\% of the most probable value between data and simulation is satisfactory for our present purposes. 
The signal-to-noise of the Tracker is so large that such a difference in the simulation has nonetheless a  
negligible effect on macroscopic quantities related to physics, such as cluster or track reconstruction efficiency. 
This will not be true once the signal-to-noise drops due to irradiation. 
Hence, the tuning needs to be a continuous effort that follows the changes in the detector operation
 as a function of the integrated luminosity. 
The comparison for the Landau distribution obtained for clusters on tracks normalized to the angle of incidence on the sensor is
shown in Fig.~\ref{fig:Landau_data_mc}. The corresponding comparison for the signal-to-noise distributions is 
shown in Fig.~\ref{fig:SN_data_mc}.
The bump that can be observed in the TOB signal and signal-to-noise distribution in the Monte Carlo is due to the fact that the Monte Carlo data 
are zero suppressed and saturate beyond 255 ADC counts, while the data used for this plot are in VR mode where 
there is no saturation effect until 1023 ADC counts. 
The effect is present also in the TIB and TEC detectors but is simply less visible in the plots. 
The discrepancy observed in the distribution of the signal charge in the case of the TEC detectors is due primarily  to residual differences in the tracks impact angle in data and Monte Carlo and to the crude approximation of having in the simulation a single number to describe the interstrip 
capacitance coupling for the whole Tracker. 

\subsection{Capacitive Coupling}

The capacitive coupling is defined as the fraction of signal charge that
is transferred from a signal strip (crossed by the minimum ionizing
particle) to each of its neighbors.  
An important part of the simulation of the detector response is the
simulation of the capacitive coupling of a strip with its nearest neighbour.
The value is a configurable parameter of the simulation and at the time of this analysis, the default value 
used for the peak mode of operation was 7\%, a preliminary value obtained from previous, 
low statistics, studies with cosmic muons~\cite{MTCC}. 
In order to measure properly the capacitive coupling, 
one needs to disentangle it from other effects of charge sharing that are 
position dependent (such as diffusion, track inclination, Lorentz angle). 
A response function that can be used to extract the value of the capacitive coupling has been 
devised, that applies to the cases where the other charge sharing effects are minimized, 
that is for perpendicular tracks ($XZangle \le 0.1$\,rad): 

$$\eta(symm)  = \frac{Q_{left}+Q_{right}}{2 \cdot Q_{seed}}$$

where $Q_{seed}$ is the charge of the highest strip in the cluster and $Q_{left}$ and $Q_{right}$ the charges of its neighbors. 

The measurement is performed on the Virgin Raw datasets that contain the information of the charge 
(positive or negative) for all the strips after pedestal subtraction.
This allows one to define the eta functions also for single strip
clusters using the charge of neighboring strips, even when below threshold. 
Clusters containing up to three strips have been considered for this analysis.  
This distribution is well modeled by a Gaussian plus a tail at positive value. 
The tail is due to the residual charge sharing and the width is due to the noise of the side strips.
The $\eta$ function is correlated in an analytical way to the value of 
the coupling: 
$$CC = \frac{\eta}{1+2\cdot \eta}$$

\begin{figure}[hbtp]
\centering
\begin{minipage}{0.4\textwidth}
    \includegraphics[width=0.8\textwidth, angle=90]{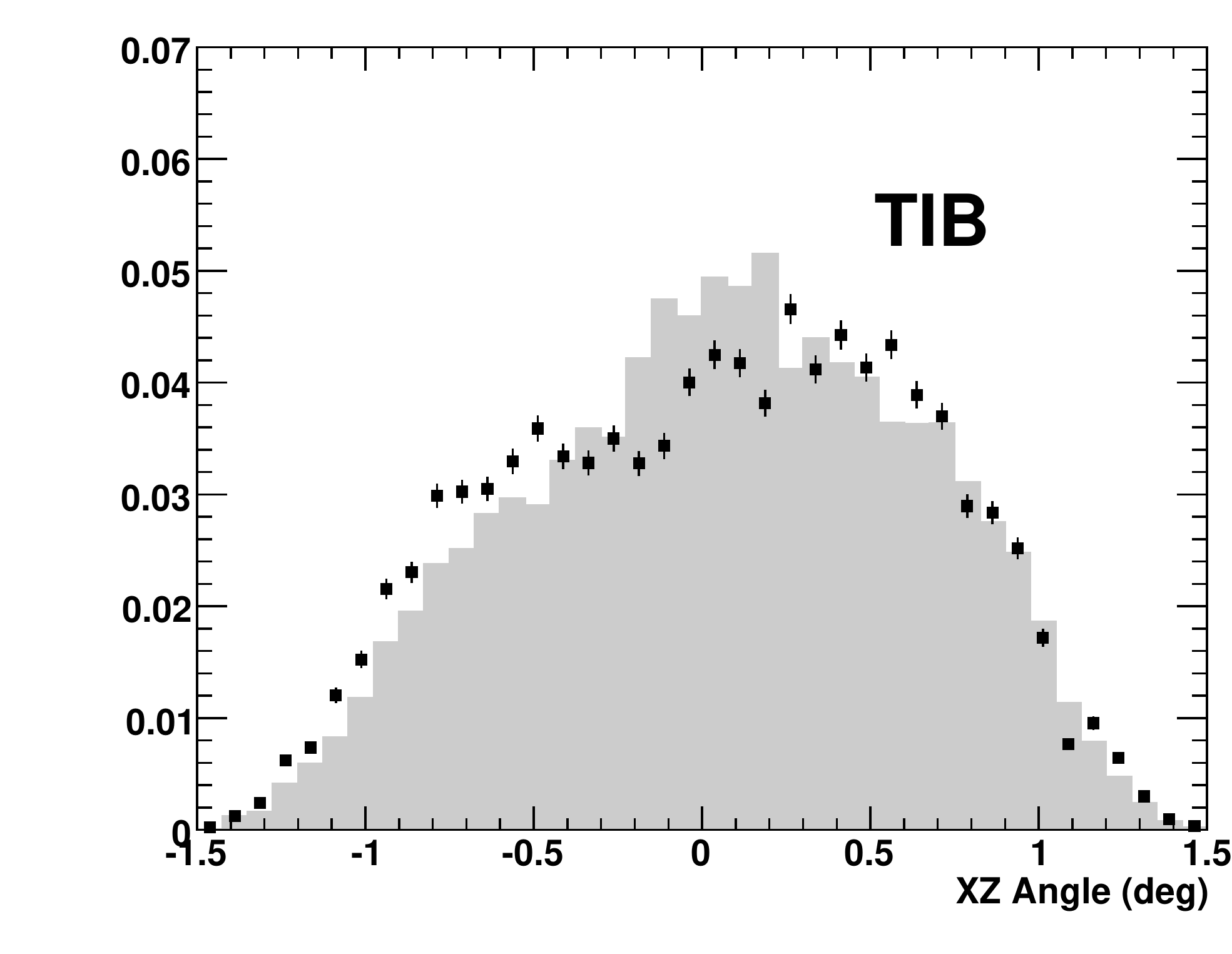}
\end{minipage}
\begin{minipage}{0.4\textwidth}
    \includegraphics[width=0.8\textwidth, angle=90]{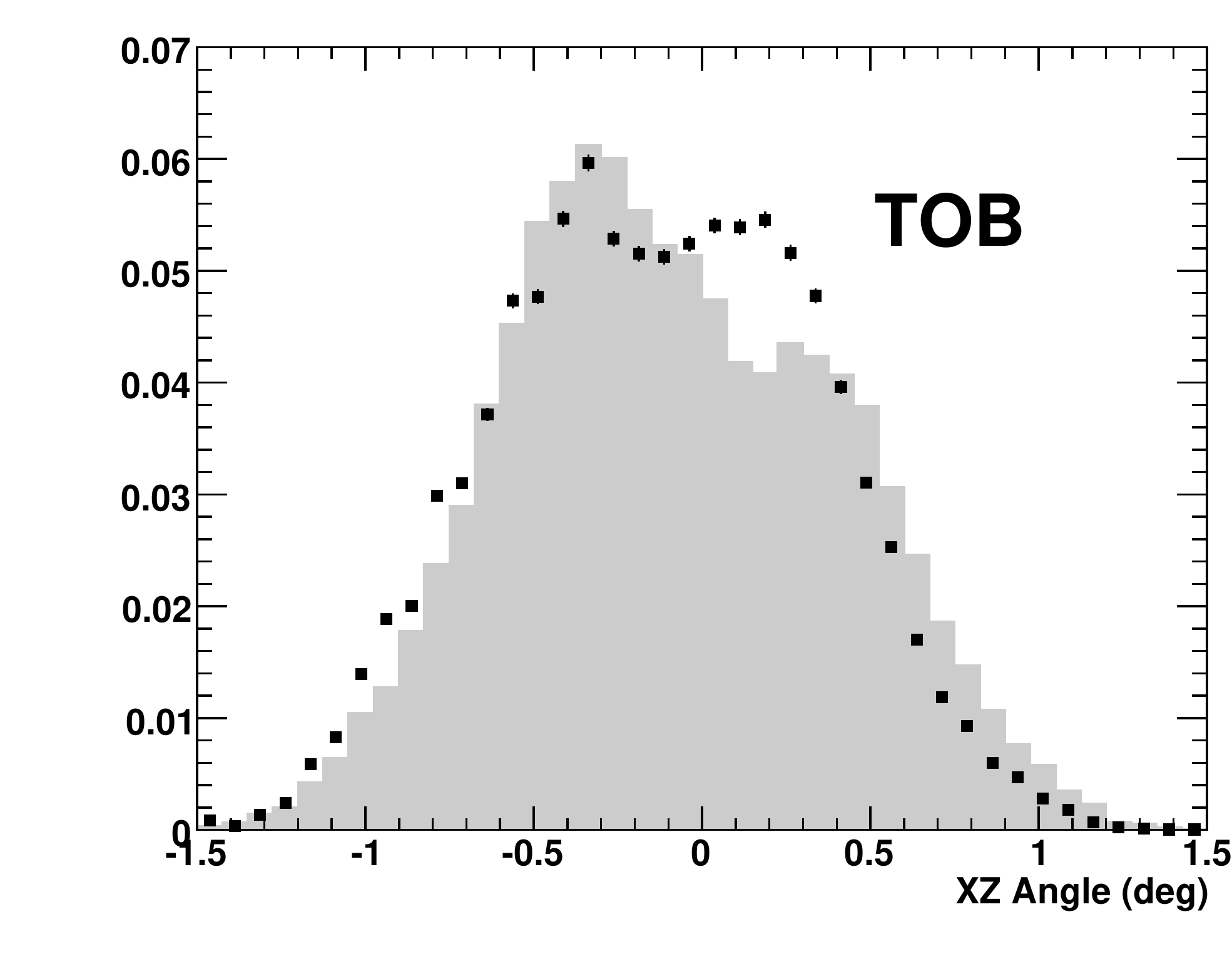}
\end{minipage}
    \caption{\sl Local angle of incidence (XZ) for tracks on sensors in data (points) and MC (histogram) for TIB(left) and TOB(right)}
    \label{fig:loc_angles}
\end{figure}

\begin{figure}[h]
\begin{minipage}{0.5\textwidth}
\centering
    \includegraphics[height=7cm,angle=90]{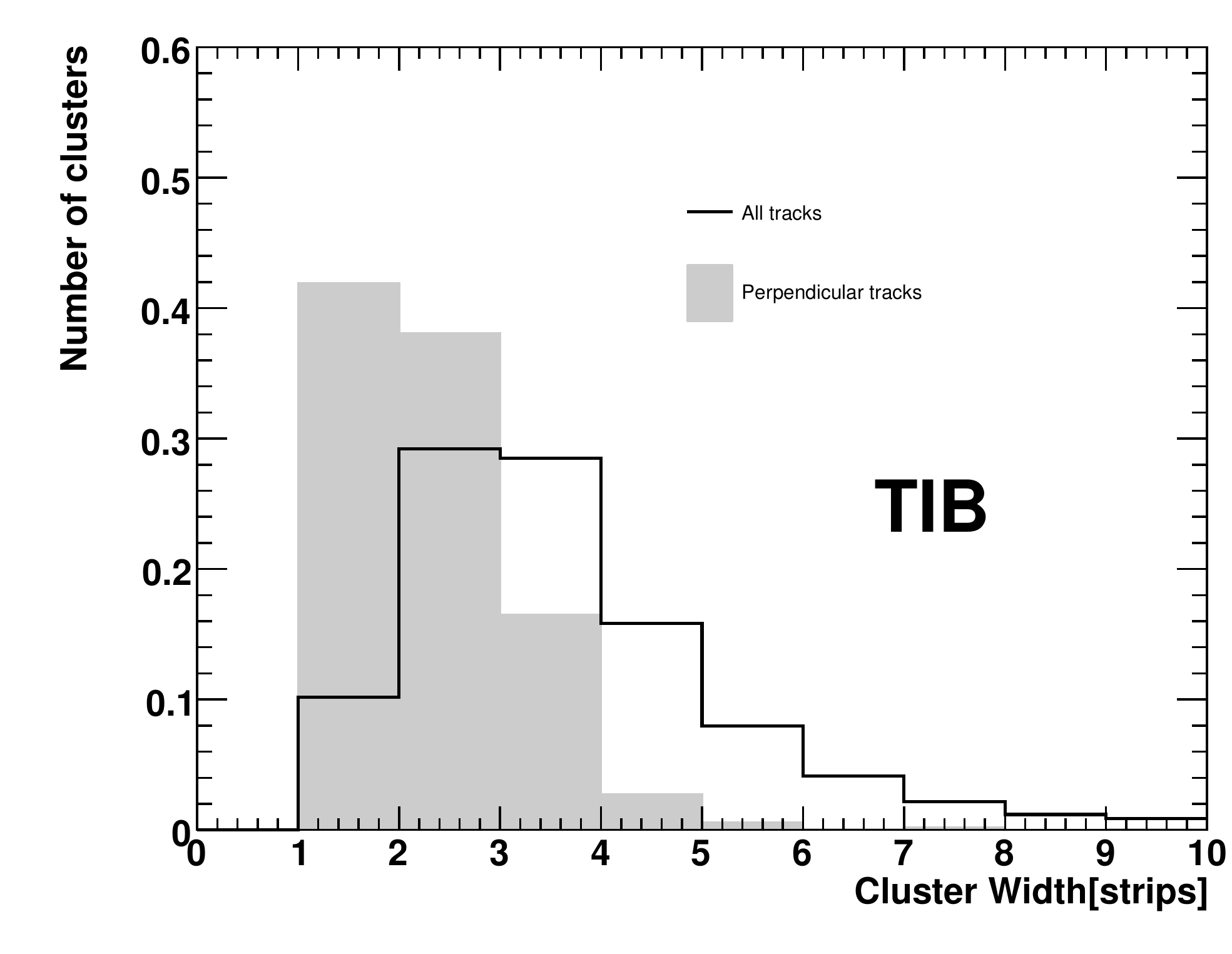}
\end{minipage}
\begin{minipage}{0.5\textwidth}
\centering 
   \includegraphics[height=7cm, angle=90]{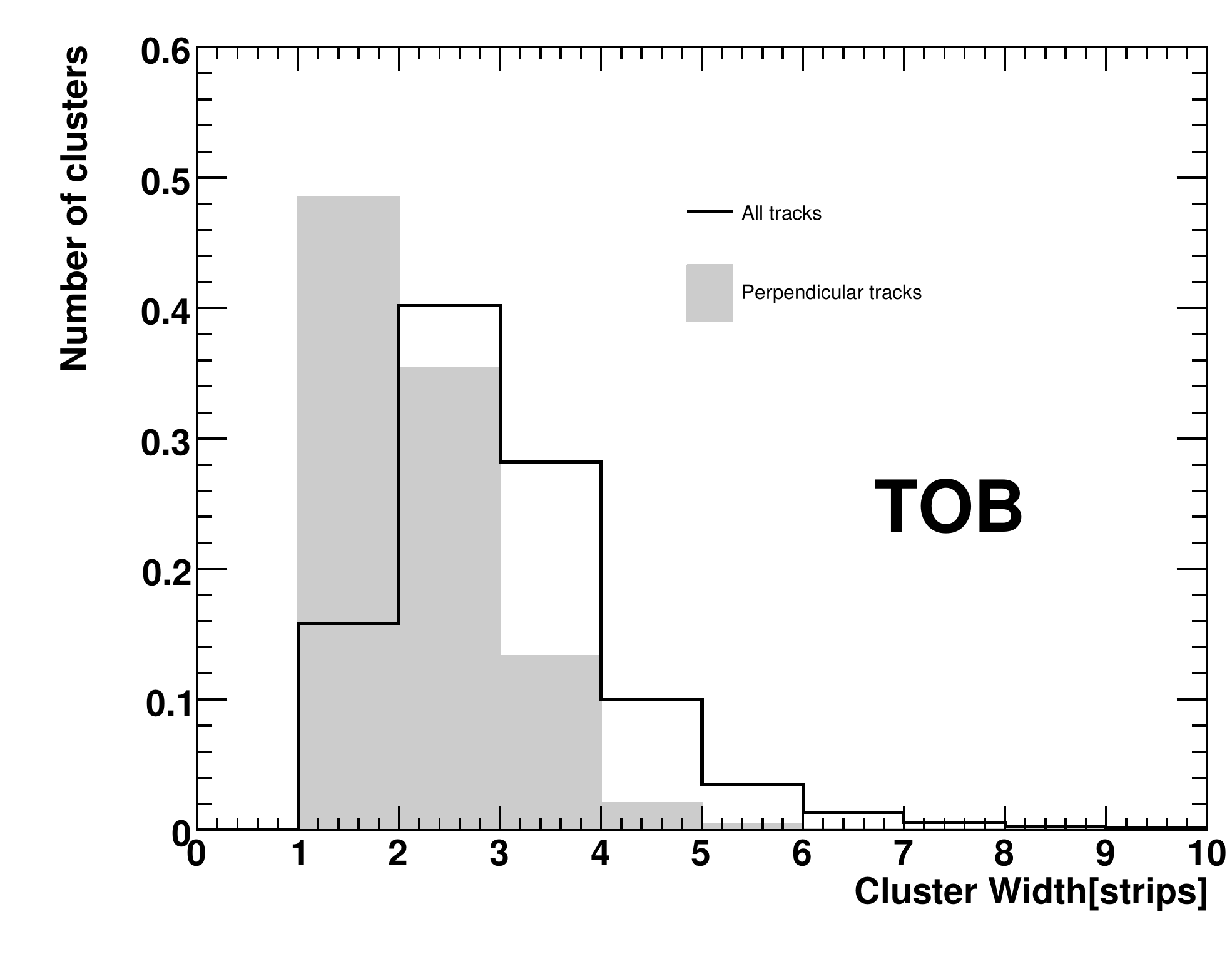}
\end{minipage}
\caption{\sl Cluster width for TIB(left) and TOB(right) hits for all tracks and perpendicular tracks} 
    \label{fig:cluster_size}
\end{figure}

Fig.~\ref{fig:loc_angles} shows the 
distribution of the ``XZ'' impact angle (defined in Section~5) between a track and a sensor in data and Monte Carlo 
(in Configuration TC). The difference in the angular distribution is due to the  a shift in the position of the scintillator 
triggers between simulation and data. 
The distribution for the cluster width in data before and after the 
perpendicularity requirement is shown in Figure~\ref{fig:cluster_size}. 

\begin{figure}[p]
\begin{minipage}{0.5\textwidth}
\includegraphics[height=7cm]{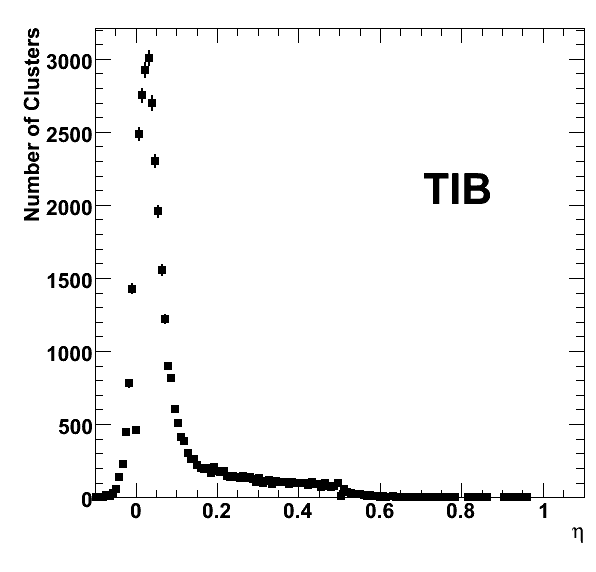}
\end{minipage}
\begin{minipage}{0.5\textwidth}
   \includegraphics[height=7cm]{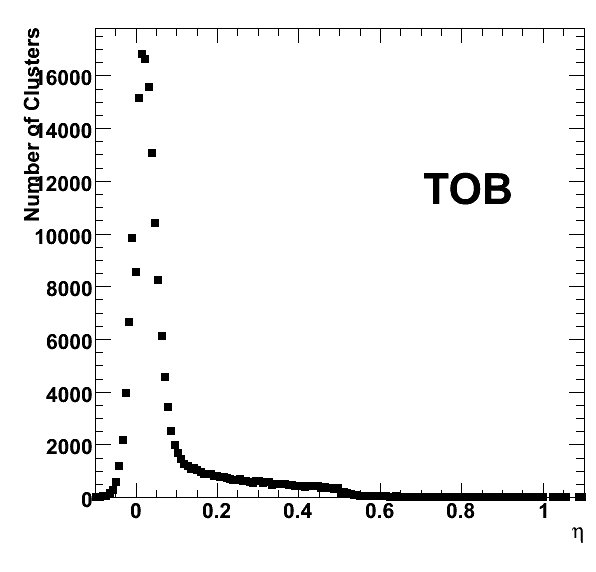}
\end{minipage}
    \caption{\sl $\eta$ function for TIB(left) and TOB(right) clusters. 
The data are from Virgin Raw runs where the information of the strips below the clustering threshold is
    available.}
    \label{fig:eta_TIB_VR}
\end{figure}

\begin{figure}[p]
\begin{minipage}{0.5\textwidth}
    \includegraphics[height=7cm,width=7cm]{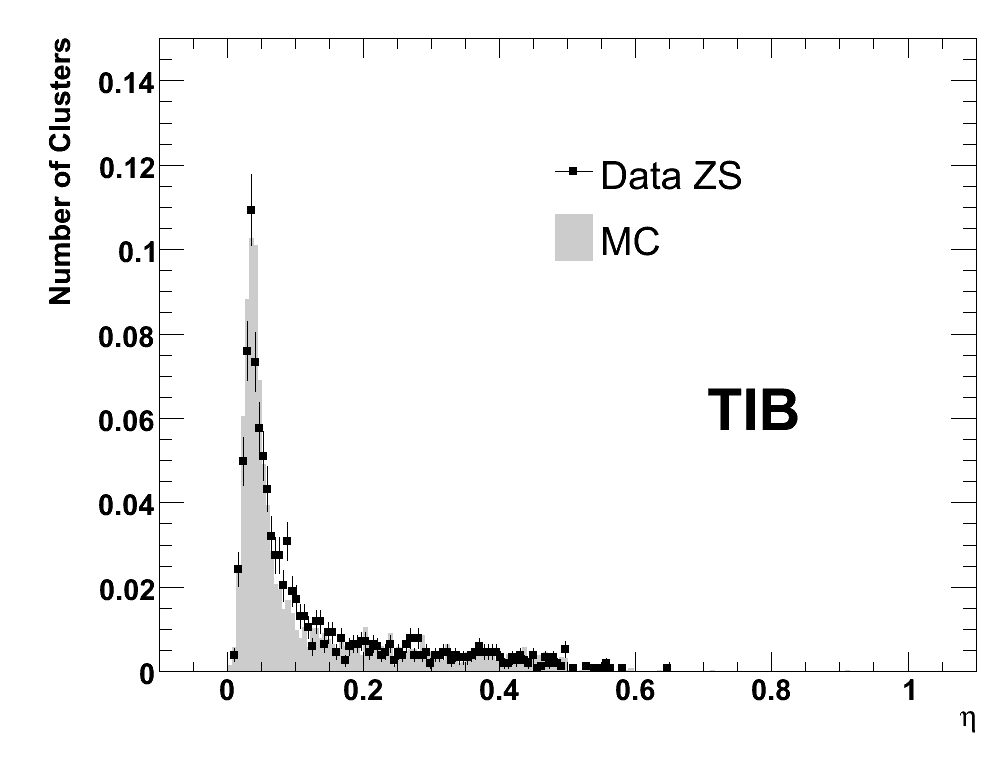}
\end{minipage}
\begin{minipage}{0.5\textwidth}
    \includegraphics[height=7cm,width=7cm]{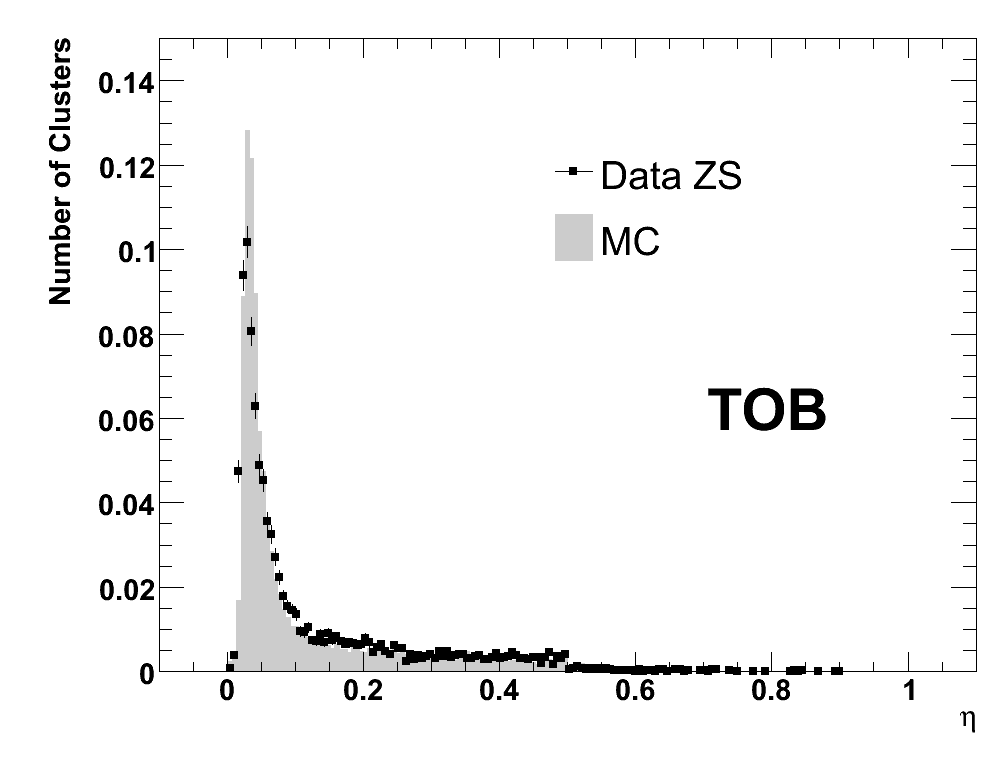}
\end{minipage}
    \caption{\sl $\eta$ function for TIB(left) and TOB(right) clusters
The data are from Zero Suppressed runs compared with Monte Carlo with 3\% capacitive
    coupling.}
    \label{fig:eta_TIB_ZS}
\end{figure}

The distributions for the symmetric $\eta$ function obtained from the Virgin Raw data are
shown in Fig.~\ref{fig:eta_TIB_VR} for the
TIB and TOB respectively.
From the position of the peak of the distribution of the coupling estimator
a preferred value of 3.0\% for the capacitive coupling is obtained. 
Given the strong dependence of the capacitive coupling on trigger timing shown by previous studies~\cite{CRACK}, 
and the non-standard timing configuration in the Sector Test, 
the results from the cosmic data do not allow to extract a precise measurement.
Nonetheless, applying the best value of the capacitive coupling to the Monte Carlo the 
precision of the tuning can be seen comparing simulation to real data taken in ZS mode,  
see Fig.~\ref{fig:eta_TIB_ZS}.



\begin{figure}[h]
\begin{minipage}{0.5\textwidth}
\centering
    \includegraphics[height=7cm,angle=90]{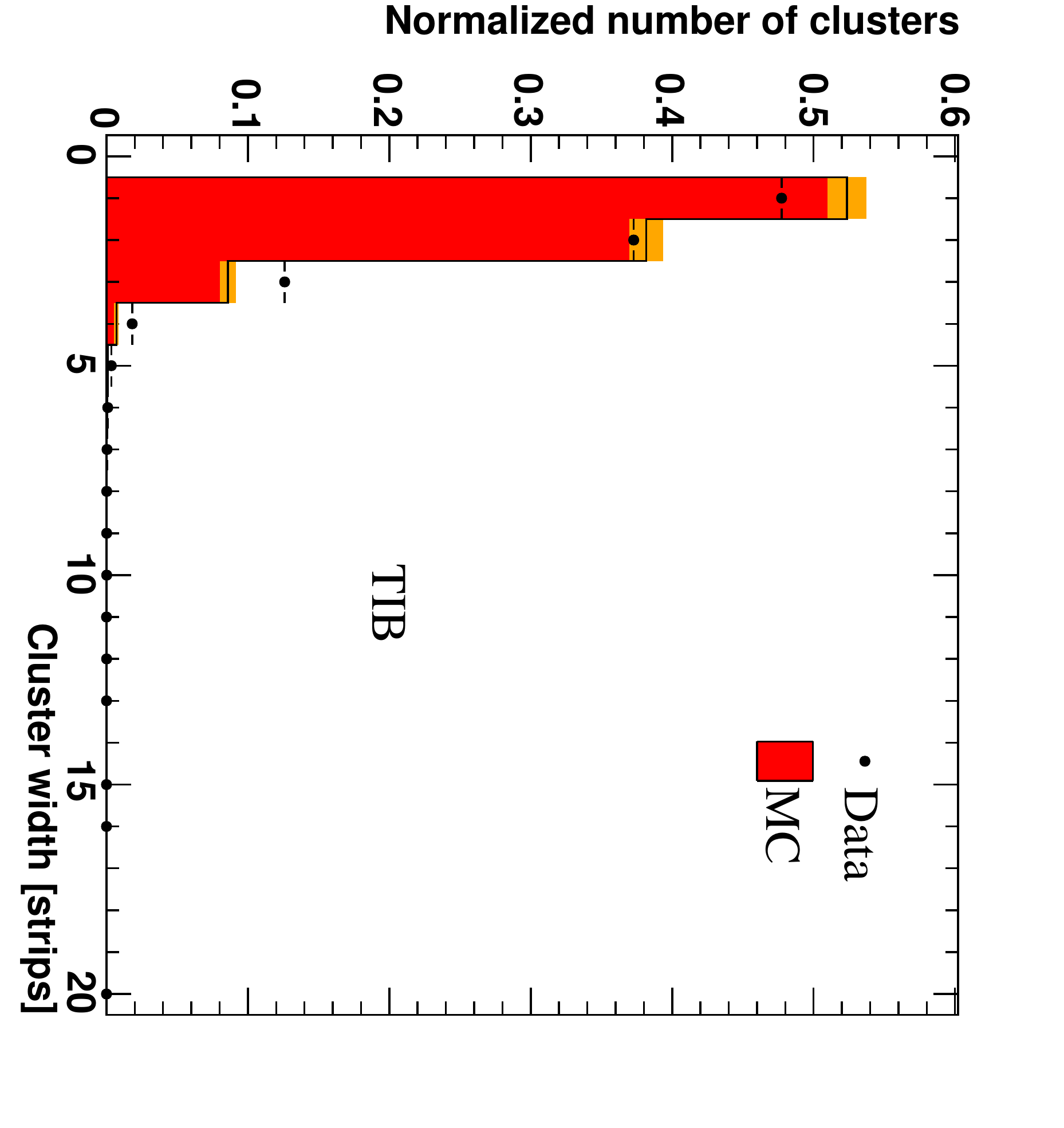}
\end{minipage}
\begin{minipage}{0.5\textwidth}
\centering 
   \includegraphics[height=7cm, angle=90]{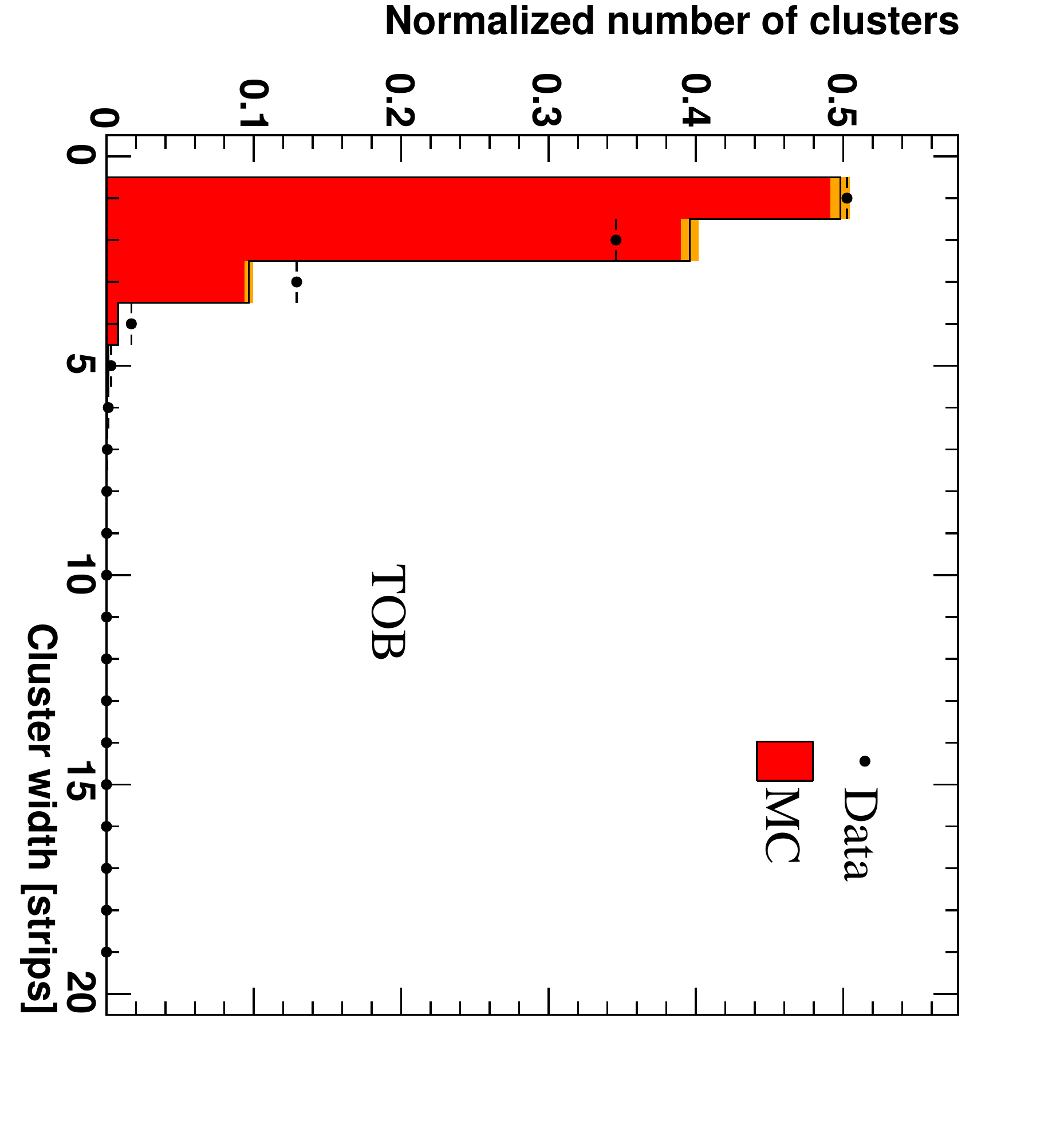}
\end{minipage}
\caption{\sl Cluster width for TIB(left) and TOB(right) hits on
perpendicular tracks. The black points represent the data, the 
histogram is obtained from simulation with a 3\% capacitive coupling where the gray
band represents the statistical uncertainty.}
    \label{fig:clusize_cut}
\end{figure}

The cluster width is strongly affected by the change in the capacitive coupling constant. Figure~\ref{fig:clusize_cut} 
shows the excellent agreement for the cluster width for perpendicular tracks in the data, after the tuning of the capacitive 
coupling in the Monte Carlo.

\subsection{Cluster width studies}

\begin{figure}[hbtp]
\centering
    \includegraphics[width=0.6\textwidth]{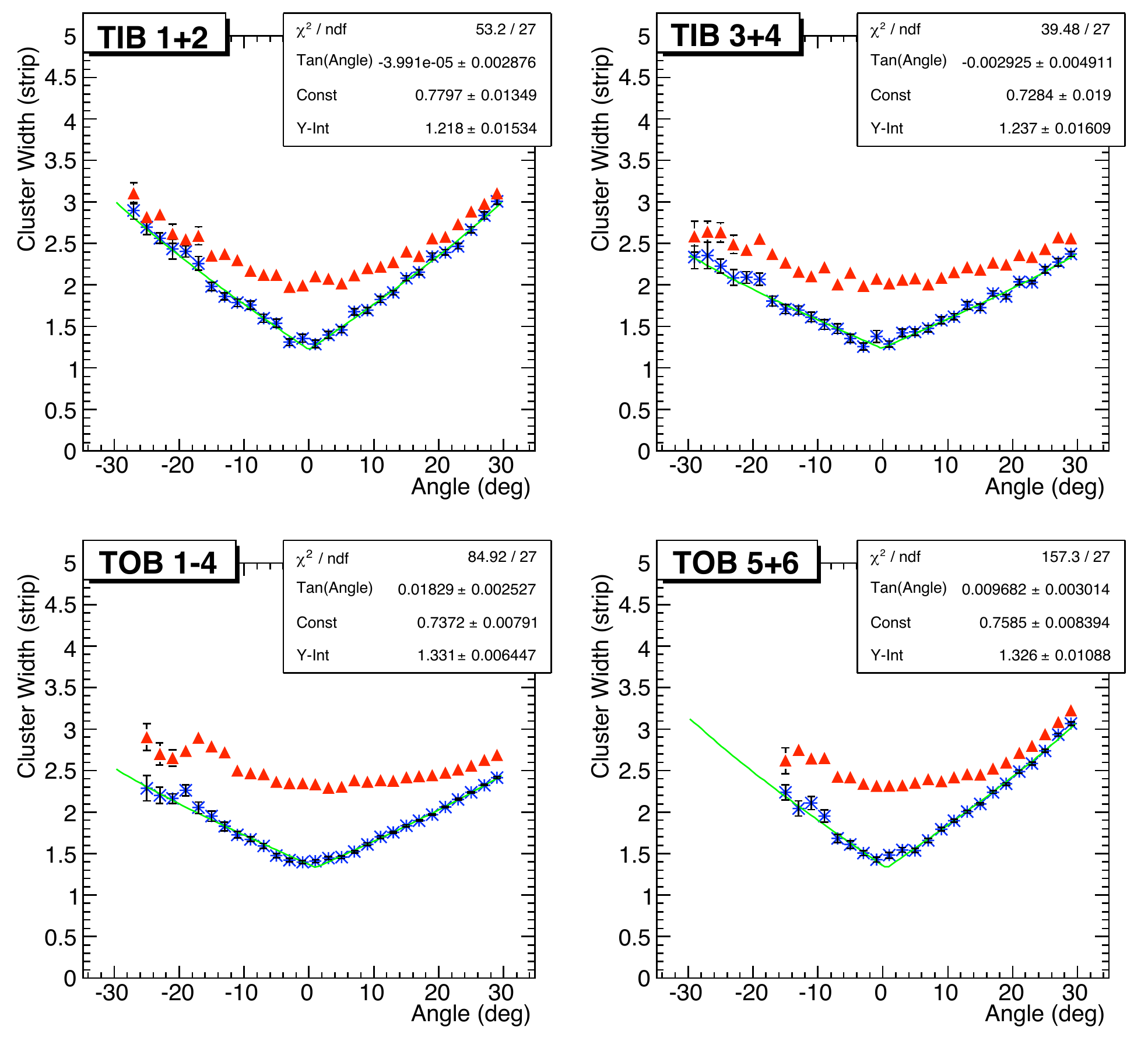}
    \caption{\sl Cluster width versus $\theta_t$ for barrel layers on
    Monte Carlo simulated data using capacitive coupling of 7\%(triangles) and
    3\% (crosses).} 
 \label{fig:LA_MC_CC}
\end{figure}

In the presence of a magnetic field the diffusion path of charge
carriers inside the bulk silicon will be influenced by the Lorentz
force, which modifies the shape of the charge distribution of the
clusters. The net effect is a shift in cluster position (relative to
track intersection points) and a change of the cluster width.  The
shift and widening are largely due to charge drifting across cell
boundaries, but the capacitive coupling between neighboring channels
also has an influence on the final position and cluster width. In the
case of the Sector Test, where there was no magnetic field, it is
still interesting to evaluate the behavior of the cluster width as a function of
the track incident angle perpendicular to the strip
direction (angle XZ described in Fig.\ref{fig:local_angles} in Section\,5) and compare the data and Monte Carlo
distributions. Figure~\ref{fig:LA_MC_CC} shows the effect on cluster
width of varying the impact angle when the capacitive coupling is
changed from 7\% to 3\% in the simulation. 

Since the value of the Lorentz angle might change as the silicon is 
irradiated a calibration strategy has been developed~\cite{LorentzAngle}
which will determine and monitor its value from data
and write it in the database, so that it can be retrieved by the offline analysis code,
to correct cluster positions. TIF datasets have been used to validate and 
test this software with real data for the first time. 

\begin{figure}[hbtp]
\centering
    \includegraphics[width=0.7\textwidth]{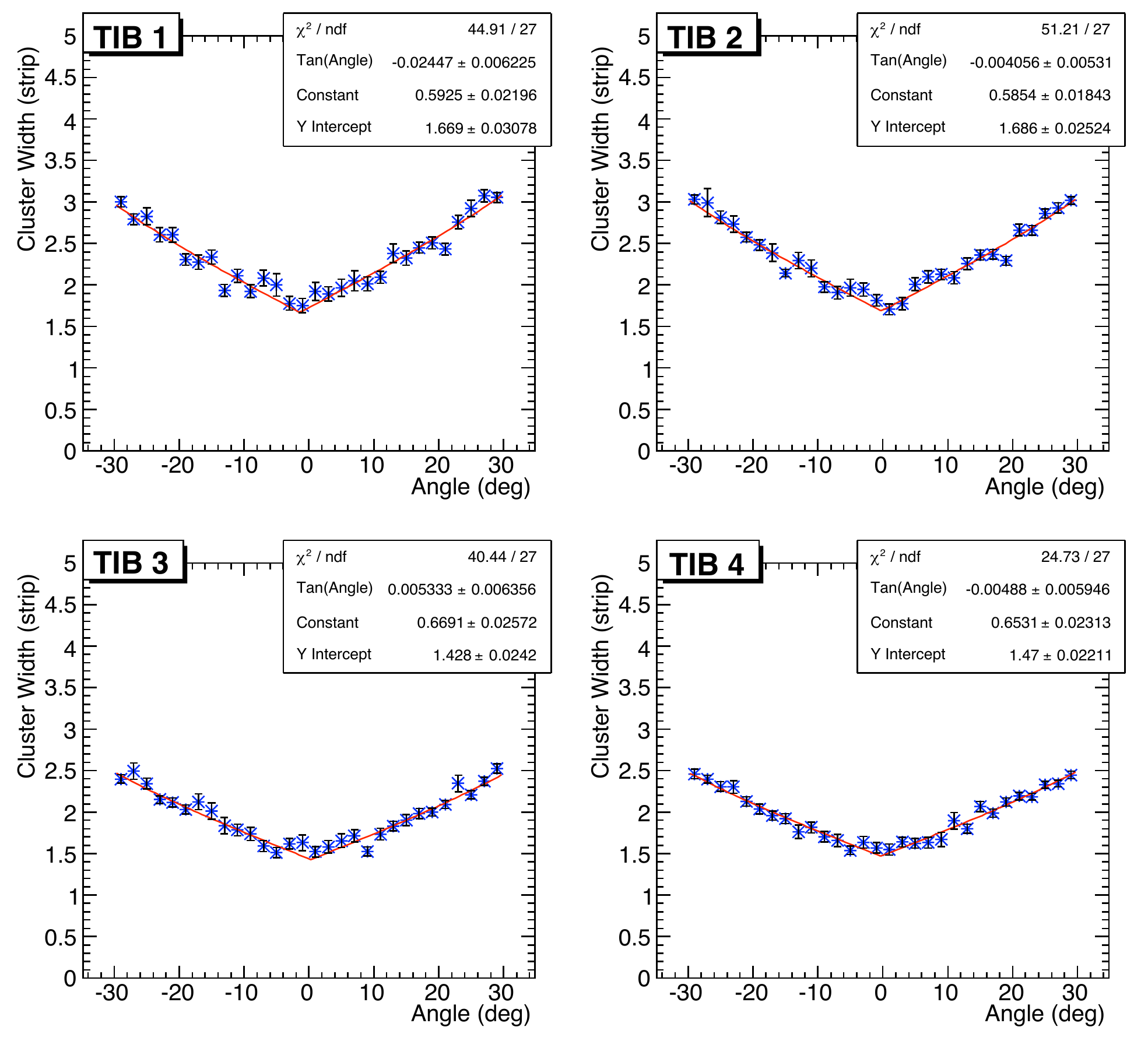}
    \caption{\sl Cluster width versus $\theta_t$ for TIB layers.} 
    \label{fig:LA_TIB_layer}
\end{figure}

\begin{figure}[hbtp]
\centering
    \includegraphics[width=0.7\textwidth]{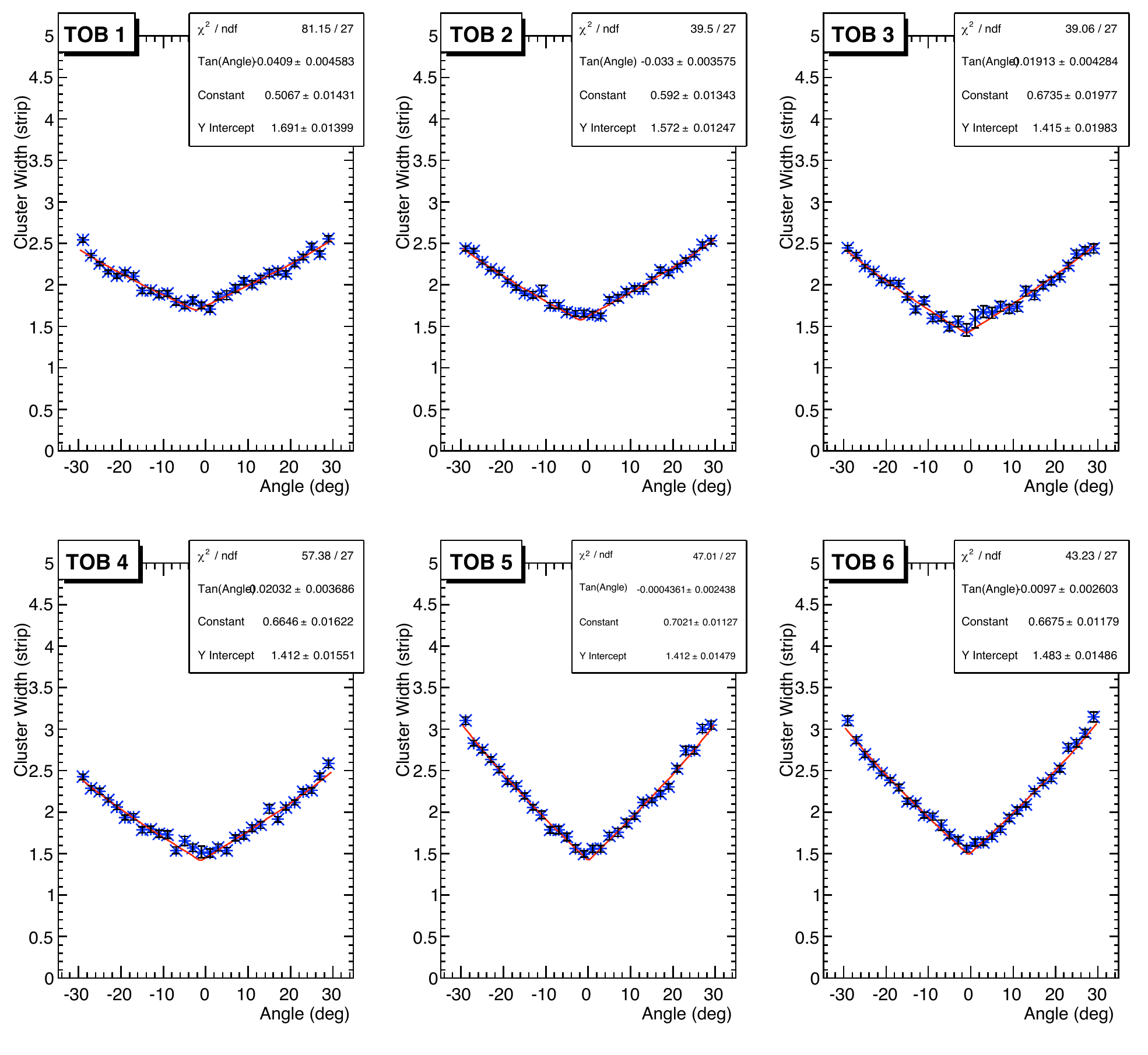}
    \caption{\sl Cluster width versus $\theta_t$ for TOB layers.} 
    \label{fig:LA_TOB_layer}
\end{figure}

\begin{figure}[hbtp]
\centering
    \includegraphics[width=0.7\textwidth]{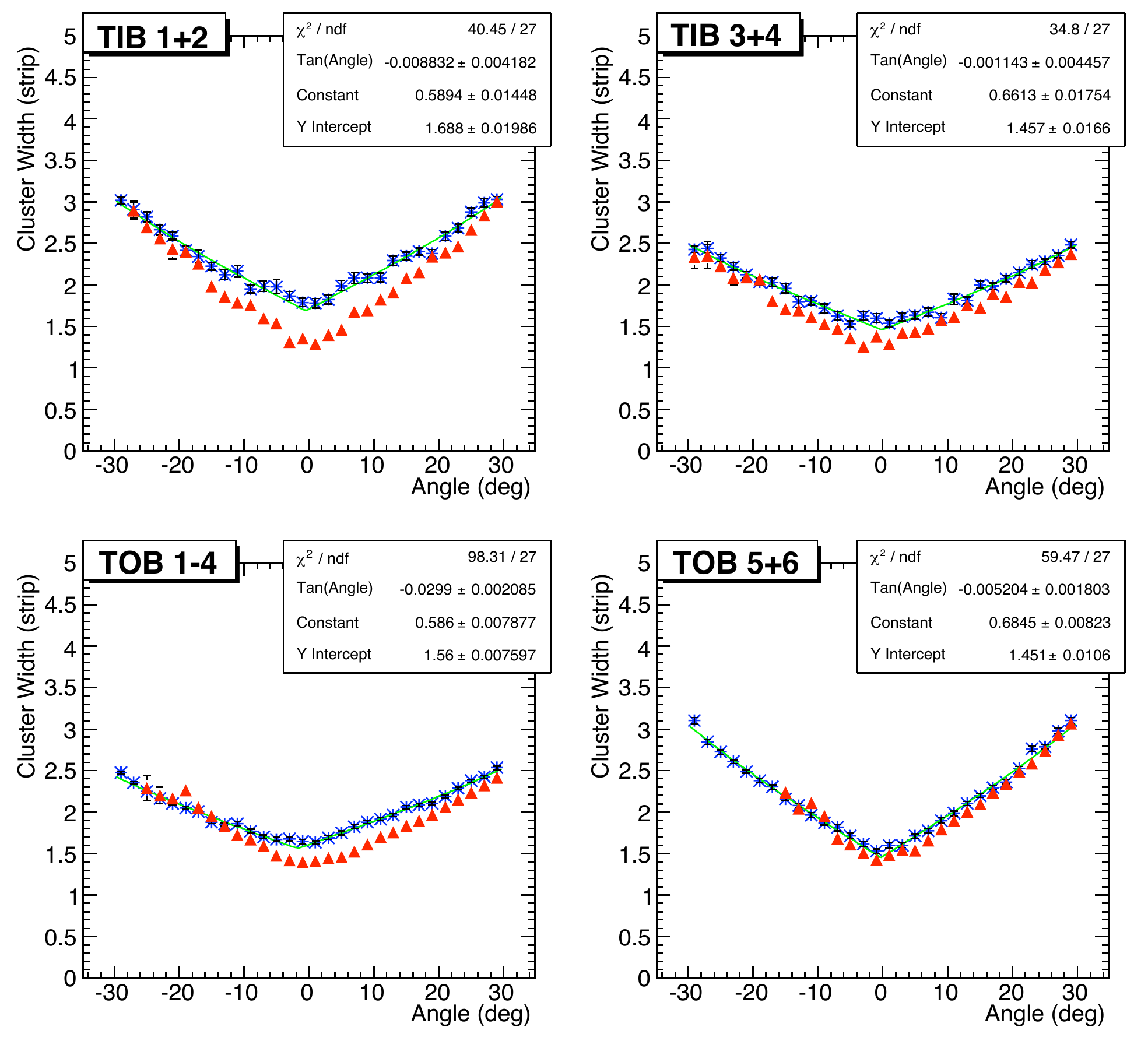}
    \caption{\sl Cluster width versus $\theta_t$ for barrel layers on
real data (crosses) compared with Monte Carlo (triangles) using capacitive coupling of 3\%.} 
 \label{fig:LA_data_MC_CC}
\end{figure}

The average cluster width for tracks
incident with an angle $\theta_t$ with respect to the detector normal
is given by:
 
$$ <cluster\, width\,> = a+| \frac{t}{p}\cdot b\cdot (\tan\theta_t - \tan\Theta_L)|$$ 

where $t$ is the detector thickness, $p$ is the pitch, $a$ and $b$ are
coefficients expressing the carrier diffusion and the
electronic capacitive coupling between nearby channels, and $\Theta_L$
is the Lorentz angle. Therefore, plots of mean cluster width as
a function of $\tan(\theta_t)$ are produced for each layer and the
Lorentz angle is obtained fitting them with the function above. 
In this fit the parameters are: the estimate of the Lorentz angle ($\tan\Theta_L$), 
the slope normalized to the ratio of thickness over pitch (that should be constant), and the average cluster width
at the minimum. 
Of course in the TIF setup there is no magnetic field and the Lorentz angle determination 
should return zero. The plots for each layer of the barrel
detectors are shown in Figs.~\ref{fig:LA_TIB_layer} and
\ref{fig:LA_TOB_layer}. The result from the data, grouped by layer with similar characteristics, 
has been overlaied by Monte Carlo result with a capacitive coupling of 3\% in Fig.\ref{fig:LA_data_MC_CC}. 
The fitted Lorentz angles are consistent with zero as expected inside the statistical uncertainties, 
and with a disagreement of 2\% at the most in some cases likely due to systematic misalignment effects.
The comparison between distributions in real and simulated data indicates also that the granularity 
of the capacitive coupling constant in the simulation should be increased from the single value present now, 
to at the level of a parameter for each layer or even for each different type of module.  


\section{Conclusions}

The CMS Silicon Strip Tracker construction was completed in March 2007 at the CERN Tracker Integration Facility. 

During the Sector Test, which spanned a period of about four months, the Tracker was operated in a stable and continuous manner for 
periods up to few consecutive weeks. Safe operations were established while running in different configurations and over 
a wide range of temperatures, from $15\,^{\circ}\mathrm{C}$ to $-15\,^{\circ}\mathrm{C}$. The 4.5~million cosmic rays events collected demonstrate the successful operation of the entire electronic chain of data acquisition system and the commissioning of such a large fraction of the detector for the first time.

The results of the data analyses here described show the excellent detector performance. 
The behavior of the noise and signal-to-noise has been very stable for all sub-detectors, better than 0.6\%.
In particular, the signal-to-noise in peak mode is larger than 26 for all layers or rings and the hit reconstruction efficiency is above 99.8\%. The fraction of bad strips is 0.2\% and of bad components is 0.6\%.

Comparing the data with the expectations from the simulation for signal, noise, cluster width and capacitive coupling has allowed a preliminary tuning of the simulation input parameters. 
The agreement between data and simulation is at a level to allow large production of datasets to be used for early physics studies to be carried out with the first collision data in 2008. 


The sector test has been a very important milestone achieved by the Silicon Strip Tracker and can be considered one of the best reference points for operating the full detector and understand its performance even after completion of the CMS experiment in 2008. 

\section*{Acknowledgments}
We thank the administrative staff at CERN and other Tracker
Institutes.  
This work was supported by: 
the Austrian Federal Ministry of Science and Research;
the Belgium Fonds de la Recherche Scientifique and Fonds voor Wetenschappelijk Onderzoek; 
the Academy of Finland and Helsinki Institute of Physics; 
the Institut National de Physique Nucl\'eaire et de Physique des
    Particules~/~CNRS, France; 
the Bundesministerium f\"ur Bildungund Forschung, Germany;
the Istituto Nazionale di Fisica Nucleare, Italy; 
the Swiss Funding Agencies;
the Science and Technology Facilities Council, UK; 
the US Department of Energy, and National Science Foundation.
Individuals have received support from the Marie-Curie IEF program (European Union) and the A.P.\,Sloan Foundation.


\end{document}